\documentclass[twocolumn]{aastex701}

\usepackage{amsmath}
\usepackage{hyperref}
\usepackage{helvet}
\usepackage{mathpazo}
\usepackage{listings}
\usepackage{multirow}
\usepackage{array}
\usepackage{float}
\usepackage{siunitx}
\usepackage{xparse}
\usepackage{etoolbox}
\usepackage{isotope}
\usepackage{url}
\usepackage{graphicx}
\usepackage{bm}
\usepackage{xcolor}

\newlength\figureheight 
\newlength\figurewidth 
\newcommand{\msun}{\ensuremath{\mathrm{M_{\odot}}}}
\newcommand{\e}[1]{\times 10^{#1}}

\DeclareSIUnit[number-unit-product = \;]
\solarmass{\msun}

\DeclareSIUnit[number-unit-product = \;]
\dyn{dyn}

\DeclareSIUnit[number-unit-product = \;]
\ergy{erg}

\hypersetup{
    colorlinks,
    citecolor=blue,
    filecolor=black,
    linkcolor=blue,
    urlcolor=black
}

\newcommand{\Cfrac}{X_\mathrm{C}}

\newcommand{\Machfuel}{\mathcal{M}_\mathrm{f}}

\newcommand{\urho}{\mathrm{g\,cm^{-3}}}

\newcommand{\FLASH}{{\sc FLASH}}
\newcommand{\MATLAB}{{\sc MATLAB}}

\newcommand{\Proteus}{{\sc Proteus}}

\newcommand{\SciPy}{{\sc SciPy}}
\shorttitle{Fuel Preheat in SN Ia}
\shortauthors{Learn, Zhiglo \& Plewa}

\begin{document}

\title{Mesoscale Turbulence in Type Ia Supernova Deflagrations II.
Evidence for Fuel Preheat Due to Flame Thermal Expansion from Lagrangian Analysis}

\correspondingauthor{Tomasz Plewa}

\author[0000-0002-2083-0453]{Ryan Learn}
\affiliation{Department of Mathematics, Washington State University Tri-Cities,
             Richland, WA 99354, U.S.A.}
\email{ryan.learn@wsu.edu}

\author[0000-0003-0416-5796]{Andrey Zhiglo}
\affiliation{Kharkiv Institute of Physics and Technology, Kharkiv, 61108, Ukraine, on leave from}
\email{azhiglo@gmail.com}

\author[0000-0002-1762-2565]{Tomasz Plewa}
\affiliation{Department of Scientific Computing, Florida State University,
             Tallahassee, FL 32306, U.S.A.}
\email{tplewa@fsu.edu}

%
%
%
%
%
\begin{abstract}
Thermonuclear flame fronts in Type Ia supernovae develop hydrodynamic instabilities that generate turbulent mesoscale flow and substantial local thermodynamic variability. Recent work by \citet{Brooker+25} demonstrated systematic shortening of fuel ignition times near Rayleigh--Taylor unstable deflagration fronts, but the physical origin remained unclear.
We investigate the thermodynamic evolution of fuel near the flame front using multidimensional simulations and Lagrangian tracer-particle analysis, focusing on flame-driven expansion and ignition-time variability in partially degenerate carbon--oxygen plasma.
Tracer trajectories show that the most strongly preconditioned fuel parcels near the flame front, which form a minority of the fuel layer, undergo approximately pressure-constrained expansion with modest density reductions and temperature increases of $\approx 2\times10^{8}$~K to $\approx7\times10^{8}$~K ($\approx13$ to $63$ per cent above the unperturbed fuel temperature) between the low- and high-density models. Despite moderate thermodynamic perturbations, the strong temperature sensitivity of carbon ignition produces substantial shortening of ignition times.
The results indicate a two-step mechanism: flame thermal expansion perturbs the fuel away from hydrostatic equilibrium, while restoration of pressure balance by weak acoustic waves redistributes internal energy from degenerate to thermal components. This raises the fuel temperature and produces localized thermodynamic preconditioning.
These results suggest that unsteady thermonuclear flames can shorten ignition times ahead of the flame, influencing fuel over distances larger than the laminar flame thickness. Whether this preconditioning leads to secondary ignition or contributes to deflagration-to-detonation transition remains untested here, as nuclear burning is disabled in the analyzed fuel.
\end{abstract}
%
%
%
%
%
%
%
%
\section{Introduction}\label{s:intro}
A recent study by \citet[][Paper I, hereafter]{Brooker+25} of Rayleigh-Taylor (RT) unstable deflagrations powering centrally-ignited Chandrasekhar mass white dwarfs revealed significant shortening of fuel ignition times just ahead of the flame front. This phenomenon is potentially of critical importance for the explosion process as, if the conditions are suitable, fuel may ignite and produce deflagrating regions isolated from the original flame. It is conceivable that in some cases those newly created deflagrations might become seeds for detonations \citep{Brooker+21}, thus offering brand new sites for deflagration-to-detonation transitions (DDTs) in SNe Ia. Because this mechanism appears generic to SN Ia deflagrations, and given the recently proposed unifying scenarios for several existing subclasses of thermonuclear supernovae, one may speculate that the inherent sensitivity of the flame-fuel interaction (FFI) might be one of the sources of the observed diversity of these events. 

The aim of this paper is to identify the underlying physical processes responsible for the observed shortening of the fuel ignition times. Given that the key participating physics is limited to buoyancy and thermonuclear combustion, one may speculate that the interaction of those two processes could result in changes to the thermodynamic state of the fuel. Furthermore, one can also expect this effect to be the strongest when the combustion and buoyancy operate in physical proximity. Because both processes involve production and transport of different types of energy, it is conceivable that the energy transport and its conversion could be the key to understanding the observed decrease in fuel ignition times. 

\citet{Zhao+22} provided the first comprehensive theory of the energy transfer in non-reactive RT instability (RTI). We adopt and further extend the original \citet{Zhao+22} formalism to SN Ia RTI-unstable deflagrations and demonstrate how the flame thermal expansion results in systematic shortening of fuel ignition times. We employ tracer particles to record thermodynamic history of the fuel parcels. This Lagrangian representation allows us to identify relative contributions of various physics processes affecting fuel thermodynamics and their mutual interactions through correlation analysis. We find that ignition-time shortening near buoyantly unstable flame fronts exhibits a density-dependent effective local scaling with flame distance, with the fitted slope steepening strongly with fuel density.

This paper is organized as follows. We first briefly discuss the computational model used to track evolution of the thermonuclear fuel as it is approached by the flame front. Next we present the adopted computational model and data analysis methods, with the results presented in the subsequent section. We close by discussing the results and offering conclusions. 
\section{Methods and initial conditions}\label{s:methods}
In this work, we study SN Ia RT-unstable deflagrations using the physics model, flame-in-a-box computational setup, and gravitational field identical to those described in Paper I. The present models are restarted from Paper I's own evolved, quasi-steady-state simulations, well after Paper I's own initial time. Owing to continual flame thermal expansion, the fuel density at this restart point is therefore already somewhat lower than Paper I's own initial density (their Table 1), because the analysis presented here covers times close to the end of the Paper I runs. Fuel densities just ahead of the flame front at the initial time of the models analyzed here are $4.9\times 10^6$ (LOD), $1.4\times 10^7$ (MED), $5.9\times 10^7\:\urho$ (HID). 

For completeness we recall the key elements of that setup: the plasma is described by the Helmholtz equation of state \citep{TimmesSwesty2000}, which includes the Coulomb corrections; the finest effective cell size is $62.5$ m; and the computational domain measures $768\times 32\times 32$ km. The initial carbon mass fraction of the fuel is $\Cfrac = 0.42$, $0.425$, and $0.33$ in the LOD, MED, and HID models, respectively (cf.\ Paper I, Table 1). Note that $\Cfrac$ is distinct from the fuel filling factor, $X_\mathrm{f}$, listed in Table~\ref{t:models}. The mean temperature of this same reference fuel, $T_\mathrm{ref}$, later used to quantify the fuel's temperature excess (Section~\ref{s:tempDist}), is $1.51\times 10^{9}$, $1.33\times 10^{9}$, and $1.17\times 10^{9}$~K in the LOD, MED, and HID models, respectively.
\begin{deluxetable}{lccccccc}
\tablewidth{0pt}
\tablecaption{Parameters of the Lagrangian analysis and mesh-based fuel diagnostics.\label{t:models}}
\tablehead{
  \colhead{Model} &
  \colhead{$t_\mathrm{i}$\tablenotemark{a}} &
  \colhead{$t_\mathrm{f}$\tablenotemark{b}} &
  \colhead{$X_\mathrm{f,i}$\tablenotemark{c}} &
  \colhead{$X_\mathrm{f,f}$\tablenotemark{d}} &
  \colhead{$\Machfuel$\tablenotemark{e}} &
  \colhead{$\Delta m_\mathrm{p}$\tablenotemark{f}} &
  \colhead{$r_\mathrm{p}$\tablenotemark{g}} \\
  \colhead{} &
  \colhead{(s)} &
  \colhead{(s)} &
  \colhead{} &
  \colhead{} &
  \colhead{} &
  \colhead{(g)} &
  \colhead{(cm)}
}
\startdata
LOD & 1.651 & 1.695 & 0.23 & 0.14 & 0.038 & $2.4\times10^{16}$ & 1133 \\
MED & 1.216 & 1.241 & 0.22 & 0.13 & 0.044 & $6.7\times10^{16}$ & 1124 \\
HID & 0.701 & 0.719 & 0.19 & 0.09 & 0.073 & $1.5\times10^{17}$ & 979  \\
\enddata
\tablenotetext{a}{Initial simulation time of the analyzed time interval.}
\tablenotetext{b}{Final simulation time of the analyzed time interval.}
\tablenotetext{c}{Initial fuel filling factor in the region initially occupied by the tracer particles.}
\tablenotetext{d}{Final fuel filling factor in the region initially occupied by the tracer particles.}
\tablenotetext{e}{Time-averaged turbulence Mach number of the fuel in the same region.}
\tablenotetext{f}{Average tracer particle mass.}
\tablenotetext{g}{Representative tracer particle radius, the radius of the sphere that contains the average particle mass at the density of the fuel upstream of the flame.}
\end{deluxetable}

To enable Lagrangian analysis of the fuel, the computer model adopted from Paper I was extended to include tracer particles \citep{Dubey+12}. The particle-instrumented data sets were obtained by continuing simulations reported in Paper I. Consequently, our models were already in a quasi-steady state at the outset of this additional set of particle-instrumented simulations. 

For the purpose of analysis of the overall fuel evolution, we generated a set of simulations, dubbed FUEL, in which, upon simulation restart, fuel-rich regions of the initial models below a plane positioned $1$ km ahead of the leading edge of the flame front were seeded with $\approx 3$ million tracer particles using a mass-based acceptance-rejection method. This way of initializing particle positions is appropriate because of radial (along the direction of gravity) fuel density variations and, more importantly, the highly convoluted structure of the fuel in the turbulent post-flame (TPF: {cf.} Paper I) region.

 A second particle-seeded set of simulations was generated to investigate in detail the particle heating process; we dub it FLT (``First Law of Thermodynamics''). This model used the same initial restart conditions as the FUEL model, but with particles seeded in a thin layer corresponding in thickness to the "heating" layer identified by \citep{Brooker+25} with its inner surface initially placed at the outer edge of the heating region. In all FLT models, the region was seeded with 300,000 particles, using the same mass-based acceptance-rejection seeding method employed for the FUEL set. The FLT data was much more finely-spaced in time than the FUEL data, allowing for the creation of subsets of simulation snapshots that could be assembled for the purpose of statistical analysis. Every one of those subsets was uniquely assigned a time "anchor", the latest simulation snapshot in the subset.

The particle initialization process in the FUEL set resulted in a mass resolution per particle between $\approx 2.4\times 10^{16}$ g and  $\approx 1.5\times 10^{17}$ g with the corresponding characteristic particle radius varying between $1133$ cm and $979$ cm, in the low and high density models as shown in Table~\ref{t:models}. The characteristic particle radius, $r_\mathrm{p}$, is the radius of the sphere that contains the particle mass at the density of the fuel upstream of the flame, which is much lower than the initial density owing to expansion and time evolution. The corresponding particle diameter, $2r_\mathrm{p}\approx 23$ m in LOD and $\approx 20$ m in HID, is smaller than the finest cell size of $62.5$ m by a factor of approximately $2.8$--$3.2$. 

Following the nomenclature introduced in Paper I, we differentiate between warm fuel present in the layers adjacent to the flame front, which we dub the flame \emph{skin} layer (FSL), and the cold fuel with temperatures close to their initial values. The warm fuel layer is defined as material with flame progress variable in the range $\phi\in[1\times 10^{-5}, 3\times 10^{-2}]$. The lower range limit of the progress variable characterizes fuel well separated from the flame front profile proper (cf.\ Figure 10 in Paper I). The upper limit ($3\times 10^{-2}$) is introduced here to ensure the thermodynamic fuel state is not directly affected by the energy deposition of the ADR flame model \citep[Model B]{Zhiglo07}: its deposition rate decreases smoothly with $\phi$, staying positive $\forall \phi>0$, instead of dropping to 0 as a step function in the original implementation \citep{Khokhlov1995}. The same threshold is used below to define the zero point of the distance-to-flame coordinate $d_\mathrm{f}$; it is a numerically motivated reference point rather than a claim about a physically sharp flame boundary.

The fuel filling factor, $X_\mathrm{f}$, and the fuel turbulence Mach numbers, $\Machfuel$, listed in Table~\ref{t:models} were measured directly on the AMR mesh at its native resolution, independently of the tracer particle data. Here fuel is material with $\phi<3\times10^{-2}$, the same threshold as above. The region considered is the slab initially occupied by the tracer particles, between their minimum and maximum $x$ coordinates, which is $111$, $101$, and $121$ km wide in LOD, MED, and HID, respectively, and spans the full extent of the domain in the transverse directions ($32$ km). The filling factor is the fuel volume fraction of the region at $t_\mathrm{i}$. In this fixed region it decreases with time by $37$, $42$, and $51$\% in LOD, MED, and HID, respectively, at the end of the analyzed time interval. The Mach number is the volume-weighted rms of the cell Mach number, $|\mathbf{v}-\bar{\mathbf{v}}|/c_\mathrm{s}$, over the fuel in the region, averaged over the snapshots of the analyzed time interval. Here $c_\mathrm{s}$ is the adiabatic sound speed in the cell, obtained from the cell density and temperature with the Helmholtz equation of state, and $\bar{\mathbf{v}}$ is the mass-weighted mean velocity of the fuel in the region at the given time. It therefore includes the coherent motion, relative to that mean, of individual fuel structures such as the sinking spikes. The Mach number is consequently an upper bound on the fluctuations internal to such structures.

Due to the advanced stage of the system evolution, the process responsible for the fuel heating has already been fully developed. Consequently, we only need to advance the models long enough for a substantial fraction of particle-seeded cold fuel to be consumed by the flame. We found that in order to ensure that our Lagrangian analysis includes a representative population of particles tracing warm fuel, as defined above, restarted simulations had to be evolved for about $44$, $25$, and $18$ ms for LOD, MED, and HID models respectively, as shown in Table \ref{t:models}. During those time periods, about $20\%$ of tracer particles went through the full evolution from the cold fuel phase to the fully burned material phase. 

The quality of particle data is limited by the accuracy of the particle mover, which involves both the interpolation of velocities from the hydro mesh and the time integrator itself, as well as the mapping of information from the hydro mesh to particle positions. Our choice of the Heun's Runge-Kutta-type time integrator formally offers second-order accuracy. This remains to be demonstrated as verification results for the \FLASH\ tracer particle module \citep{Dubey+12} are not readily available. The accuracy of mesh data mapping proved problematic, as the original implementation was prone to producing undershoots in the region near the flame front.

Our \Proteus\ implementation features a full quadratic mesh-to-particle interpolation scheme, including explicit cross-derivative (mixed-partial) terms. Interpolation coefficients are calculated using a compact stencil with neighbor cells providing data for the pure-direction terms. A curvature correction is absorbed into the constant term so the reconstruction matches the cell average rather than the point value. Each cross-derivative coefficient is extracted independently from the four diagonal neighbors in its own coordinate plane. If the quadratic result overshoots the range bracketed by the neighboring cell centers, the scheme switches to a lower-order (linear/bilinear/trilinear) interpolation over those same centers. With this design, the expected order of the tracer-particle trajectories is second order in time and third order in space. We verified this empirically on a suite of one-, two-, and three-dimensional unit tests, cf. Appendix~\ref{app:particle_verification}.

A critical component of the Lagrangian analysis employed in this study is the dependence of plasma properties, as recorded by tracer particles, on the physical distance from the flame front. This distance was estimated to every mesh cell in the computational domain using a parallel, domain-decomposed implementation of the fast sweeping method \citep[][FSM]{Zhao2004}, employing the locking sweeping method \citep{Bak2010} for computational efficiency. The 3-dimensional distance field mesh variable was computed in respect to the isosurface of $\phi = 3 \times 10^{-2}$, which we define to be the flame front proper. The distance field was then mapped to particles using the interpolation routines described above. The parallel domain decomposition was verified to converge to machine precision against the serial, non-decomposed solution on synthetic test surfaces of varying geometric complexity. The FSM solution was finally compared against distance field obtained with a reference fast-marching-method implementation \citep{Sethian1996} on the same surfaces, reproducing the expected accuracy of the first-order discretization employed.
\section{Results}\label{s:results}
We present our simulation results in terms of the overall morphology of the fuel, statistics of the ignition time and temperature as functions of the distance from the flame front, and the Lagrangian thermodynamic history of the fuel whose ignition time is reduced the most. The fuel morphology does not show a readily discernible relation with the fuel velocity divergence, which one would expect to influence the thermodynamic evolution of the fuel. We nonetheless establish a clear relation between the fuel ignition time and the proximity to the flame front, and show that this relation is controlled almost entirely by a corresponding rise in the fuel temperature. Following the trajectories of the most strongly heated particles, we find that their thermodynamic evolution is close to isobaric. This points to pressure readjustment to hydrostatic equilibrium driven by the flame's thermal expansion 
being the cause of the fuel temperature increase. 
\subsection{Fuel morphology}
\begin{figure*}
    \begin{center}
    \begin{tabular}{ccc}
        \includegraphics[width=0.6\columnwidth]{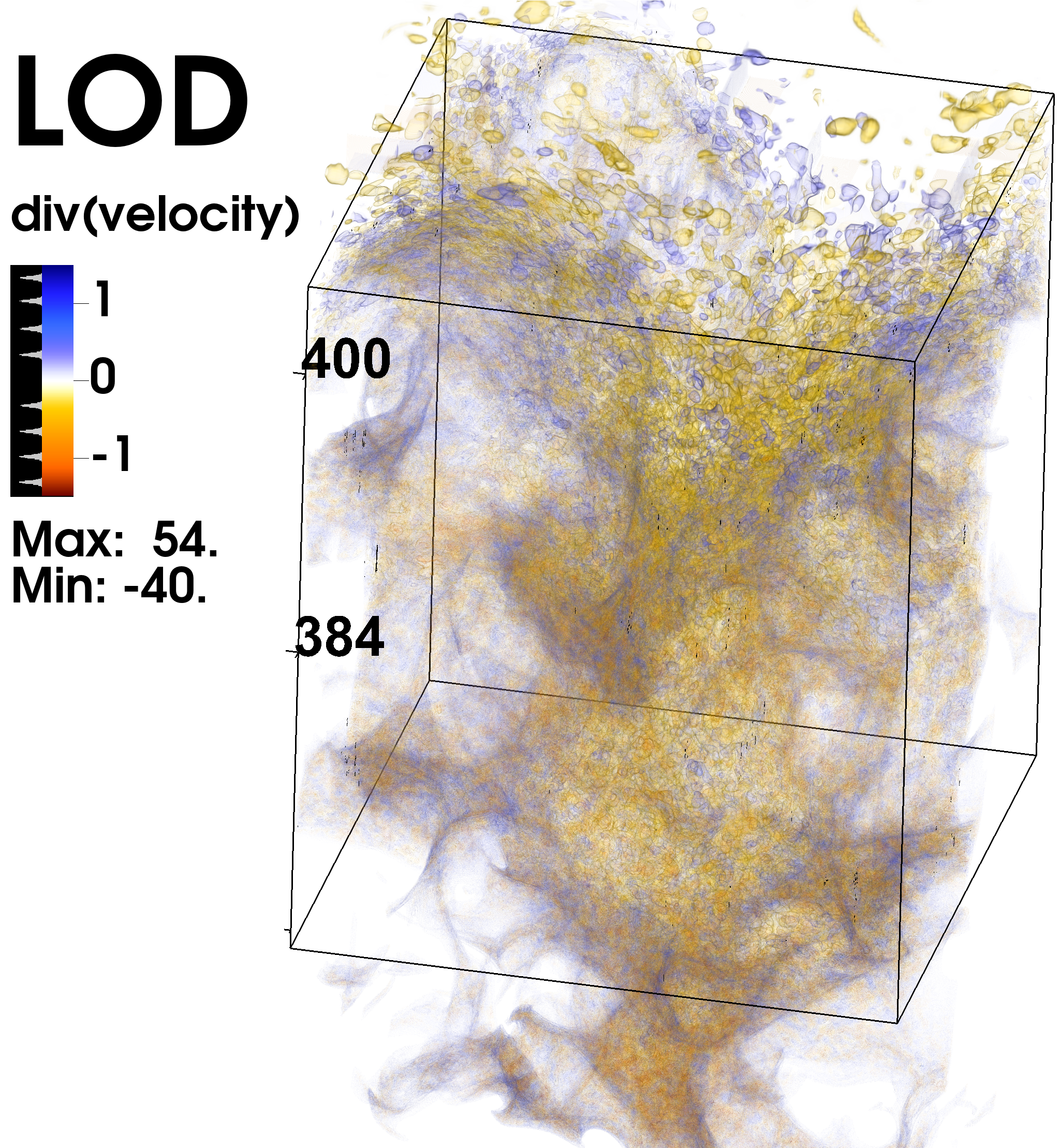} &
        \includegraphics[width=0.6\columnwidth]{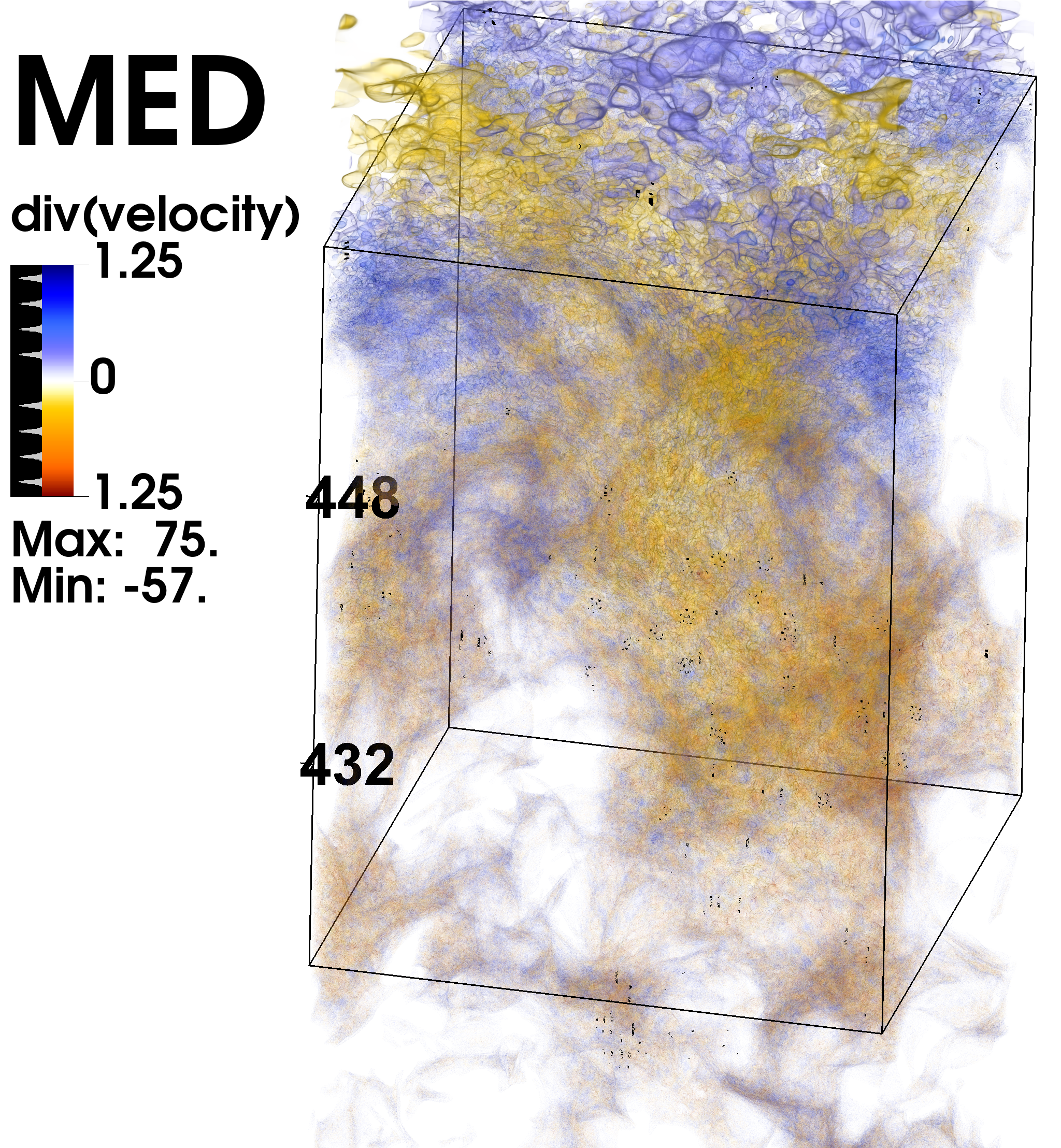} &
        \includegraphics[width=0.6\columnwidth]{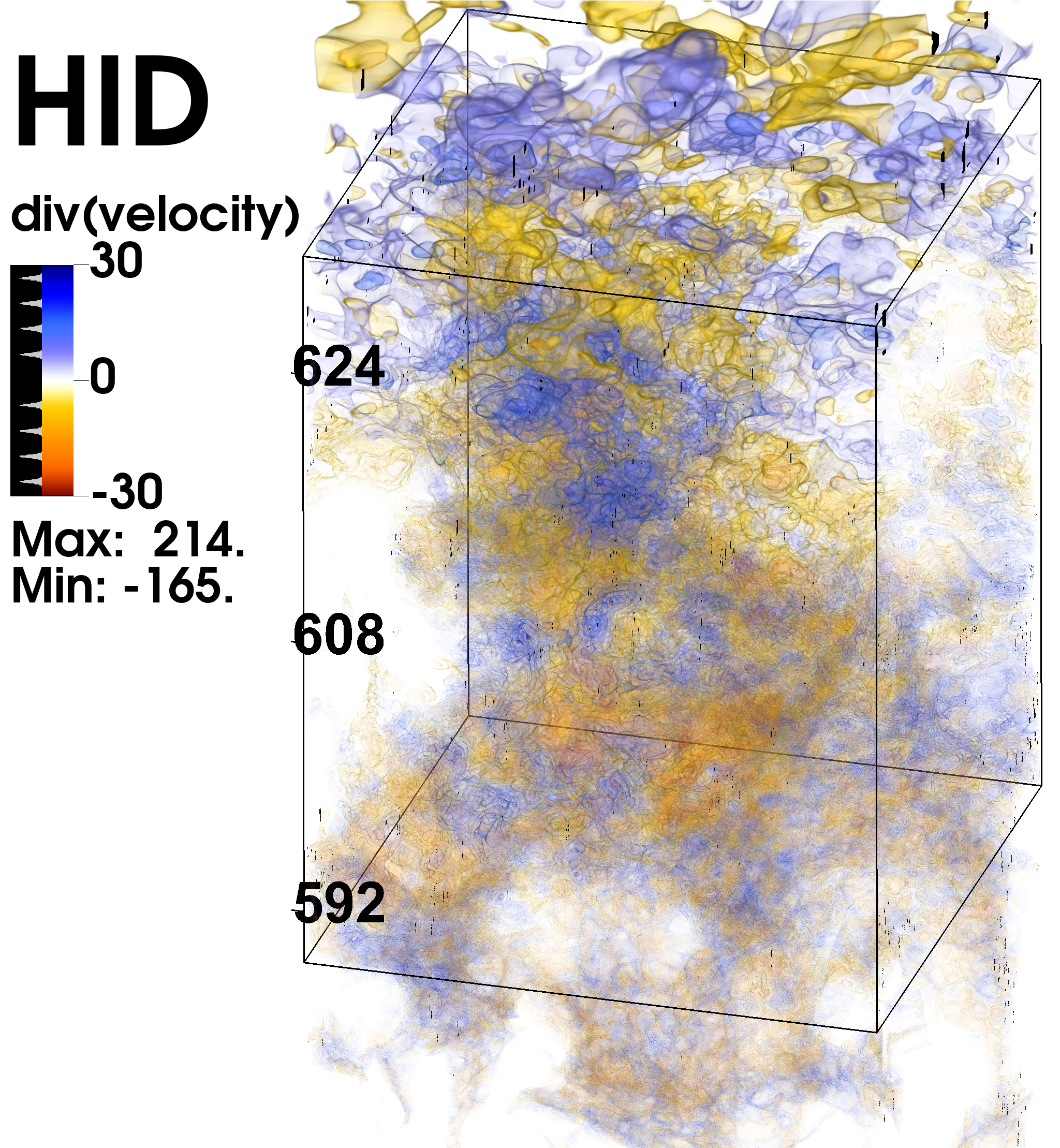}     
    \end{tabular}
    \figcaption{Divergence of the fuel velocity in the turbulent post-flame region in \citet{Brooker+25} models. The velocity divergence is depicted in the low density (LOD), medium density (MED), and high density (HID) models and shown in the left, middle, and right figure panel, respectively. The simulation data correspond to the final simulated times reported by \citet[][ see their Table 1]{Brooker+25}. The divergence field is rendered in a semi-transparent way with fuel compression indicated in the red and yellow while expansion is shown in various shades of blue. The visualized segment of data has a cross section of $32\times 32$ km with the vertical extent of $\approx 40$ km. See text for details. \label{f:fuelDivV}}
    \end{center}
\end{figure*}

Figure \ref{f:fuelDivV}
shows a rendering of the velocity divergence for the fuel in the TPF region. It suggests the fuel expansion (in blue) is typically associated with the rising flame bubbles, while the sinking fuel spikes undergo compression (in yellow and red). The degree of those effects increases with the strength of buoyancy (hence, density), as discussed in Paper I. An additional consequence of the increasing buoyancy strength, also discussed in Paper I, is more violent evolution with the typical divergence values differing by a factor of a few between the LOD and HID models. In consequence, the fuel spikes appear to have greater structural integrity in LOD and MED models while in the HID model they form more structures on smaller scales.  Given the less contiguous fuel spike structures, the regions undergoing coherent expansion or compression are several times smaller than at lower densities.
\subsection{Fuel ignition time-flame distance relation}
Our Lagrangian analysis provides additional support for the systematic decrease of the fuel ignition times as the flame front approaches, as reported originally in Paper I. In this work, we study this dependence by recording fuel ignition times along the tracer particle trajectories. The ignition time is given by \citet{DursiTimmes2006},
\begin{equation} \begin{array}{rcl} 
   \tau(\rho,T,X_C)&\hspace{-0.1cm}=&\hspace{-0.1cm}1.15\times 10^{-5}(X_C\rho_8)^{-1.9}\\[1pt]
   &\hspace{-0.1cm}&\hspace{-0.1cm}\times f_\mathrm {cp}(T)[1+1193f_\mathrm {cp}(T)]\ \mathrm{s},\\[4pt]
   f_\mathrm {cp}(T)&\hspace{-0.1cm}=&\hspace{-0.1cm}(T_9-0.214)^{-7.566}\ .
   \end{array}
   \label{e:igtm}
\end{equation}
In evaluating Equation~\ref{e:igtm} we use the local carbon mass fraction recorded by each particle. In regions away from the flame skin layer, the carbon abundance equals the initial fuel carbon abundance, $\Cfrac$, while inside the FSL it decreases toward the flame front following the ADR model prescription.

\begin{figure*}
    \centering
        \begin{tabular}{@{\hspace{-12mm}}c@{\hspace{-6mm}}c@{\hspace{-6mm}}c@{}}
        \includegraphics[height=0.24\textheight]{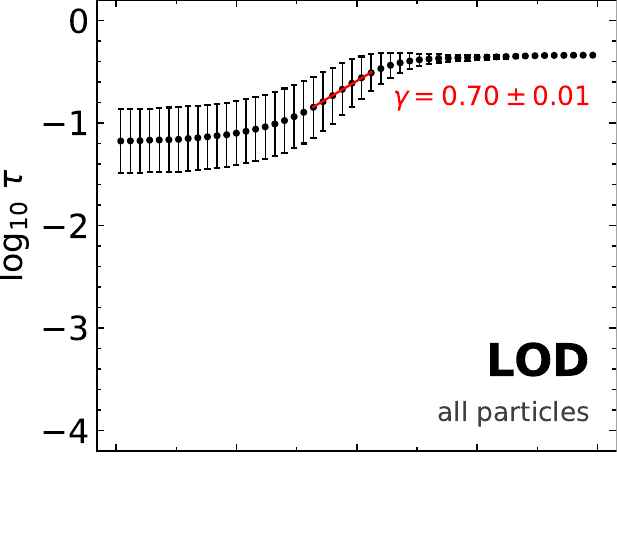}      &
        \includegraphics[height=0.24\textheight]{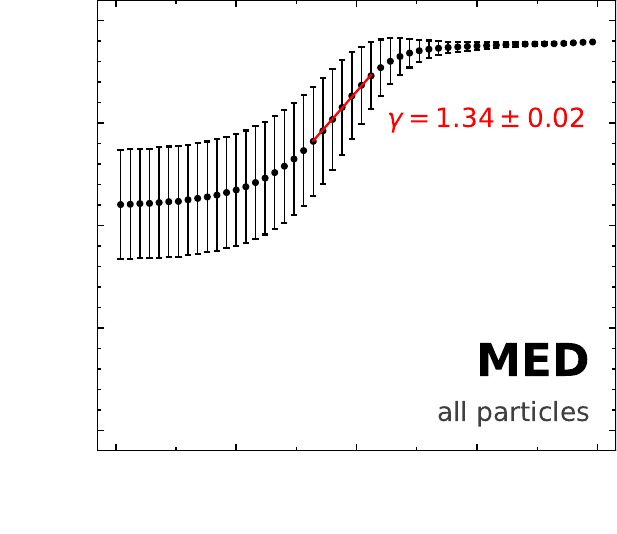}    &
        \includegraphics[height=0.24\textheight]{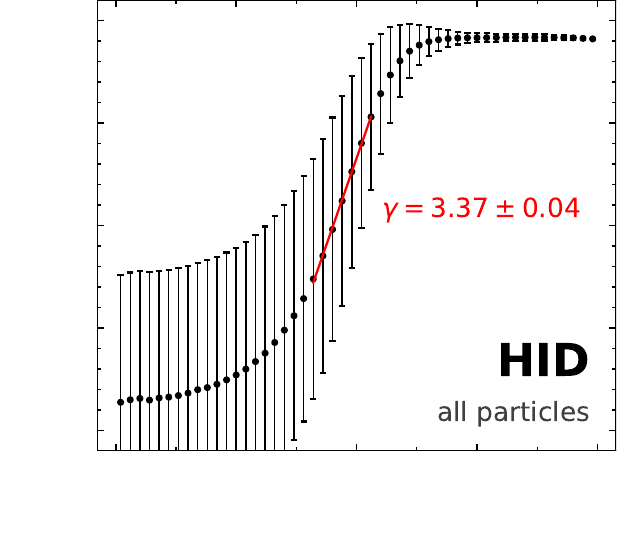}     \\[-6mm]
        \includegraphics[height=0.24\textheight]{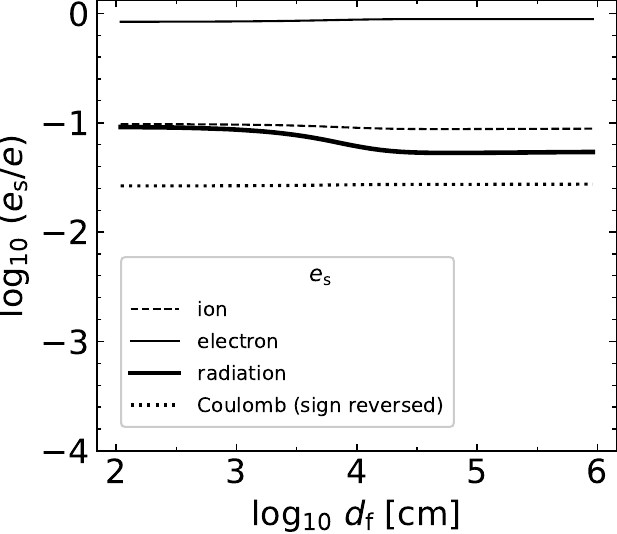}   &
        \includegraphics[height=0.24\textheight]{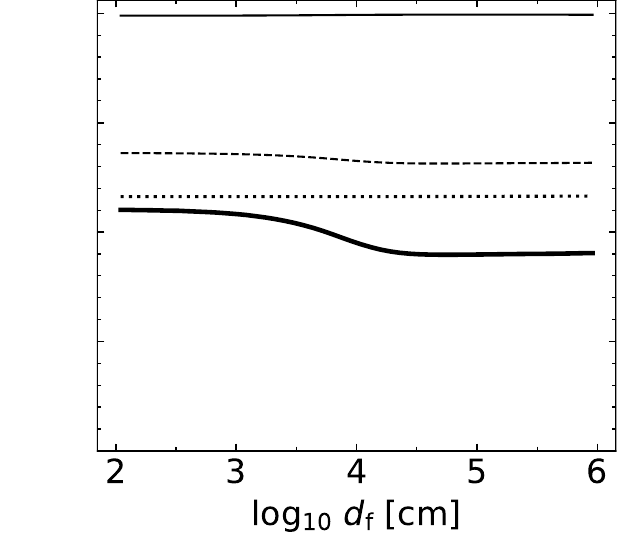} &
        \includegraphics[height=0.24\textheight]{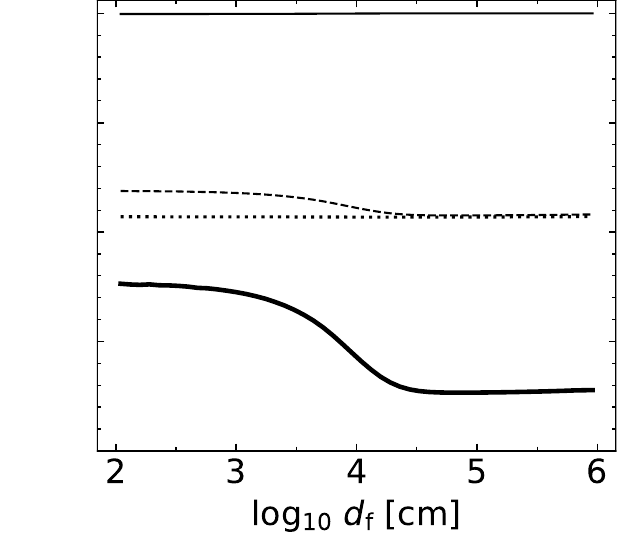}
        \end{tabular}
    \figcaption{Fuel ignition time and internal energies dependence on the distance from the flame front in model RTI-unstable deflagrations. These are shown in LOD, MED, and HID models in left, middle, and right figure panels, respectively. The averages were computed for populations of tracer particles found in 50 logarithmically spaced bins in distance from the leading edge of the flame front, out to the maximum distance of 10 km. The leading edge of the flame front is defined as the iso-surface $\phi=3\times 10^{-2}$ of the flame progress variable.
    (top row)
    The mean values of $\log_{10}\tau$ start decreasing for distances below about 4.5~dex, and follow an approximate power-law dependence on the distance within the fit window $3.64\leq\log_{10}d_\mathrm{f}\leq 4.12$ (see text). The relation breaks below $\approx 3.25$--$3.3$~dex, where interpolation stencils used to construct the distance field increasingly extend beyond the flame surface itself (see text). The associated error bars indicate the standard error of the binned mean.
    (bottom row)
    Mean relative contributions of ions ($e_\mathrm{i}$, dashed), electrons  ($e_\mathrm{e}$, thin solid), radiation ($e_\mathrm{r}$, thick solid), and Coulomb interactions ($e_\mathrm{C}$, dotted) to the plasma internal energy. The ordinate, $e_\mathrm{S}/e$, is the ratio of the given component to the total internal energy. The data were obtained using the same analysis method as above for the ignition time. A relative contribution of radiation shows the greatest increase as the distance to the flame shortens. For  presentation in logarithmic scale the absolute value of Coulomb contribution is shown; the Coulomb energy term itself is negative.  
    See text for details.
    \label{f:igtmDist}}
\end{figure*}

The recorded ignition times are assigned to 50 bins spaced logarithmically in distance from the flame front proper. Because the ignition time distributions are strongly skewed, we characterize the binned data by statistics computed in log-scale with the mean of $\log_{10}\tau$ (i.e., the geometric mean of $\tau$) and the standard error of that mean. The means of ignition times steadily decrease as the distance to the flame front, $d_\mathrm{f}$,  shortens. These are shown in Figure \ref{f:igtmDist} 
for LOD (left panel in the figure), MED (middle panel), and HID model (right panel). The power-law scaling in $\tau(d_\mathrm{f})$ is consistently observed in all models for flame front distances $3.64\lesssim \log_{10}{d_\mathrm{f}} \lesssim 4.12$.

We quantify the rate of ignition time shortening with an exponent, $\gamma = \mathrm{d}\,\log_{10}(\tau)/\mathrm{d}\log_{10}(d_\mathrm{f})$, by a weighted least-squares fit of the binned mean $\log_{10}\tau$ against $\log_{10} d_\mathrm{f}$. We selected the fit window to be  $\log_{10}d_\mathrm{f}\in[3.64,4.12]$, where the slope of $\log\tau$ vs $\log d_\mathrm{f}$ is the largest. The fit window is positioned comfortably clear of the near-front region $<3.3$~dex where the power law relation breaks down.

The fits yield $\gamma = 0.70 \pm 0.01$ (LOD), $\gamma = 1.34 \pm 0.02$ (MED), and $\gamma = 3.37 \pm 0.04$ (HID). The scaling exponent thus displays a strong dependence on density, increasing by roughly a factor of five between the LOD and HID models. This fit uses the full FUEL dataset introduced in Section~\ref{s:methods} ($51$/$100$/$72$ snapshots for LOD/MED/HID, each with the complete particle field), with each of the 7 bins spanning the fit window populated, across all snapshots, by between $\approx 6.4\times 10^{5}$ and $\approx 1.7\times10^{6}$ (LOD), $\approx 1.6\times10^{6}$ and $\approx 4.1\times10^{6}$ (MED), $\approx 9.4\times10^{5}$ and $\approx 2.6\times10^{6}$ (HID) particle-snapshot entries. Given bin populations of this size, we do not expect the fitted exponents to be sensitive to random subsampling, and the quoted uncertainties are the formal errors of the weighted least-squares fit itself, using each bin's own standard error of the mean as weight.

The above exponents should be interpreted as effective, intermediate-distance slopes rather than universal scale-free exponents. The local-slope curves are not flat over the full distance range, and the inferred value depends on the fit interval. Three considerations set that interval: it is approximately centered on each model's own peak local slope, where the heating transition is steepest and the underlying mechanism most active. The inner interval edge excludes the region where the interpolation floor takes over, a numerical rather than physical boundary (see below). The adopted interval width is a practical choice, wide enough for a stable fit without drifting far from the peak. The quoted uncertainties do not include systematic uncertainty associated with fit-window choice or temporal correlations among snapshots. A local-slope diagnostic shows the underlying transition region differs by model -- broad and shallow for LOD/MED, narrow and sharp for HID, the latter sitting close to the inner interpolation-floor boundary. The local-slope diagnostic itself is shown directly in Appendix~\ref{app:localslope} (Figure~\ref{f:localslope}): in all three models, the fit window brackets a real, well-defined peak rather than sitting in the near-front region excluded above, directly demonstrating the fit window is not biased by proximity to that region.

The trend in the shortening of ignition time changes abruptly closer to the flame front. Two factors contribute to this change. First, the carbon mass fraction in the FSL, prescribed by the ADR flame model \citep[][with modifications described in Paper I]{Khokhlov1991a, Khokhlov1995}, decreases toward the flame front, which acts to increase the ignition time and thus to flatten the shortening trend. Second, the trend change sets in at a distance of $\approx 2$ mesh cells from the flame front proper, where the mesh-particle interpolation scheme returns monotonized values. These are bounded from below by the shortest ignition time found in the region. The interpolated values themselves are valid, but the resulting flattening of the ignition-time trend is an artifact of the interpolation rather than a physical feature. Because the trend change occurs at the $\approx 2$-cell interpolation scale, we attribute the flattening primarily to this numerical floor rather than to carbon depletion. The ignition-time reductions found here, $0.9$ (LOD) to $3.6$ dex (HID) at the innermost reliably resolved bins, are smaller than the two to more than five orders of magnitude reported in Paper~I. The difference has two causes. First, the flame proper is defined differently. Here it is a reconstructed flame surface, with the distance to it given by a distance function, whereas Paper~I used the interpolated flame progress variable. Second, the two analyses cover different evolutionary times, as the present intervals lie near the end of the Paper I runs.
\subsection{Fuel temperature--flame distance relation}\label{s:tempDist}
\begin{figure*}
    \centering
        \begin{tabular}{ccc}
        \includegraphics[width=0.48\textwidth]{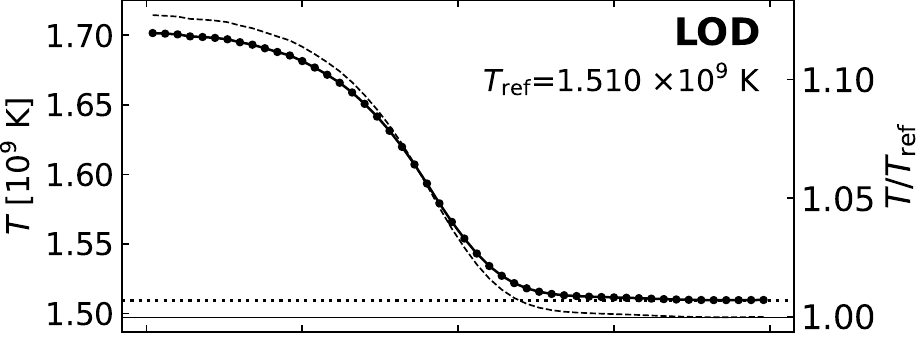}       &
        \includegraphics[width=0.48\textwidth]{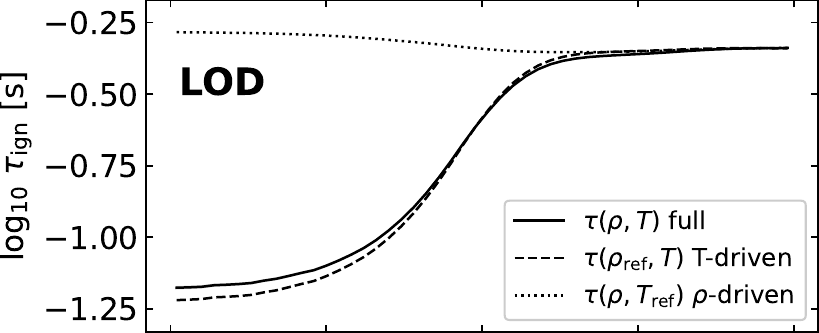}      \\
        \includegraphics[width=0.48\textwidth]{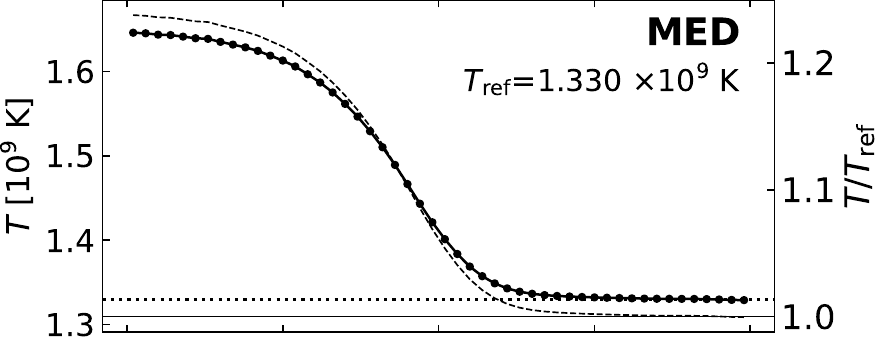}    &
        \includegraphics[width=0.48\textwidth]{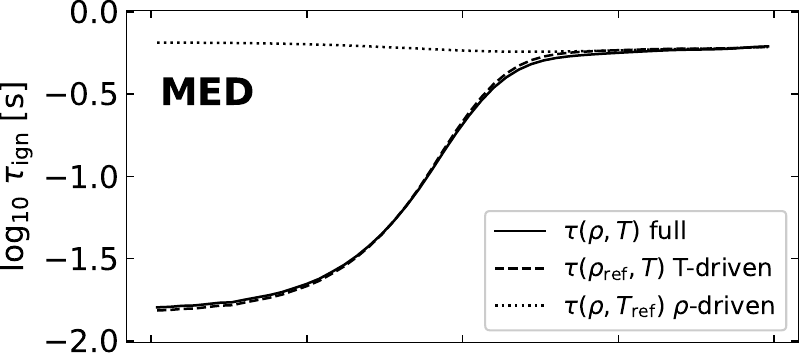}   \\
        \includegraphics[width=0.48\textwidth]{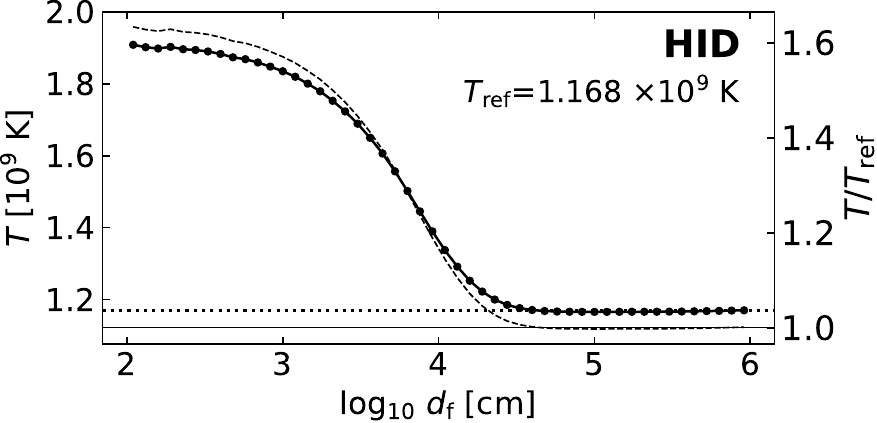}    &
        \includegraphics[width=0.48\textwidth]{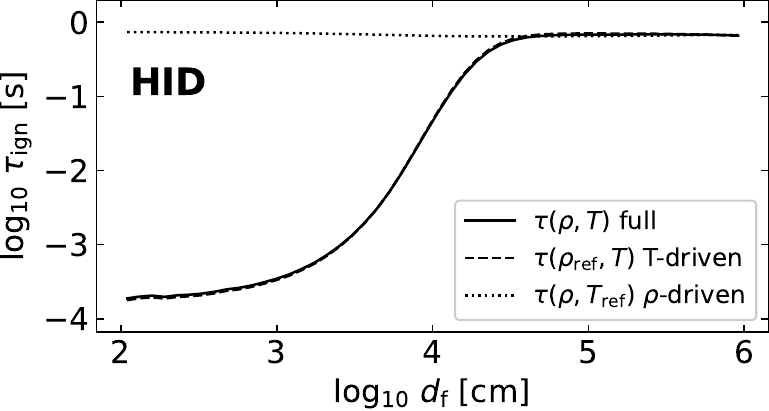}
    \end{tabular}
    \figcaption{Fuel temperature--flame distance relation and the ignition-time consistency check in the LOD (top row), MED (middle row), and HID (bottom row) models, for the same particle population and distance binning as in Figure~\ref{f:igtmDist}.
    (left column) Binned mean fuel temperature, $T$ (solid line with symbols), with the unperturbed fuel temperature, $T_\mathrm{ref}$, marked by the horizontal dotted line and quoted in each panel. The dashed curve (right vertical axis) shows the dimensionless ratio $T/T_\mathrm{ref}$, with the thin solid horizontal line marking unity.
    (right column) $\log_{10}\tau_\mathrm{ign}$ computed via Eq.~\ref{e:igtm} based on the measured $\left(\rho, T, \Cfrac\right)$ (solid); the measured temperature at the reference density $\rho_\mathrm{ref}$ (dashed); and the measured density at the reference temperature $T_\mathrm{ref}$ (dotted). The near-coincidence of the solid and dashed curves, and the flatness of the dotted curve, demonstrate that the temperature rise alone accounts for the ignition-time reduction. See text for discussion.
    \label{f:tempDist}}
\end{figure*}

The variable that controls the ignition time in Equation~\ref{e:igtm} is the fuel temperature. Figure~\ref{f:tempDist}
shows the binned mean fuel temperature, $T$, and its ratio to the unperturbed fuel temperature, $T/T_\mathrm{ref}$, as functions of the distance to the flame front, obtained with the same binning as in Figure~\ref{f:igtmDist}. Here $T_\mathrm{ref}$ is the reference fuel temperature introduced in Section~\ref{s:methods}, and the temperature excess over the unperturbed fuel is $\Delta T = T - T_\mathrm{ref}$. The observed temperature rise mirrors the ignition-time trend with $\Delta T$ growing from $\approx 10^{7}$~K near $\log_{10} d_\mathrm{f} = 4.5$ to $\approx 1.3\times 10^{8}$ (LOD), $\approx 2.1\times 10^{8}$ (MED), and $\approx 4.9\times 10^{8}$~K (HID) at $\log_{10} d_\mathrm{f} = 3.5$, ultimately reaching $\approx 2.0\times 10^{8}$, $\approx 3.2\times 10^{8}$, and $\approx 7.3\times 10^{8}$~K in the innermost reliably resolved bins. Relative to the unperturbed fuel, these excursions amount to about $13$, $24$, and $63$ percent, respectively.

Figure~\ref{f:tempDist} also demonstrates the internal consistency of the ignition-time analysis. Evaluating Equation~\ref{e:igtm} with the measured $(\rho, T, \Cfrac)$ along the binned profiles reproduces the ignition-time relation of Figure~\ref{f:igtmDist}. Repeating the evaluation with the density held fixed at its far-field reference value reproduces the full reduction to within $0.05$ dex in all three models, whereas holding the temperature fixed at $T_\mathrm{ref}$ removes essentially the entire effect: the density variation alone accounts for at most $0.06$ dex of the observed $0.9$ (LOD) to $3.6$ dex (HID) reduction at the innermost reliably resolved bins. The multi-decade shortening of the ignition time is therefore caused almost entirely by the temperature rise, with the density and composition changes playing a negligible role.

These dependencies illustrate how the spatial scaling connects to the underlying thermodynamic relations established above. It can be elucidated with the help of the chain rule applied over the fit interval,
\begin{equation}
\begin{split}
\frac{\mathrm{d}\ln\tau}{\mathrm{d}\ln d_\mathrm{f}}
&=
\\
&
\frac{\partial\ln\tau}{\partial\ln T}\,\frac{\mathrm{d}\ln T}{\mathrm{d}\ln d_\mathrm{f}}
+
\frac{\partial\ln\tau}{\partial\ln\rho}\,\frac{\mathrm{d}\ln\rho}{\mathrm{d}\ln d_\mathrm{f}}
+
\frac{\partial\ln\tau}{\partial\ln X_C}\,\frac{\mathrm{d}\ln X_C}{\mathrm{d}\ln d_\mathrm{f}} .
\end{split}
\label{e:scaling}
\end{equation}
The analysis presented in the previous paragraph shows that the first term dominates the observed change in the ignition time over the resolved heating region, while the density and composition terms are small.
This decomposition re-expresses, in the language of the fit itself, why the ignition time-distance relation closely tracks the temperature-distance relation.
%
%
%
\subsection{Lagrangian view of FSL thermodynamics}\label{s:lagrangian}
Thermodynamic behavior of the FSL fuel requires close scrutiny in order to identify the mechanism responsible for shortening of ignitions times. One may expect that the responsible effect most cleanly manifests itself in the case of fuel that shows the greatest rate of the ignition time reduction. To represent that material, we selected subsets of 500 particles within each model that underwent the most significant ignition time reduction over the course of the analyzed time interval. We refer to these as the fastest-heated particles (FHP).

\begin{figure*}
    \centering
        \begin{tabular}{ccc}
        \includegraphics[height=0.33\textwidth]{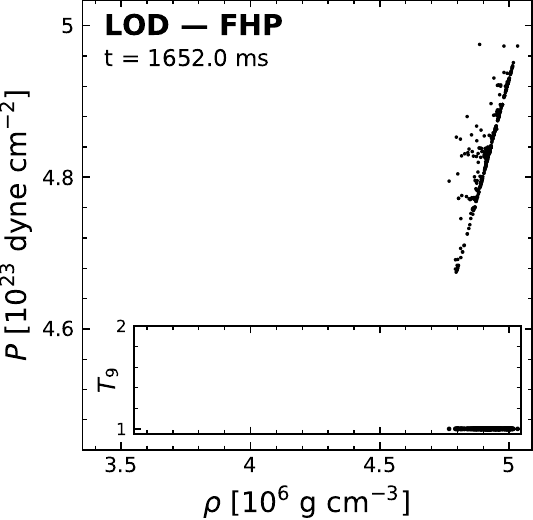}       &
        \includegraphics[height=0.33\textwidth]{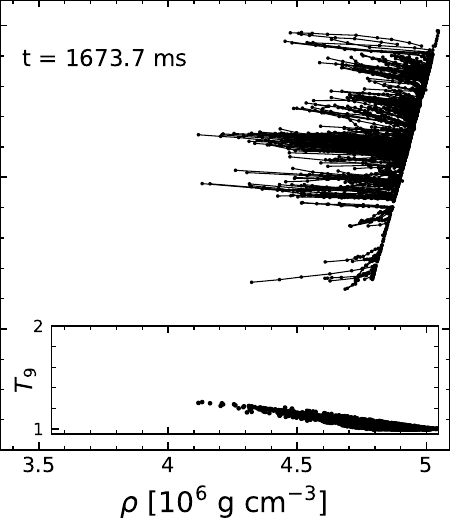}     &
        \includegraphics[height=0.33\textwidth]{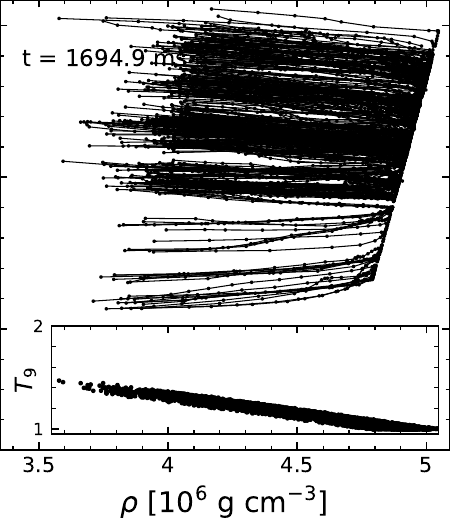}      \\
        \includegraphics[height=0.33\textwidth]{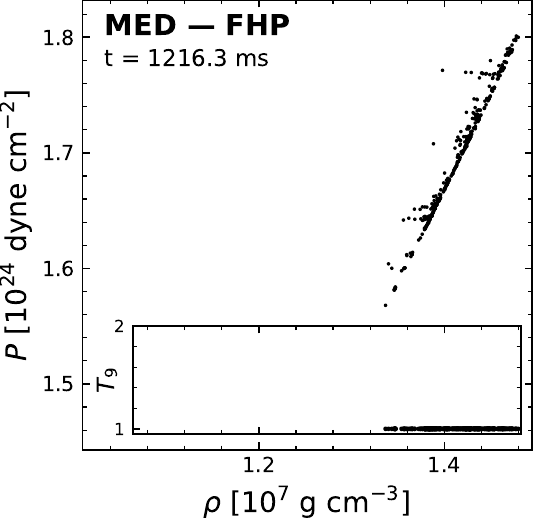}    &
        \includegraphics[height=0.33\textwidth]{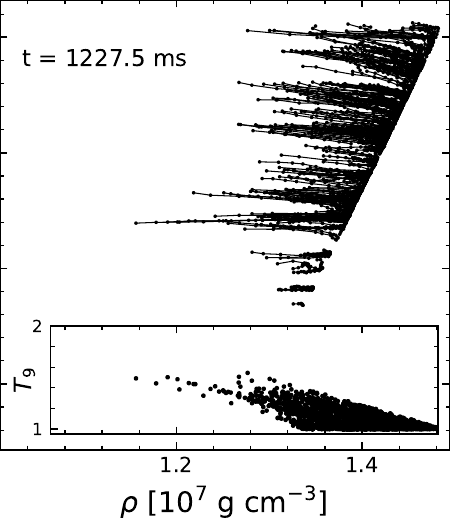}  &
        \includegraphics[height=0.33\textwidth]{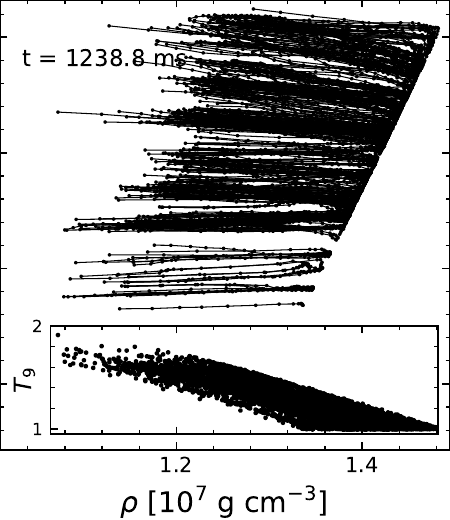}   \\
        \includegraphics[height=0.33\textwidth]{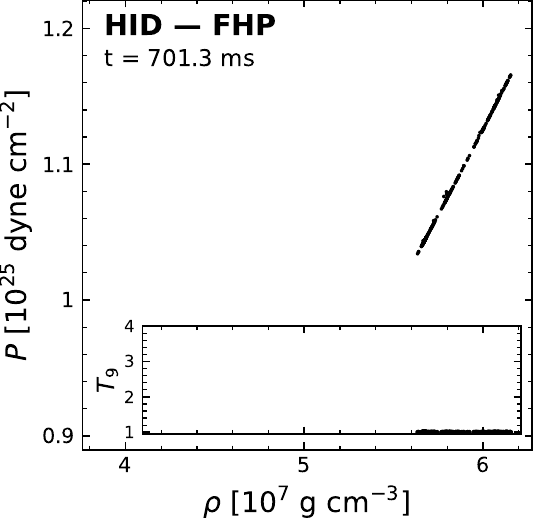}    &
        \includegraphics[height=0.33\textwidth]{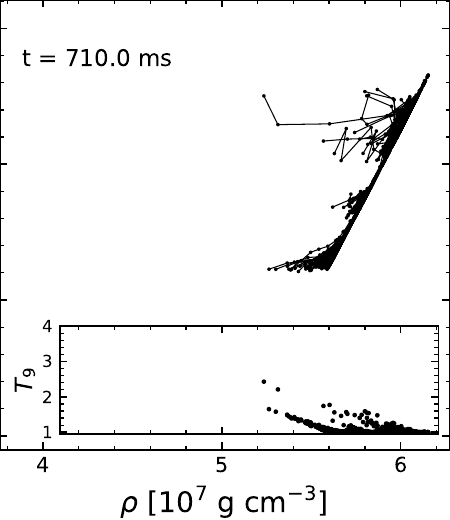}  &
        \includegraphics[height=0.33\textwidth]{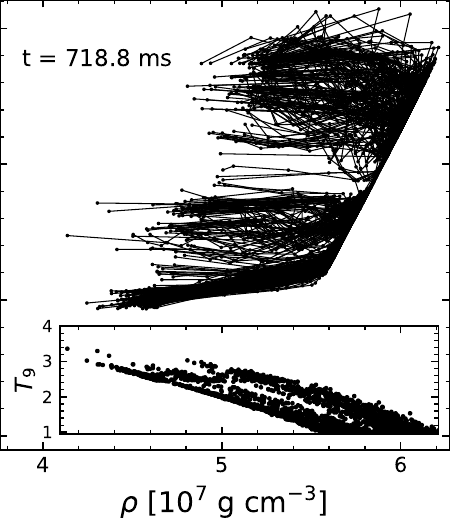}
    \end{tabular}
    \figcaption{Thermodynamic trajectories of fuel tracer particles in model SN Ia deflagrations. The top, middle, and bottom rows correspond to the LOD, MED, and HID models, respectively. The columns correspond to initial, intermediate, and final analyzed evolution times. The trajectories of the FHP members are shown in terms of density, $\rho$, and pressure, $P$, with line segments connecting the corresponding thermodynamic states of individual particles at successive times. The vertical axis scale is common across time slices. In the temperature insets vertical axes show the temperature, $T_9=T/(10^9\:\mathrm{K})$. The density scale labeling is common for the main windows and insets. The density axes correspond to the evolved fuel states at the beginning of the analyzed intervals (Section~\ref{s:methods}), rather than to the initial densities of Paper~I. Note a rapid decrease of the fuel densities recorded at FHP locations at nearly constant pressure, indicative of nearly isobaric evolution of the fuel. See text for discussion.
    \label{f:trajectories}}
\end{figure*}

We contrast the FHP sample with two control populations in Section~\ref{s:correlations}. Figure \ref{f:trajectories}
depicts the evolution of the FHP thermodynamic states in terms of the density and pressure along the horizontal and vertical axes. The left figure panel depicts the state of those fluid elements at the initial time of the analyzed period ($t_i$ in Table \ref{t:models}). We note that at that time the  FHP thermodynamic states are largely limited to a narrow band reflecting the unperturbed fuel conditions (degenerate plasma).
Because the fuel state is close to hydrostatic equilibrium, different values of pressure correspond to different heights and therefore a different location of particles across the flame brush.

At intermediate times, as shown in the middle column in Figure \ref{f:trajectories}, one can identify a significant fraction of the FHP sample showing a decrease in density at nearly constant pressure. Nearly all the particles undergo such a density decrease at the final times, as shown in the right figure panels. The nearly isobaric evolution of the fuel continues throughout the analyzed time period with the median density of the FHP sample decreasing by $\approx 18$, $15$, and $18$ per cent for the LOD, MED, and HID models, respectively (right panels of the figure), at pressure changes of a few per cent. The temperature insets show the direct, individual-particle-level counterpart of this evolution: temperature rises rapidly and nearly monotonically as density decreases at each analyzed time, consistent with the statistical closure demonstrated quantitatively in Figure~\ref{f:tempDist} (Section~\ref{s:tempDist}).

\subsection{Spatial structure of the FHP sample}\label{s:fhpStructure}
The pressure distribution of the FHP sample provides indirect information -- through the hydrostatic equilibrium condition -- about the vertical position of fuel parcels. The arrangement of particle tracks in Figure \ref{f:trajectories} suggests that the FHP sample is not uniformly distributed inside the FSL, but rather is composed of potentially several physically compact clusters of tracer particles. This grouping of particles is most clearly discernible in the HID model (lower right panel in Figure \ref{f:trajectories}) where two large distinct groups of particles can be visually identified. The compactness of the particle clusters is studied in Section~\ref{s:clustering}.
\section{Discussion}\label{s:discussion}
The mechanism of the decrease of the fuel ignition times in the proximity to the flame, reported originally in Paper I and confirmed through our Lagrangian flow diagnostics in Section \ref{s:results}, deserves close scrutiny in terms of participating physics. Potentially related to this may be the dynamic interaction of the flame front with the fuel due to a combination of burning-induced buoyancy and flame thermal expansion. We discuss these aspects below. 
\subsection{Fuel heating mechanism}
\label{s:DiscussionHeating}
In order to identify the mechanism of fuel ignition time shortening with the distance to the flame front, we analyzed the thermodynamic evolution of tracer particles that show the greatest rates of ignition time decrease. The mechanism operates in two steps. First, the thermal expansion of the flame accelerates the upstream fuel and perturbs it away from hydrostatic equilibrium. The resulting modest, near-isobaric decrease of the fuel density is the initiating perturbation, not the heating itself. Second, the fuel responds by restoring the hydrostatically stratified background pressure, a process mediated by weak acoustic waves. Because the degenerate pressure component decreases with the density, restoring the pressure requires a redistribution of internal energy from the degenerate to the thermal components of the plasma. 
 
Along any path of decreasing density with $\mathrm{d}P\approx 0$ (the hydrostatic pressure variation length scale is much larger than the height of the expansion region) the plasma temperature must increase. This is because $(\partial P/\partial T)_\rho > 0$ and $(\partial P/\partial \rho)_T > 0$. This behavior is generic to any equation of state. What distinguishes the strongly degenerate fuel is the amplification. In this case, the thermal contribution to the pressure is initially only a small fraction of the total (degeneracy-dominated) pressure and restoring pressure balance after even a modest expansion requires a disproportionately large temperature excursion. The size of the temperature response is therefore not set by the size of the initiating perturbation: the density decrease can be small, and the temperature response per unit density change grows with the degeneracy of the fuel (Table~\ref{t:isobaric}). The steep temperature sensitivity of Equation~\ref{e:igtm} then converts this excursion into an order-of-magnitude change of the ignition time; in a non-degenerate gas the same density drop would be thermodynamically unremarkable.

Throughout this discussion we use the following decomposition of the internal energy provided by the Helmholtz equation of state \citep{Clayton83,Weiss+04}: the electron--positron contribution, which dominates the pressure budget and is largely set by the zero-temperature degenerate floor of Fermi--Dirac statistics; the ion, radiation, and Coulomb components. We quantify the electron-positron $P_\mathrm{e}$ component deviation from $T=0$~K value by evaluating the Helmholtz equation of state at the reference fuel density at a temperature of $1.7\times 10^{9}$~K, representative of the heated FSL fuel. The electron degeneracy parameter $\eta_\mathrm{e} = E_\mathrm{F}/k_\mathrm{B}T$ is thus $7.4$ (HID), $3.5$ (MED), and $1.6$ (LOD), and the corresponding thermal increase of the electron pressure --- i.e., the fraction of $P_\mathrm{e}$ not accounted for by the zero-temperature degenerate floor --- is $7$, $18$, and $35$ per cent, respectively. The zero-temperature-floor approximation is therefore good for HID, fair for MED, and modest for LOD. The electron term dominates the total pressure budget in all three models ($87$--$95$ per cent at this temperature), so the operational split into $T=0$~K floor and thermal part remains a reasonable leading-order characterization throughout. This has an immediate consequence for the fuel ignition times as evidenced by their decrease as the flame front approaches, as discussed in Section \ref{s:results}.

From the perspective of the ignition time evolution, based on Equation~\ref{e:igtm}, the ignition time can decrease only due to rising density, temperature, or carbon abundance. Of those three parameters, carbon abundance can only decrease in our model, and in particular in regions close to the flame front. Figure \ref{f:trajectories} indicates that during the evolution, the density gradually decreases for the FHP parcels. This leaves the temperature as the factor responsible for the observed reduction in ignition time. This is confirmed directly by Figure~\ref{f:tempDist}, which shows that Equation~\ref{e:igtm} at the reference density, so that only the measured temperature varies along the profile, accounts for the whole ignition-time reduction to within $0.05$ dex in all three models, whereas evaluating it at the reference temperature, so that only the measured density varies, leaves the ignition time essentially unchanged (at most $0.06$ dex of a $0.9$--$3.6$ dex effect). The measured fuel temperature itself rises by $\approx 2\times 10^{8}$ to $\approx 7\times 10^{8}$~K ($\approx 13$ to $63$ per cent above the unperturbed value) from the LOD to the HID model. We stress that the region considered here is not in thermal contact with the flame and there is no nuclear burning in this fuel. Therefore some thermodynamic process must be responsible for temperature increase and the observed shortening of fuel ignition times.

FHP pressure-density trajectories (cf.\ Figure \ref{f:trajectories}) suggest a very nearly isobaric evolution of this fuel population, with only a few percent pressure change observed. Quantitatively, the median density decrease of $15$--$18$ per cent at a median pressure change of at most about two per cent corresponds to an effective slope $|\mathrm{d}\ln P/\mathrm{d}\ln\rho| \approx 0.04$--$0.1$ (per-particle medians $-0.04$, $-0.07$, and $+0.11$ for LOD, MED, and HID). This is far below the adiabatic exponent $\Gamma_1 \approx 4/3$--$5/3$. The recorded paths are nearly isobaric and strongly non-adiabatic, and therefore involve a secular increase of the fuel entropy. We attribute the entropy increase to dissipation of the weak acoustic perturbations continuously generated by the unsteady flame, which repeatedly traverse the fuel ahead of the front. The evolution, however, cannot be isobaric in its strict thermodynamic sense as the fuel outflow is driven by pressure perturbations associated with the flame thermal expansion. Rather, the plasma state is constrained by hydrostatic equilibrium with pressure perturbations due to burning communicated by acoustics on a time scale much shorter than the flame crossing time scale of the FSL region.

The pressure, density, and temperature distributions underlying this near-isobaric argument are not sensitive to the specific $N=500$ FHP sample-size choice: repeating the selection at $N=250$ and $N=1000$, for all three models, leaves the pressure distribution essentially unchanged in LOD and HID (median shift $<0.4\%$) and shows a small residual sensitivity in MED ($\approx1.3\%$ at $N=1000$). Density and temperature show a small, monotonic trend with $N$ (typically $1$--$4\%$ per doubling, largest in HID at $\approx4\%$). This is consistent with FHP's own construction as a rank-ordered selection of the most extreme heating events because a larger sample necessarily admits progressively less extreme particles. None of these shifts approaches a magnitude that would affect the sufficiency of $N=500$ for identifying the heating mechanism itself, which is governed by the most extreme cases that defines the FHP population.

This confirms that acoustic-mediated processes, not nuclear burning, are responsible for the fuel temperature increase. In this process, the hydrostatic background pressure is restored by acoustics while the fuel pressure value is adjusted through a redistribution of internal plasma energies such that the loss of degenerate pressure component is compensated by the increase of the thermal components. Figure~\ref{f:tempDist} traces this redistribution directly in terms of its observable outcome, the fuel temperature, as a function of distance from the flame front. This complements the energy-partition view presented next.

The aforementioned behavior is evident when considering the distribution of internal energy components in the FSL (Figure \ref{f:igtmDist}). The internal energy of cold fuel, located at large distance from the flame front (near the right edge in the lower row figure panels) is dominated by the degenerate electrons with ions contributing from $\approx 10$\% in the LOD model to $\approx 2$\% in the HID model. At those large distances from the flame, the radiative component contribution in the HID case is essentially negligible, increases to about 1\% in the MED model, and a few per cent in the LOD model. The situation is dramatically different at short distances with the radiative component average contribution increasing by a factor of 2 to 10 between the LOD and HID models. We emphasize that the radiation component acts here as a diagnostic of the temperature rise rather than as its driver. This is because $e_\mathrm{rad}\propto T^4$ is the steepest-in-temperature term of the internal energy, it shows the largest fractional response to any temperature change. In absolute terms, however, the radiation contribution remains a minor fraction of the internal energy throughout the FSL. The causal variable controlling Equation~\ref{e:igtm} remains the temperature itself. We study the internal energy redistribution process in detail using a dedicated model as follows.
\subsubsection{Confirmatory one-dimensional model}\label{s:odcd}
We systematically study the behavior of fuel in the FSL region using a one-dimensional confirmatory diagnostic (ODCD) model. This model is not intended to reproduce quantitative amplitudes and only aims to isolate thermodynamic response mechanisms. Although simplified relative to the 3D models, it retains the essential physics — the same equation of state, hydrostatic stratification, and acoustic response — while isolating the redistribution process from advection, nuclear burning and the resulting thermal expansion, as well as a range of multidimensional effects such as unsteady flame propagation, mixing, and turbulence. This allows us to probe the system behavior in a controlled manner.

We use a 1D Cartesian domain extending 20 km with gravity pointing in the $-x$ direction. The mesh is uniformly refined with resolution matching that of our 3D models. The initial conditions consist of a flame front in hydrostatic equilibrium positioned at the center of the computational domain. To conserve the total mass, we impose closed (zero-flux) boundary conditions. The flame speed and energy deposition source term are both set to zero, leaving only the advection part of the ADR model active. At the beginning of the evolution, the density in the post-flame region with $\phi > \phi_\mathrm{pf}$ is lowered by a fixed fraction $f_{\rho\mathrm{h}} < 1$. In the experiments presented here $\phi_\mathrm{pf} = 0.995$, $f_{\rho\mathrm{h}} = 0.95$, with the density perturbation several times smaller than that produced by thermal expansion of the fully activated ADR flame.

The imposed density perturbation produces a pressure jump at the flame front and a slight force imbalance across the post-flame region that does not affect the subsequent evolution with  the fuel remaining in hydrostatic equilibrium. Once the evolution begins, the initial pressure discontinuity drives a compressive wave into the ash and a rarefaction wave into the fuel. With boundaries closed, both waves eventually reflect and collide, and this pattern of reflections and collisions continues throughout the simulation. Its interaction with the FSL mimics acoustic perturbations generated by an unsteady RT-unstable supernova flame. In the process, fuel and ash undergo rapid, low-amplitude perturbations, with the sound crossing time of the domain on the order of a few milliseconds. For the above conditions, the average maximum Mach number in the ODCD model increases with the fuel density from $\approx 0.009$ in the LOD model to $\approx 0.017$ in the HID case. These values are given for orientation only. They are not a quantitative counterpart of the rms fuel turbulence Mach numbers measured directly on the mesh in the 3D simulations (cf.\ Table~\ref{t:models}), which are substantially higher. The two statistics differ in kind (a time-averaged maximum of a closed-boundary purely acoustic field in one case vs.\ a volume-weighted rms of a turbulent velocity field in the other), and only the qualitative statement that both are small is intended. The ODCD acoustic field is not meant to reproduce the turbulent velocity environment of the 3D models quantitatively and should not be read as a proxy for it.

\begin{figure*}
    \centering
        \begin{tabular}{ccc}
        \includegraphics[height=0.225\textwidth]{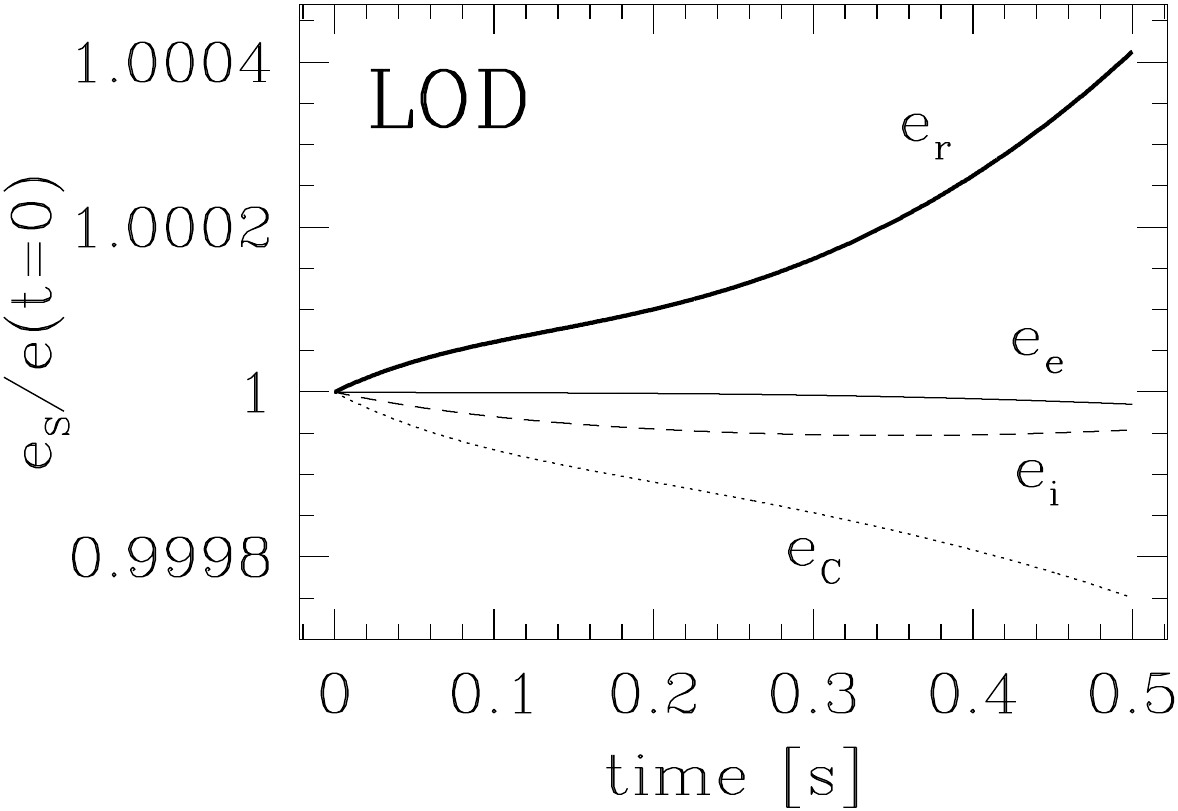} &
        \includegraphics[height=0.225\textwidth]{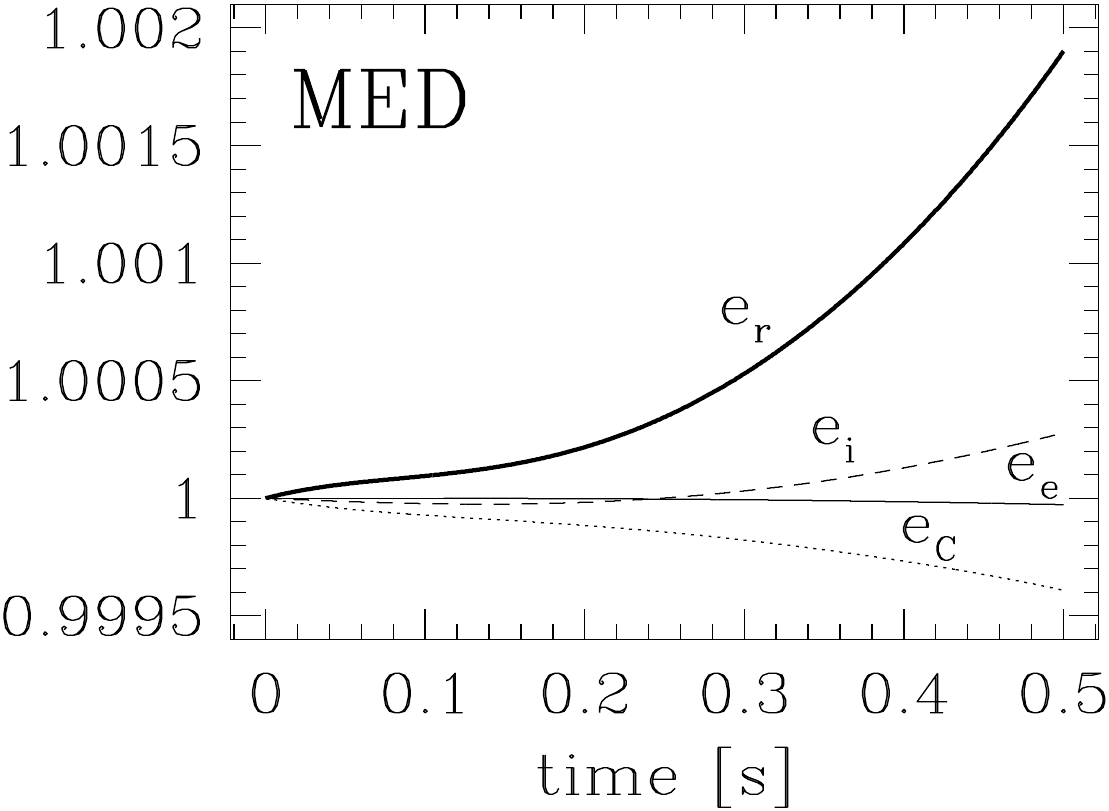} &
        \includegraphics[height=0.225\textwidth]{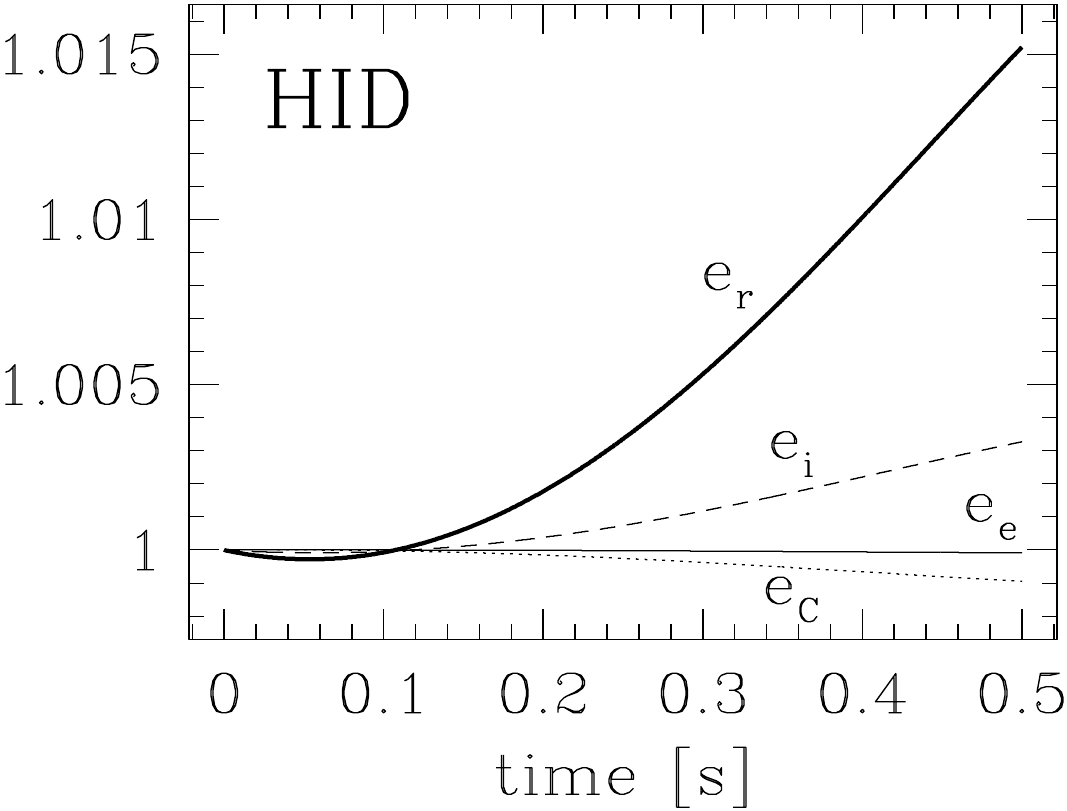} \\
    \end{tabular}
    \figcaption{Time evolution of relative internal energies in the ODCD model for nominal initial conditions of the LOD (left panel), MED (middle panel), and HID (right panel) model. The results are shown for internal energy due to ions (dashed line), electrons (thin solid), radiation (thick solid), and Coulomb interactions (dotted). Each energy component is normalized to its initial value. For clarity of presentation, a high-frequency oscillatory component was removed using a low-pass filter. See text for discussion.
    \label{f:deints1D}}
\end{figure*}

Relative contributions of ions, electrons, radiation, and Coulomb interactions to the internal energy budget of the FSL region were averaged over a layer 1.3 km thick beginning 0.15 km below the initial flame location, chosen such that the mass inside the monitored region remains approximately constant during the evolution. As the monitored region is periodically swept by waves, the amounts and relative contributions of the various internal energy components oscillate around their mean values. Because we are interested in systematic effects, rapid oscillations were removed using a low-pass filter. In all models, the radiation fraction of the internal energy increases over time, the electron contribution decreases, and the ion and Coulomb contributions vary with fuel density (Figure \ref{f:deints1D}). 
This confirms that acoustic perturbations can redistribute internal plasma energy among its components when the system evolution is constrained by hydrostatic equilibrium.

The ODCD results agree \emph{qualitatively} with the results of the Lagrangian analysis (cf.\ bottom row of Figure \ref{f:igtmDist}) with radiative component gradually increasing its contribution to the internal energy as the distance to the flame decreases. The temporal scaling of the ignition time discussed here is not an independent measurement from the ODCD simulation itself. This is because, as we noted above, ODCD's own flame speed and energy-deposition source term are both set to zero, and so no propagating flame exists inside it from which a time-slope could be measured. Instead, it is the spatial power-law exponent $\gamma$ already reported in Section~\ref{s:results} (Figure~\ref{f:igtmDist}), re-expressed on a nominal time axis using the ADR flame model's own predefined, held-constant subgrid-scale flame speed $v_{f,\mathrm{SGS}}$ \citep{Brooker+25}. Further assuming that the flame distance is proportional to the time remaining until flame arrival, $d_\mathrm{f}\propto t_\mathrm{a}$, where $t_\mathrm{a}$ decreases as the flame approaches a given parcel, we can write $\log_{10}d_\mathrm{f} = \log_{10}v_{f,\mathrm{SGS}} + \log_{10}t_\mathrm{a}$. This substitution preserves the exponent exactly,
\begin{equation}   
\mathrm{d}\log\tau_\mathrm{ign}/\mathrm{d}\log t_\mathrm{a} = \mathrm{d}\log\tau_\mathrm{ign}/\mathrm{d}\log d_\mathrm{f} = \gamma,
\end{equation}
independent of the value of $v_{f,\mathrm{SGS}}$ used. The slope with respect to the time remaining until flame arrival is therefore identical, by construction, to the exponent $\gamma$ reported above.

For physical orientation, at this same $v_{f,\mathrm{SGS}}$ the heating region itself ($d_\mathrm{f}\lesssim10^{4.5}$~cm, the threshold used throughout this analysis) would be crossed in $\approx35$, $26$, and $21$~ms for LOD, MED, and HID, respectively -- an order-of-magnitude timescale for the heating process itself, distinct from the much shorter acoustic transit time invoked elsewhere in this subsection. We again stress that only this trend, i.e.\ the steepening of the scaling with increasing fuel density, is claimed to agree between the ODCD model and the three-dimensional analysis. The ODCD model is not designed to reproduce quantitative amplitudes. We also note that, with the closed boundary conditions adopted here, the ODCD acoustic field is a repeatedly reflected wave pattern rather than the open, radiating field of an RT-unstable flame. The redistribution result should therefore be read as a demonstration that hydrostatically constrained acoustic perturbations are sufficient to drive the internal-energy redistribution, not as a quantitative model of the flame acoustics.
\subsubsection{Thermodynamic consistency
\label{s:consistency}}
Our analysis relies on the internal energy histories recorded along tracer trajectories. We therefore test whether the inferred heating is thermodynamically self-consistent rather than an artifact of the particle integration or the equation of state inversion. To this end we verify for a representative sample of tracer particles that the recorded change in specific internal energy is consistent with the first law of thermodynamics,
\begin{equation}
\mathrm{d}e = T\mathrm{d}S - P\mathrm{d}V + \dot{Q}_\mathrm{e,nuc}\,\mathrm{d}t,
\end{equation}
where $e$ is the specific internal energy, $T$ the temperature, $S$ the specific entropy, $P$ the pressure, $V$ the specific volume, and $\dot{Q}_\mathrm{e,nuc}$ the specific nuclear energy generation rate (identically zero for the fuel-region particles analyzed here, where no burning occurs). We quantify the level of consistency through the symmetric relative residual,
\begin{equation}\label{e:dflt}
\Delta_\mathrm{flt} = \frac{\left|\mathrm{d}e - (\bar{T} \mathrm{d}S - \bar{P} \mathrm{d}V + \dot{Q}_\mathrm{e,nuc}\Delta t)\right|}{ \frac{1}{2}\left[ \left|\mathrm{d}e\right| +\left|\bar{T} \mathrm{d}S - \bar{P} \mathrm{d}V + \dot{Q}_\mathrm{e,nuc}\Delta t\right|\right]}
\end{equation}
where $\mathrm{d}e$ is the internal energy change recorded directly along the trajectory and $\bar{T}\mathrm{d}S - \bar{P}\mathrm{d}V + \dot{Q}_\mathrm{e,nuc}\Delta t$ is the change reconstructed from the equation of state using the time interval-averaged temperature $\bar{T}$ and pressure $\bar{P}$, together with any nuclear energy release $Q_\mathrm{e,nuc}$ over the interval $\Delta t$ (identically zero for the fuel-region particles analyzed here, where no burning occurs). The two estimates are evaluated over model outputs separated by a representative interval of $\approx 0.25$ ms, which varies slightly by model (Table~\ref{t:cadence}). A value of $\Delta_\mathrm{flt}\ll 1$ indicates that the recorded energy change is consistent with the first law of thermodynamics at the given cadence, while $\Delta_\mathrm{flt} \sim 1$ flags either a genuine inconsistency or a residual dominated by numerical noise.

\begin{figure*}
    \centering
    \begin{tabular}{ccc}
        \includegraphics[height=0.33\textwidth]{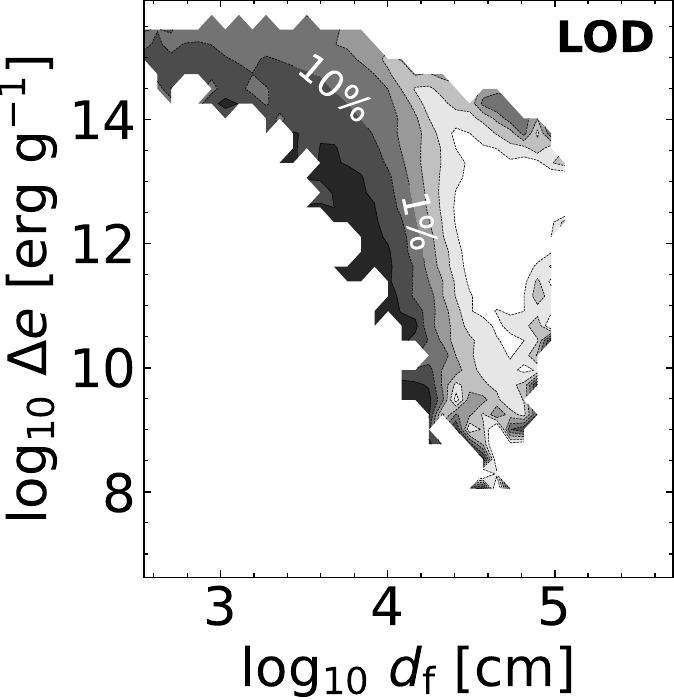} &
        \includegraphics[height=0.33\textwidth]{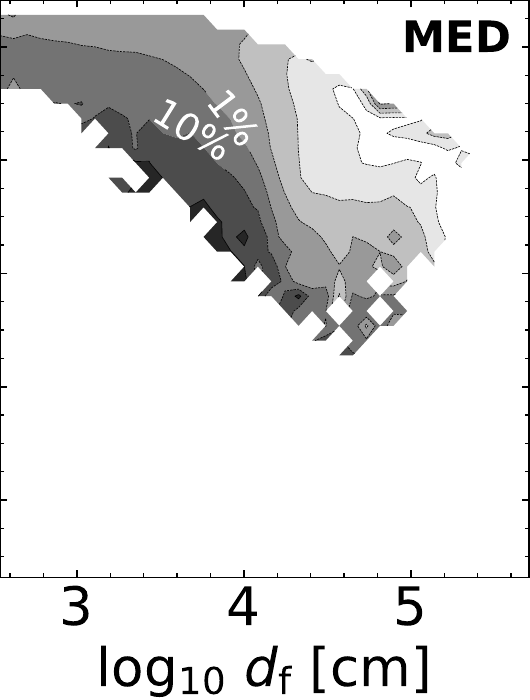} &
        \includegraphics[height=0.33\textwidth]{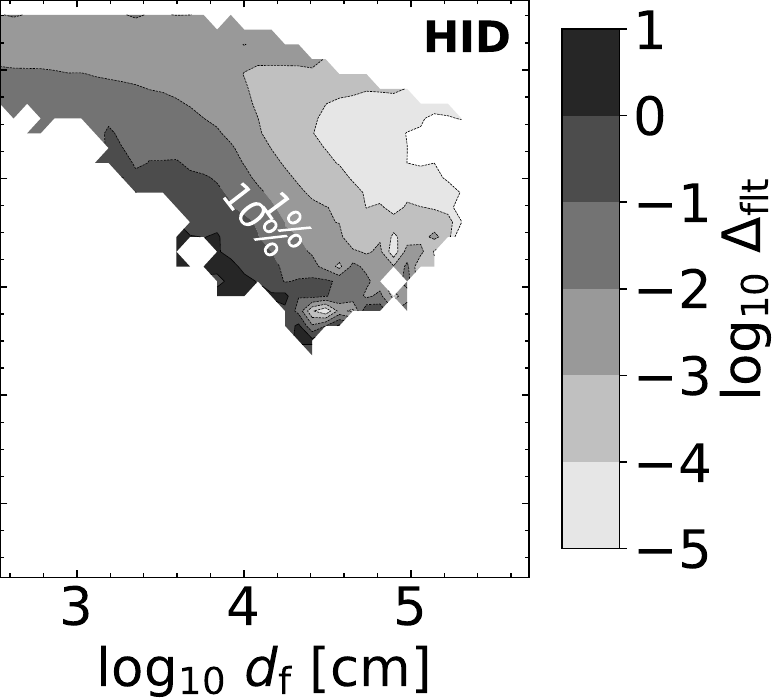}
    \end{tabular}
    \figcaption{First-law self-consistency of the recorded fuel trajectories in all three models (LOD, MED, HID; left to right). The thermodynamic consistency indicator, $\Delta_\mathrm{flt}$, is shown for all tracer particles as a function of distance to the flame front, $d_\mathrm{f}$, and the fuel internal energy gain, $\Delta e$. All quantities are shown in log-scale. Regions of parameter space not occupied by any particles are shown in white. All three panels use the anchor sets of Table~\ref{t:cadence}. See text for discussion.
    \label{f:thermo}}
\end{figure*}

Figure \ref{f:thermo}
shows the distribution of $\Delta_\mathrm{flt}$ for all tracer particles in the HID model as a function of the distance to the flame front, $d_\mathrm{f}$, and the internal energy gain, $\Delta e$.

The residual is small over most of the occupied parameter space, rising toward the small-$\Delta e$ corner. Most importantly, there is no region in which a large energy change coincides with a large first-law residual. The largest values of $\Delta_\mathrm{flt}$ are confined to particles with the smallest $\Delta e$, where the energy change approaches the noise floor set by the finite output frequency and the limit of interpolation accuracy of thermodynamic quantities to particle locations. In that regime the small denominator in Equation \ref{e:dflt} inflates the relative residual, while the absolute discrepancy in $\Delta e$ itself stays small. The strongly heated fuel located at short distances from the flame, which drives the ignition time shortening, shows the best consistency. The recorded trajectories are therefore thermodynamically self-consistent in the regime most important to this work. We note that this check is necessary, but not by itself sufficient to establish the physical nature of the heating. This is because the first law holds along any quasi-static path, and dissipation enters through the entropy term rather than as a violation of the first law. The physical interpretation of the heating rests instead on the near-isobaric character of the recorded paths, the ODCD experiment, and the correlation structure discussed below.

Our choice of the time differencing interval warrants additional scrutiny. At the model sound speed of $\approx 5500$ km s$^{-1}$, the sound crossing time of the heating layer, $d_\mathrm{f} \lesssim 10^{4.5}$ cm, is $\approx 0.06$ ms. This is roughly four times shorter than the $\approx 0.25$ ms interval over which $\Delta_\mathrm{flt}$ is evaluated above. The first-law check therefore averages over several acoustic transits of the FSL, and the closure characterizes the slow, systematic, hydrostatically constrained evolution of the fuel. Had the diagnostic time interval been shorter, our consistency metric would probe the rapid acoustic oscillations that mediate the pressure adjustment rather than the systematic effect. The fuel turbulence Mach numbers measured in the region initially occupied by the tracer particles are low ($0.038$--$0.073$, Table~\ref{t:models}). The fuel velocity perturbations therefore remain modest, the interval-averaged $\bar{T}$ and $\bar{P}$ are expected to faithfully represent the thermodynamic path, and it is unlikely that the small residuals are solely due to aliasing of acoustic signals.

\begin{figure*}
    \centering
    \begin{tabular}{@{\hspace{0mm}}c@{\hspace{2mm}}c@{\hspace{2mm}}c@{}}
        \includegraphics[height=0.335\textwidth]{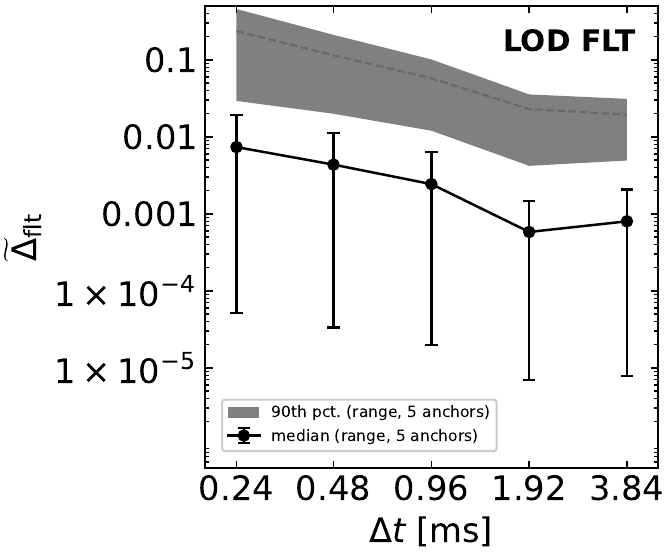}   &
        \includegraphics[height=0.335\textwidth]{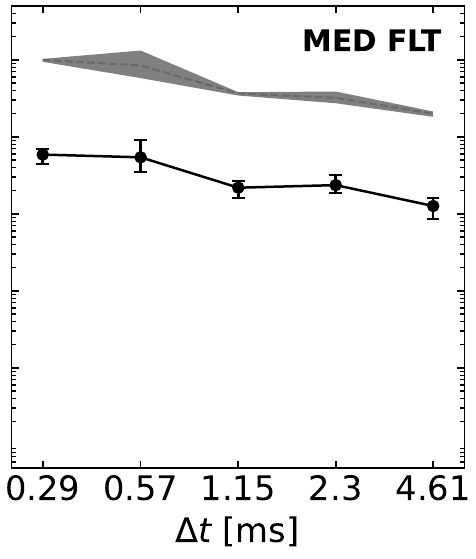} &
        \includegraphics[height=0.335\textwidth]{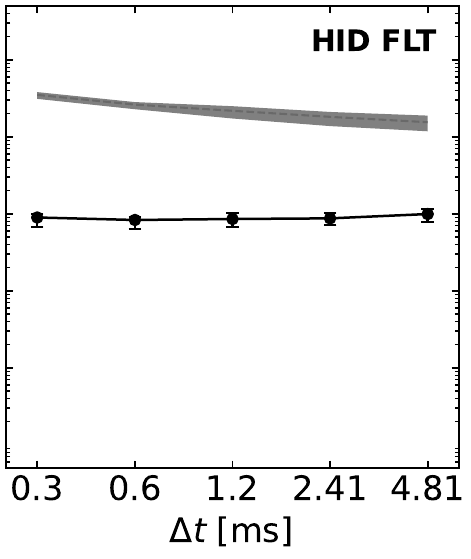}
    \end{tabular}
    \figcaption{Median and 90th percentile of $\widetilde{\Delta}_\mathrm{flt}$ as a function of the differencing interval $\Delta t$, for the LOD, MED, and HID models (left to right). Filled circles show the median and error bars its range over the five anchors; the dashed line shows the 90th percentile and the shaded band its range over the five anchors. This is the figure-form complement to Table~\ref{t:cadence}. All three panels use the anchor sets of Table~\ref{t:cadence}.
    \label{f:cadence_fig}}
\end{figure*}

To make the robustness of the closure with respect to the differencing interval explicit, we recomputed $\Delta_\mathrm{flt}$ for a range of output cadences in a dedicated set of particle-instrumented continuation runs of all three models. For a particular model, a series was identified by an anchor snapshot --- a reference model output. To eliminate a potential selection effects, 4 additional anchors (for a total of 5 anchors) were chosen at simulation times separated sufficiently to ensure independence of anchor sets. For each of those additional anchors, a set of dependent snapshots was identified using the adopted cadence intervals. Table~\ref{t:cadence} provides parameters of anchor sets for each model. Each model's own achieved differencing interval corresponds to its own particle-instrumented output frequency and differs across models: HID spans a factor of $16$, from $0.30$ to $4.8$ ms, and MED and LOD span closely similar ranges, $0.29$ to $4.6$ ms and $0.24$ to $3.84$ ms, respectively.
\begin{deluxetable}{ccccccc}
\tablewidth{0pt}
\tablecaption{Median first-law residual as a function of the differencing interval for all three models.
\label{t:cadence}}
\tablehead{
  \colhead{}              &
  \multicolumn{2}{c}{LOD} &
  \multicolumn{2}{c}{MED} &
  \multicolumn{2}{c}{HID} \\
  \colhead{Position}                                                      &
  \colhead{$\Delta t$ (ms)} & \colhead{$\widetilde{\Delta}_\mathrm{flt}$ (\%)} &
  \colhead{$\Delta t$ (ms)} & \colhead{$\widetilde{\Delta}_\mathrm{flt}$ (\%)} &
  \colhead{$\Delta t$ (ms)} & \colhead{$\widetilde{\Delta}_\mathrm{flt}$ (\%)}
}
\startdata
1 & 0.24 & 0.23 & 0.29 & 0.58 & 0.30 & 0.09 \\
2 & 0.48 & 0.14 & 0.58 & 0.50 & 0.60 & 0.08 \\
3 & 0.96 & 0.08 & 1.15 & 0.21 & 1.20 & 0.09 \\
4 & 1.92 & 0.02 & 2.30 & 0.23 & 2.41 & 0.09 \\
5 & 3.84 & 0.03 & 4.60 & 0.12 & 4.81 & 0.10
\enddata
\tablecomments{Each model's own $\Delta t$ is a representative value for its own achieved output cadence. The anchor-averaged median values, $\widetilde{\Delta}_\mathrm{flt}$ (Equation~\ref{e:dflt}), are computed across each model's own 5 independent analysis anchors, for each of the 5 tested cadence positions. Every one of these median values, across the full model-by-cadence-position grid, falls between $\approx7\times10^{-4}$ and $\approx2$ per cent, with no cadence-driven breakdown for any model over the tested range.}
\end{deluxetable}
Every value across the full grid, for all three models, corresponds to $\Delta_\mathrm{flt}$ between $\approx 7\times10^{-4}$ and $\approx 2$ per cent. We found no cadence-dependent breakdown of the first-law consistency anywhere in the tested range, for any model. This demonstrates that the consistency shown at one differencing interval is not an artifact of that particular choice. In the LOD model, we observe a largely consistently decreasing trend except a minimum near $\Delta t \approx 1.9$~ms. The best MED consistency is at the widest $\Delta t$ tested while HID is essentially flat across its own full range. The figure-form result for all three models is shown in Figure~\ref{f:cadence_fig}.
\subsubsection{Spatial clustering of the FHP sample}\label{s:clustering}
The FHP sample is selected on the outcome, the strongest reduction of the ignition time, and its members are therefore unlikely to be independent draws from the fuel. This subsection examines whether they are spatially clustered. The answer determines how the uncertainties of the correlation coefficients discussed in Section~\ref{s:correlations} have to be estimated.

\paragraph{Banding of the trajectories and spatial compactness.} It is likely that the selection criterion of the FHP sample results in spatial correlations between its members, which demonstrate themselves as discrete banding of particle trajectories in Figure \ref{f:trajectories}. Such banding is particularly conspicuous in the HID trajectory plots (bottom row of panels in Figure \ref{f:trajectories}) in which case, one can identify three distinct groups of particles. Two of those groups, with initial pressures around $1.05\times 10^{25}$ and $1.07\times 10^{25}$ $\mathrm{dyne}\ \mathrm{cm}^{-2}$, appear to evolve differently from the third group of particles (initial pressure around $1.13\times 10^{25}$ dyne cm$^{-2}$), with their pressure showing the greatest relative decrease across all the models. Recall that in our model the pressure is related to the vertical position of the plasma sample and thus the banding reflects spatial closeness of particles in terms of their height.

\begin{figure*}
    \centering
    \begin{tabular}{ccc}
        \includegraphics[height=0.255\textwidth]{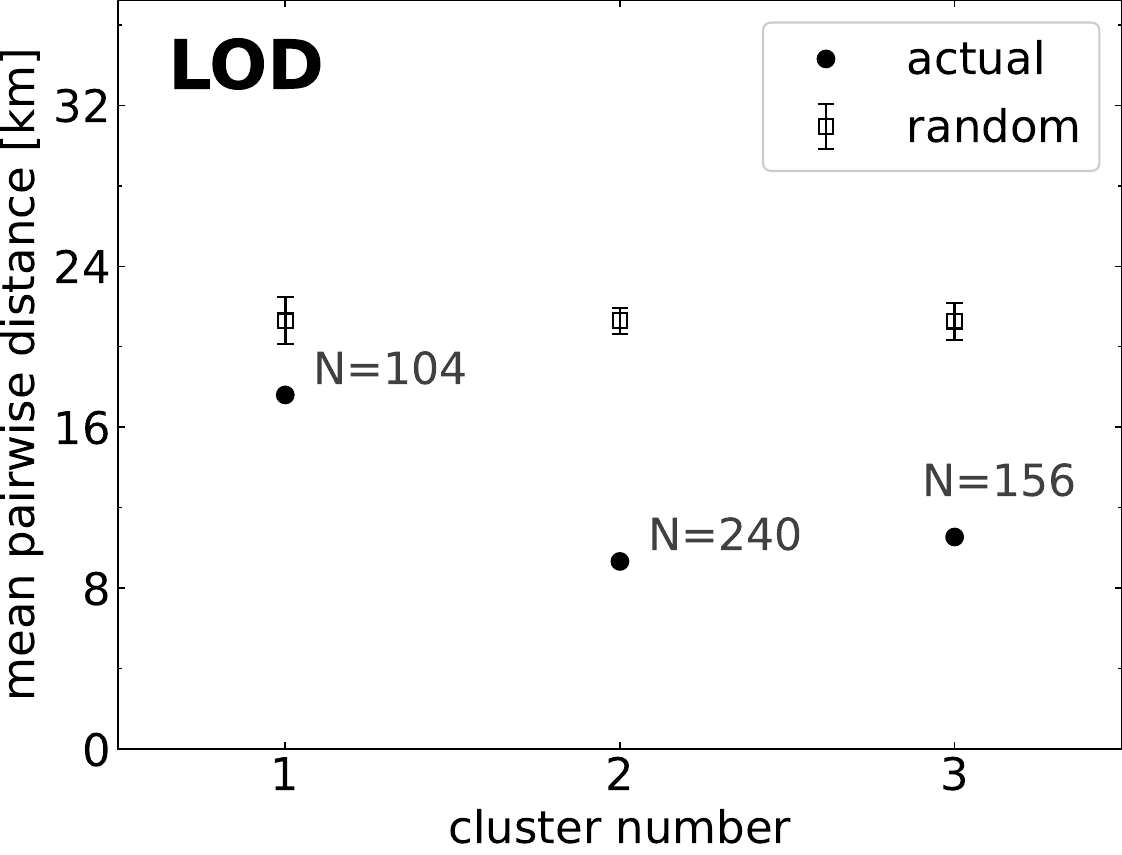}      &
        \includegraphics[height=0.255\textwidth]{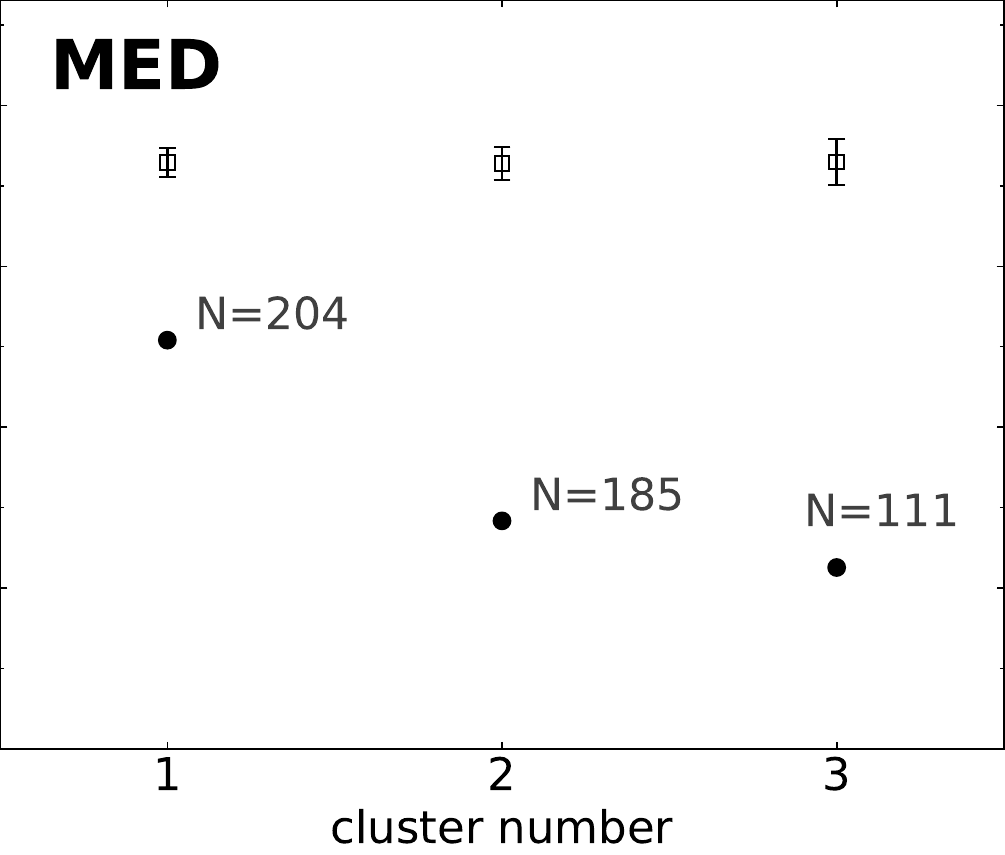}    &
        \includegraphics[height=0.255\textwidth]{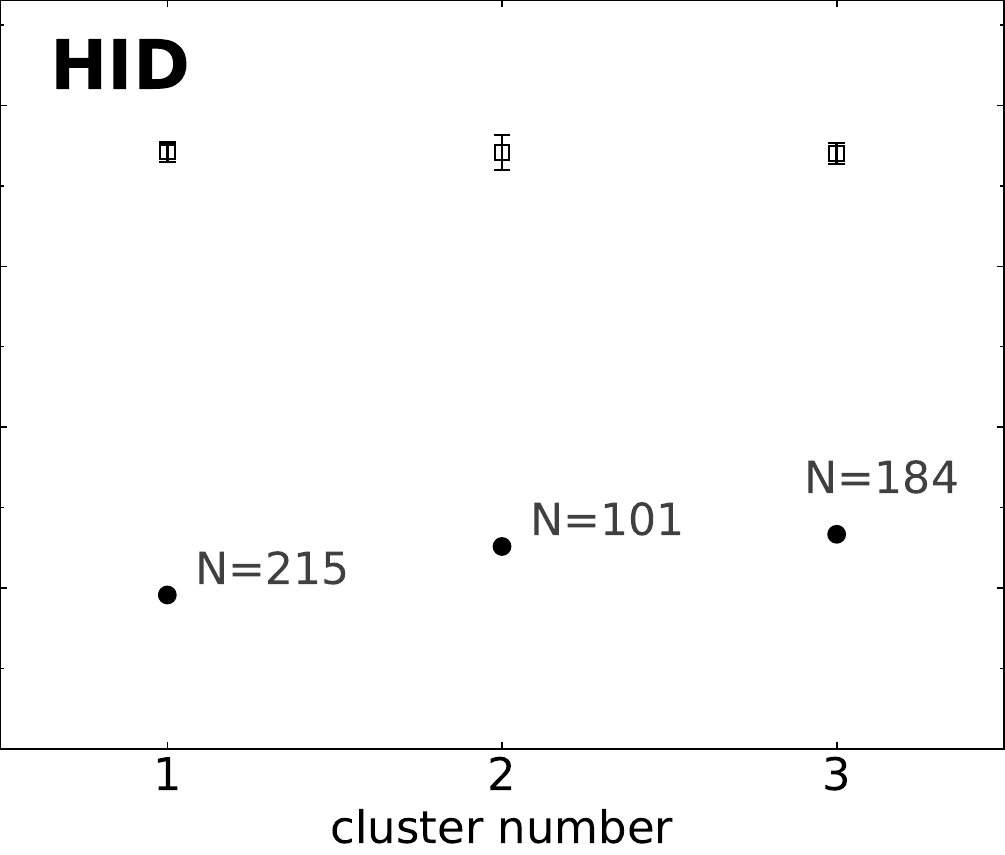}
    \end{tabular}
    \figcaption{Spatial compactness of the pressure-defined FHP clusters (cf.\ Figure~\ref{f:trajectories}) for the LOD, MED, and HID models (left to right). Filled circles show the actual mean pairwise 3D distance within each cluster, with the clusters numbered in order of increasing mean pressure. Open squares show the same statistic for a random ensemble in which cluster membership is shuffled among the same particles, with error bars giving the standard deviation of that ensemble. Labels give the number of particles, $N$, in each cluster; the standardized effect sizes (SES) are discussed in the text. Every cluster is more compact than its null ensemble ($1000$ permutations of cluster membership, exact $p\le1/1001$).
    \label{f:spatial_corr}}
\end{figure*}

To test whether fuel particles with similar initial pressure also occupy similar locations, we first partitioned the population of $500$ tracked FHP fuel particles of each model into three clusters by one-dimensional $k$-means clustering of the initial pressure. We numbered the clusters in order of their increasing average pressure. For each cluster we computed the mean pairwise 3D Euclidean distance, $\bar{d}$, between the initial positions of its members. Next, to assess whether such compactness could arise without any connection between pressure and position, we generated a null ensemble by pooling all particles, randomly reassigning them to groups with the same sizes as the actual clusters (that is, shuffling the cluster assignments among the same particles), and recomputing $\bar{d}$ for each group. This was repeated $N_\mathrm{p}=1000$ times for each model. The significance of each cluster is the one-sided exact permutation $p$-value, $p=(N_\mathrm{null}+1)/(N_\mathrm{p}+1)$, where $N_\mathrm{null}$ is the number of null realizations with $\bar{d}_\mathrm{null}\le\bar{d}_\mathrm{actual}$ \citep{DavisonHinkley1997,NorthCurtisSham2002}. Because $N_\mathrm{null}=0$ for all nine clusters (three per model), every cluster reaches the smallest value the test can return, $p=1/(N_\mathrm{p}+1)\simeq10^{-3}$. This is a resolution limit set by $N_\mathrm{null}$ and not a measure of how extreme the result is. We therefore also report the standardized effect size, $\mathrm{SES}=(\langle\bar{d}_\mathrm{null}\rangle-\bar{d}_\mathrm{actual})/\sigma_\mathrm{null}$ \citep{GotelliMcCabe2002}, where $\langle\bar{d}_\mathrm{null}\rangle$ and $\sigma_\mathrm{null}$ are the mean and standard deviation of the null distribution. All nine clusters are more compact than the null ensemble (Figure~\ref{f:spatial_corr}), with $p\le1/1001$. The result also holds after a Bonferroni correction \citep{Dunn1961} for the nine tests ($p_\mathrm{adj}\le9/1001\simeq0.009$).

Since a single significance statement cannot distinguish the clusters from one another, the SES is the informative measure. It ranges from $3.2$ to $46.2$, exceeding the conventional threshold of $2$ in all nine clusters. It varies by more than an order of magnitude, from the weakest cluster in LOD to the strongest in HID (the model that is most visibly banded in Figure~\ref{f:trajectories}). The populations comprise particles that remain unburned through the final analyzed snapshot, and the result therefore describes this retained subset and not the full initial FHP selection. In hydrostatic fuel the pressure encodes height, and compactness of pressure-defined clusters is thus expected at least in the vertical direction. The test shows that the clusters are localized groups of particles, but it does not show that they are compact laterally, which was not tested. We also note that members of the same cluster are not independent, which is taken into account in the uncertainty estimates of Section~\ref{s:correlations}.
\subsubsection{Interactions of parameters}\label{s:correlations}
This subsection is organized as follows. The correlation matrices are introduced first, followed by their sensitivity to the particle population, which includes the comparison with the control populations. The uncertainty of the coefficients and their interpretation in terms of the heating mechanism are considered next, and the comparison with Paper~I concludes the subsection.

\begin{figure*}
    \centering
        \begin{tabular}{@{}c@{\hspace{-2mm}}c@{\hspace{-2mm}}c@{\hspace{-2mm}}c@{}}
        \includegraphics[height=0.260\linewidth]{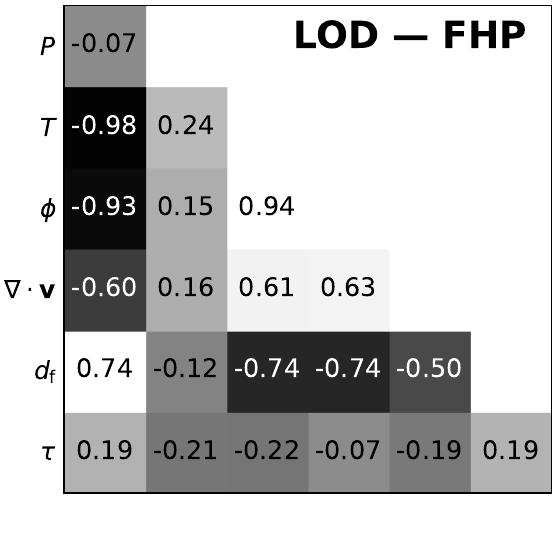}   &
        \includegraphics[height=0.260\linewidth]{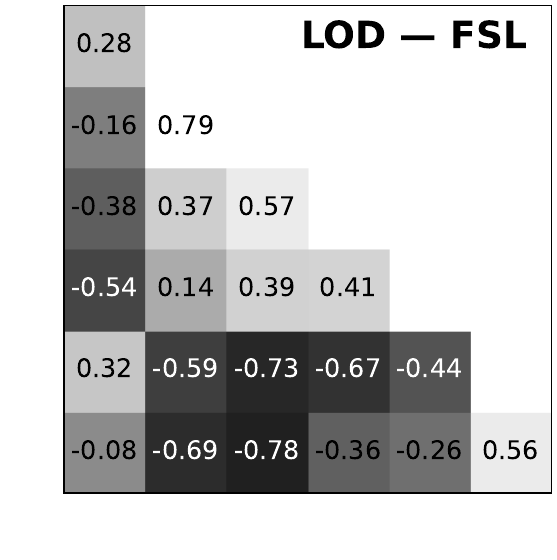} &
        \includegraphics[height=0.260\linewidth]{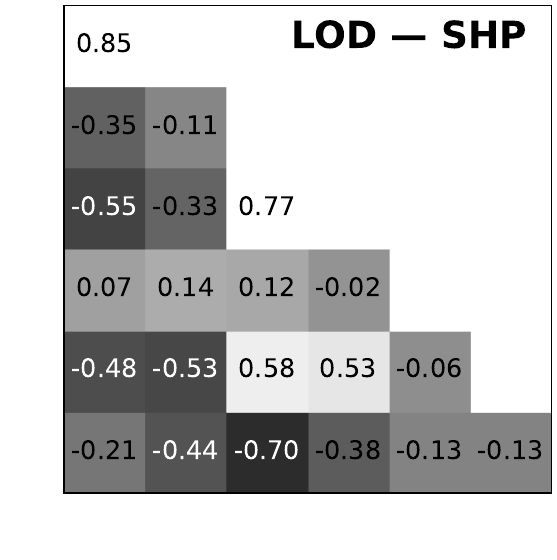}  &
        \includegraphics[height=0.260\linewidth]{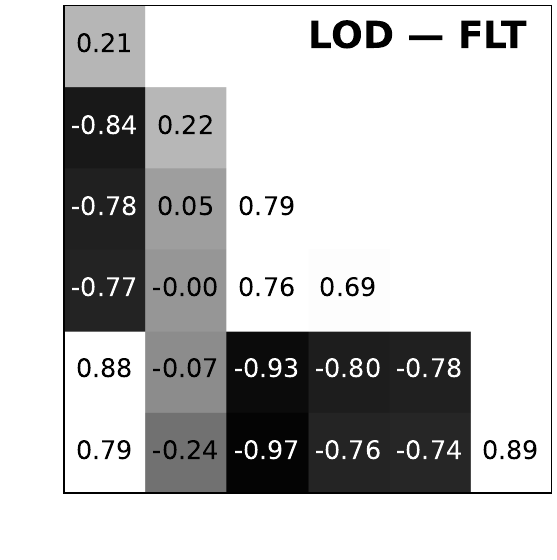}    \\[-4mm]
        \includegraphics[height=0.260\linewidth]{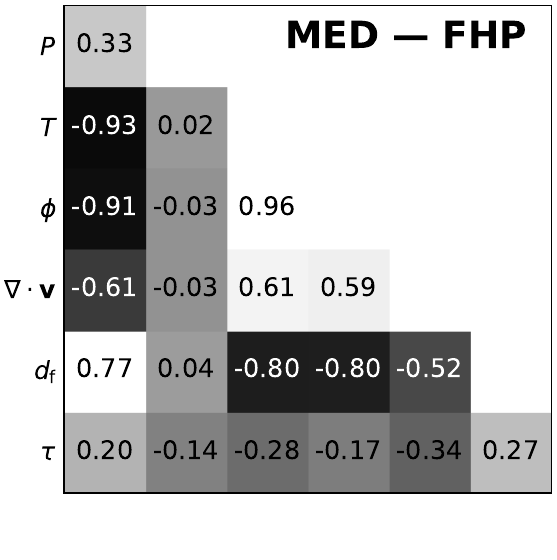}   & 
        \includegraphics[height=0.260\linewidth]{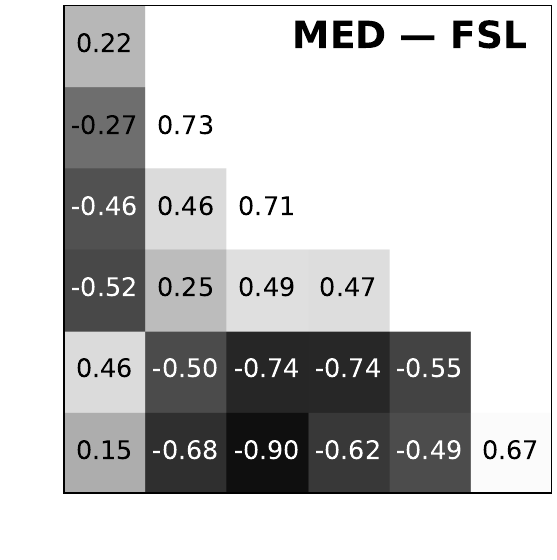} &
        \includegraphics[height=0.260\linewidth]{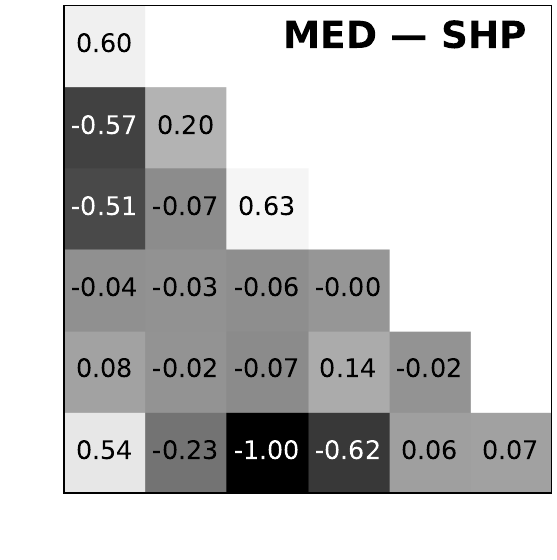}  &
        \includegraphics[height=0.260\linewidth]{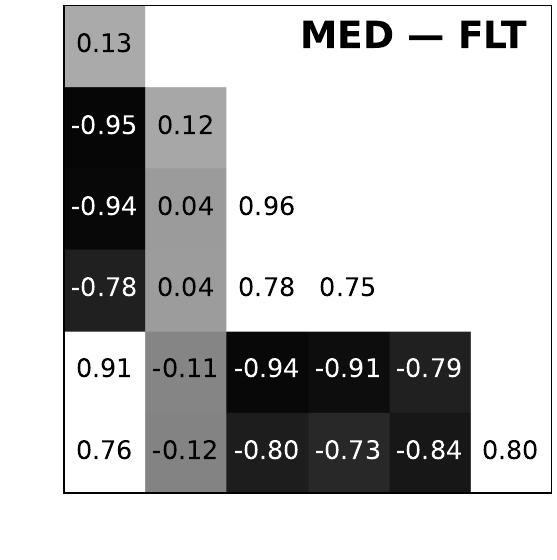}    \\[-4mm]
        \includegraphics[height=0.260\linewidth]{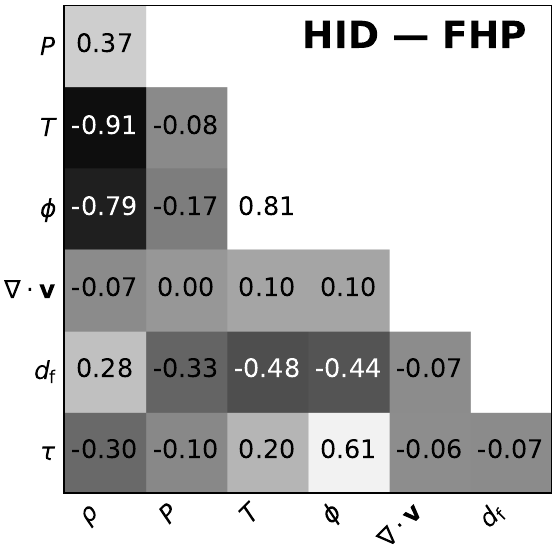}   & 
        \includegraphics[height=0.260\linewidth]{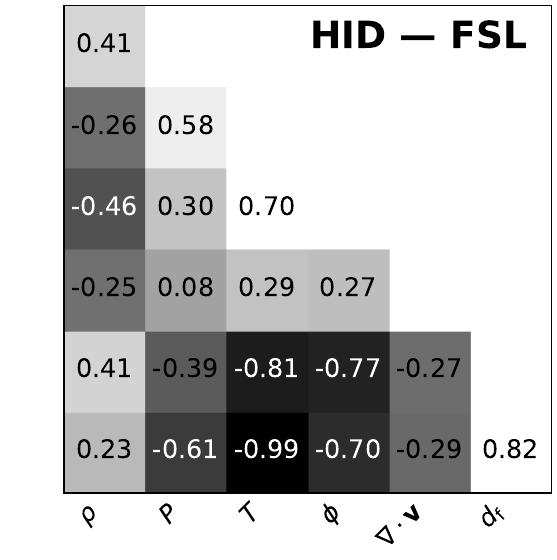} &
        \includegraphics[height=0.260\linewidth]{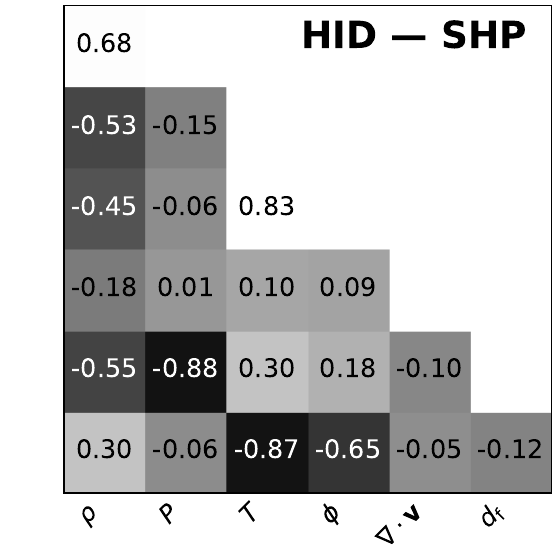}  &
        \includegraphics[height=0.260\linewidth]{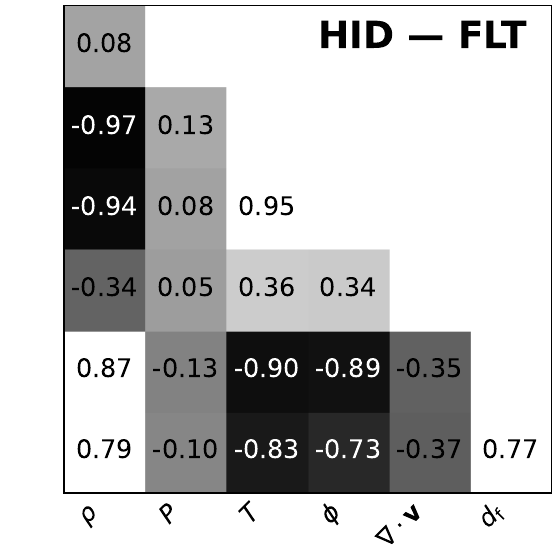}
        \end{tabular}
    \figcaption{Mutual dependencies of a subset of physical properties, expressed in terms of Spearman correlation coefficients. The coefficients are presented in matrix format for the LOD (top row), MED (middle row), and HID (bottom row) models, and for the FHP, random FSL, SHP (the Slowest-Heated Particles), and FLT populations in the first, second, third, and fourth column, respectively. The rows follow the LOD, MED, and HID model ordering used in Figure~\ref{f:trajectories}. The fourth column shows dependencies for the full FLT particle population, which is not subject to the ignition-time-outcome selection defining FHP and SHP populations. The dependencies shown are between the density, $\rho$, pressure, $P$, temperature, $T$, flame progress variable, $\phi$, velocity divergence, $\bm{\nabla \cdot v}$, distance to the flame front, $d_\mathrm{f}$, and the carbon ignition time, $\tau$. The correlation coefficient values are shown in the corresponding matrix cells. The matrix cells are colored with different shades of gray from strongly negative correlations in black, to weak correlations in medium gray, to strong positive correlations in white. See text for discussion.
    \label{f:correlations}}
\end{figure*}

\paragraph{Correlation matrices.} Interactions between various physical properties of the RTI flame model in the FSL region described in Section \ref{s:consistency} can be illustrated with the help of the Spearman correlation coefficients calculated for various particle populations of interest. Figure \ref{f:correlations} 

shows correlation coefficients computed at the simulation's final snapshot, for particle populations selected using each particle's own full-trajectory ignition-time reduction, in the form of a correlation matrix. Individual matrix cells are gray-shaded according to the correlation coefficient value, with dark gray for strongly negative correlations and white for strongly positive correlations. Coefficient values are shown for the LOD, MED, and HID models in the top, middle, and bottom row, respectively, while the four columns show the FHP, random FSL, SHP (the Slowest-Heated Particles, i.e.\ those undergoing the least ignition-time reduction), and FLT populations, left to right.

\paragraph{Sensitivity on the population.} Based on the similar structure of the correlation matrices, the mutual interactions between different physical quantities appear to be broadly similar across all models, with the exception of the FHP dataset. For the primary variable of interest, ignition time, only pressure and velocity divergence correlations retain a consistent sign across all three models. In contrast, correlation coefficients for the remaining four variables ($\rho$, $T$, $\phi$, $d_\mathrm{f}$) each differ in sign for HID relative to LOD and MED.  This is not unexpected and reflects the selection methodology of the FHP set and the use of a single snapshot at the end of our simulation time to construct these correlation matrices. Because the FHP sample is selected on the outcome and the key quantities are strongly mediated by the flame distance, raw bivariate coefficients in this sample cannot by themselves be read as structural relationships.

To separate direct from distance-mediated dependencies we computed partial Spearman coefficients controlling for $d_\mathrm{f}$, and we repeated the full correlation analysis for two control populations, a random sample of FSL particles and SHP, as well as for the FLT population. For the random FSL sample the raw ignition time-density correlation is weak and inconsistent in sign across models. For the SHP population it is more strongly positive and also inconsistent. Both random FSL and SHP populations nevertheless show substantially stronger and more consistent correlations of ignition time with the temperature and the flame progress variable than FHP, more comparable in magnitude to those of the FLT population.

\begin{figure*}
    \centering
    \begin{tabular}{@{\hspace{-12mm}}c@{\hspace{-10mm}}c@{\hspace{-10mm}}c@{}}
    \includegraphics[height=0.350\textwidth]{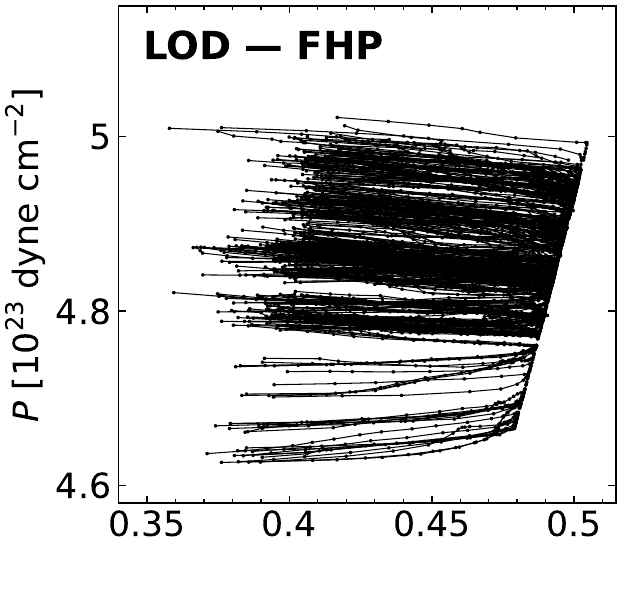}      &
    \includegraphics[height=0.350\textwidth]{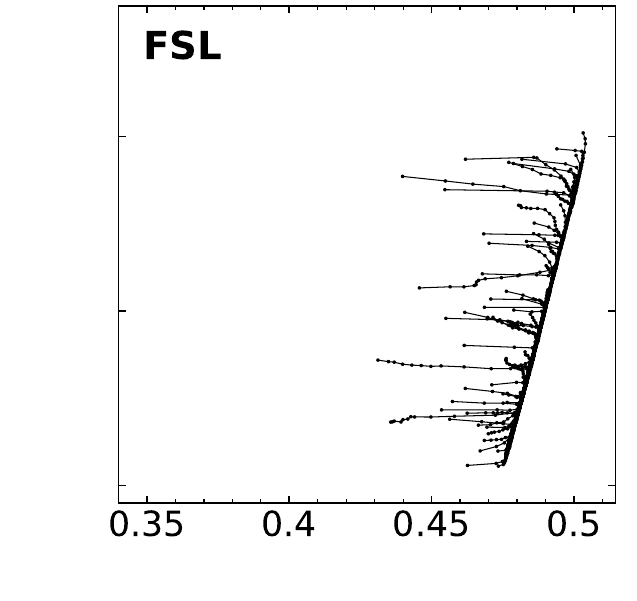}    &
    \includegraphics[height=0.350\textwidth]{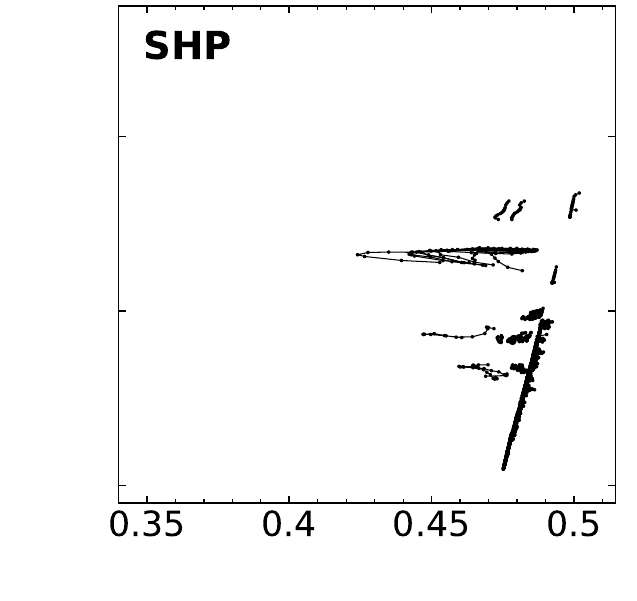}     \\[-5mm]
    \includegraphics[height=0.350\textwidth]{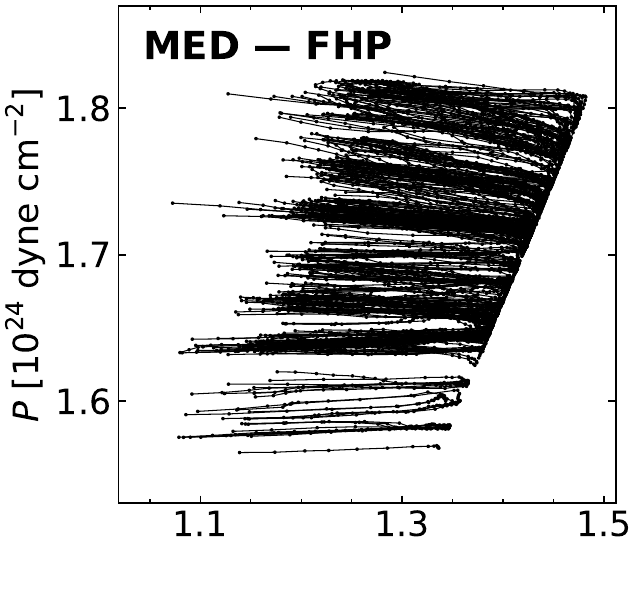}   &
    \includegraphics[height=0.350\textwidth]{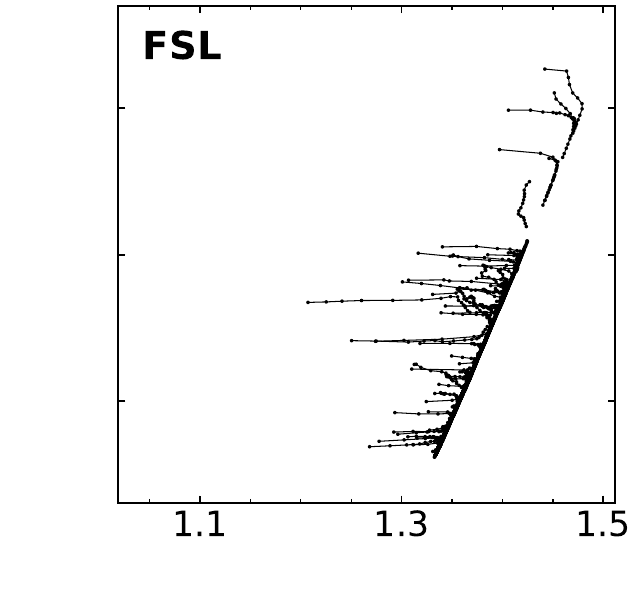} &
    \includegraphics[height=0.350\textwidth]{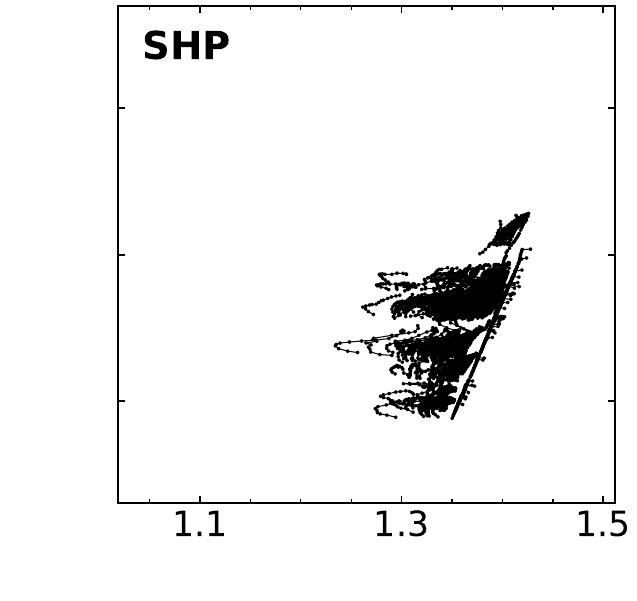}  \\[-5mm]
    \includegraphics[height=0.350\textwidth]{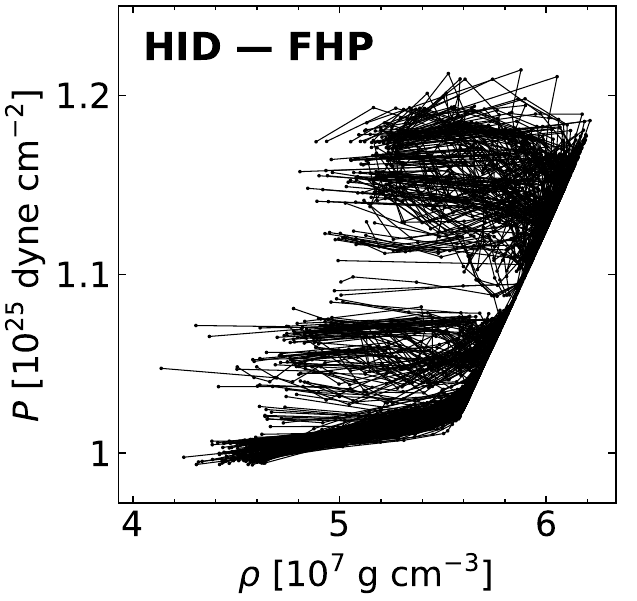}   &
    \includegraphics[height=0.350\textwidth]{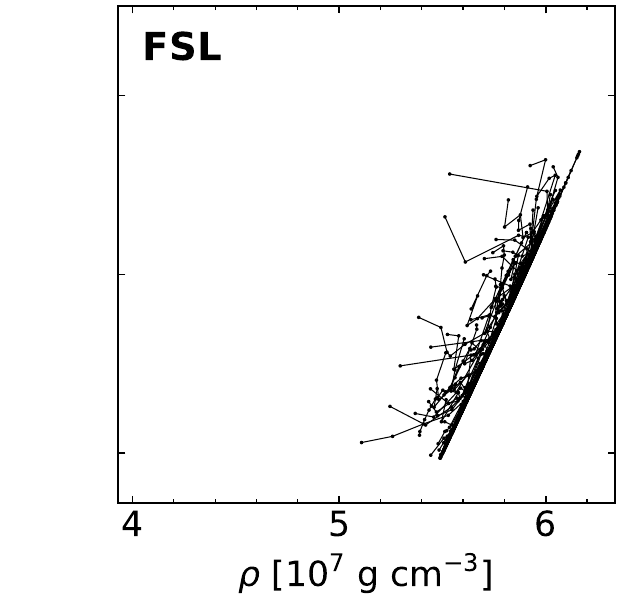} &
    \includegraphics[height=0.350\textwidth]{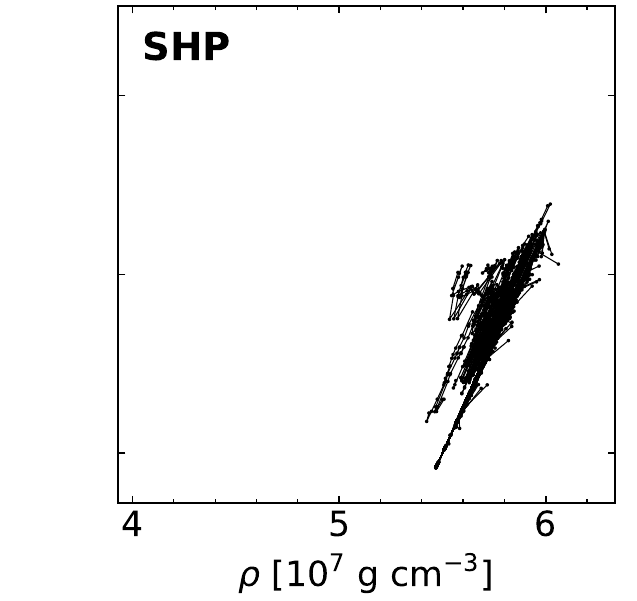}
    \end{tabular}
    \figcaption{Density--pressure trajectories (cf.\ Figure~\ref{f:trajectories}) for the FHP sample (left column) compared against two control populations: a random sample of FSL particles (``FSL,'' middle column) and SHP (``SHP,'' right column), for the LOD, MED, and HID models (top to bottom).
    \label{f:populations}}
\end{figure*}

Figure~\ref{f:populations}
shows the density--pressure trajectories of the three populations in the same representation as Figure~\ref{f:trajectories}. The FHP densities decrease at nearly constant pressure in all three models, whereas most of the random FSL and SHP trajectories stay close to a narrow band along which density and pressure change together, and the density excursions of those that depart from it are smaller. The near-isobaric expansion is therefore strongest in the sample selected for the largest ignition-time reduction and weakest in SHP, and the FHP sample should not be regarded as representative of the fuel as a whole. How common such trajectories are among all FSL particles is quantified in Table~\ref{t:isobaric} and discussed below. The weaker and less consistent FHP correlations (Figure~\ref{f:correlations}) can be explained as follows.

The ignition time is a highly nonlinear function of temperature (cf.\ Equation~\ref{e:igtm}) that exhibits steep dependence at low-to-moderate temperature and flattens as temperature rises. An FHP particle, selected specifically for having already undergone the largest ignition time reduction by its own termination point, is by construction already at or near that same termination point by the time the shared reference snapshot is reached. Both its temperature and ignition time have already saturated, so the remaining variation in temperature among FHP's own members produces almost no variation in ignition time.

On the opposite extreme are slowest-heated particles (SHP) which are still early in their own heating at the shared reference snapshot. Those particles sit on the steep part of the ignition time curve where temperature variation still produces meaningful changes in ignition time. Random FSL particles are drawn without reference to trajectory progress, naturally span a mix of steep- and flat-curve particles at the shared snapshot, and preserve a strong correlation. Last but not least, a partial correlation controlling for $d_\mathrm{f}$ does not restore FHP's own ignition time-temperature relationship either, with the (LOD, MED, HID) correlations reported nearly equal ($-0.12$, $-0.11$, $+0.19$). Those weak correlations are consistent with a weak ignition time-temperature dependence following from the $\tau(T)$ curve's own shape at FHP's particular temperature range rather than a distance-mediated confound. In other words, the extreme-tail selection of FHP attenuates its correlations relative to the other populations which are less selection-restricted.

A substantially more consistent picture between models emerges from the FLT population, which is not selected on the ignition-time outcome and shows the same-signed, several-times-stronger ignition time correlations in all six properties analyzed across all three models, e.g.\ with temperature below $-0.8$, flame progress variable below $-0.7$, and flame distance below $-0.7$. We regard the FLT-based correlations as the more reliable structural signal for this reason while the FHP-based results as secondary and confirmatory where the two agree, rather than as the primary quantitative evidence.

\paragraph{Prevalence of the initiating perturbation and size of the response.} We quantified how common the near-isobaric expansion is among all FSL particles by comparing the thermodynamic state of every FSL particle at the last and the first analyzed snapshot.

In the present context, isobaric does not mean that the pressure is strictly constant. The fuel is hydrostatically stratified and is crossed by weak acoustic waves. In consequence, the pressure of a plasma parcel follows the local hydrostatic background and varies slightly. What distinguishes the FHP particles (Figure~\ref{f:trajectories}) is that their density decreases substantially while their pressure hardly changes: the pressure changes are much smaller than the density changes. This separates them from two other kinds of paths. A parcel that changes its height in the stratified fuel changes its density and pressure together, with $\Delta\ln P/\Delta\ln\rho\approx 1.5$ (the stratification band of Figure~\ref{f:populations}), and an adiabatic change involves a ratio of order unity, comparable to the adiabatic exponent. Based on these properties we define a particle as following a nearly isobaric expansion path if its density decreased by at least a given amount and its pressure changed by no more than a quarter of the logarithmic density change, $|\Delta\ln P| \le 0.25\,|\Delta\ln\rho|$. The factor of a quarter is a heuristic: it lies well below the slopes of both the stratification band and adiabatic paths, and above the near-zero slopes of the FHP trajectories. We require a density decrease of at least $10$ per cent and, to test the sensitivity of the result, also of at least $2$ per cent, since the density changes of the strongly preconditioned particles in the denser models turn out to be small.

The great majority of the FSL population changes their properties very little over the analyzed window (Table~\ref{t:isobaric}) with its median density and temperature remaining essentially unchanged. Only a small minority of the FSL particles, a fraction that increases with the density of the model, has its ignition time reduced by at least an order of magnitude. An even smaller fraction, well below one per cent, shows a near-isobaric expansion by at least $10$ per cent. The FHP sample, a minute fraction of the FSL, lies in this tail: nearly all of its members expand by at least $10$ per cent (none by more than about $29$ per cent), with pressure changes of a few per cent at most. The near-isobaric expansion in Figure~\ref{f:trajectories}, and the strong preconditioning that follows from it, are therefore properties of a small fraction of the FSL, not of the fuel layer as a whole.

The expansion needed for strong preconditioning is small, and it becomes smaller with increasing fuel density (Table~\ref{t:isobaric}). In LOD nearly all of the strongly preconditioned particles, those with an ignition-time reduction of at least an order of magnitude, expand by at least about $5$ per cent at nearly constant pressure, and most of them by at least $10$ per cent. In MED, and especially in HID, a large share of them expands by only a few per cent or less. At the same time, their temperature rises by a larger factor. The ratio of the logarithmic changes of temperature and density, $\Delta\ln T/\Delta\ln\rho$, of these particles grows in magnitude with the density of the model, and so does that of FHP. This trend is consistent with the degeneracy amplification discussed in Section~\ref{s:DiscussionHeating}: the thermal contribution to the pressure is a progressively smaller fraction of the total in the more degenerate fuel, so a smaller expansion requires a larger temperature excursion to restore the pressure. A small expansion can therefore be sufficient for strong preconditioning in the denser models, and the density change alone is not a measure of the heating.
\begin{deluxetable}{lccc}
\tablewidth{0pt}
\tablecaption{Prevalence of near-isobaric expansion and the temperature response in the FSL, from the last relative to the first analyzed snapshot.\label{t:isobaric}}
\tablehead{
  \colhead{Quantity} & \colhead{LOD} & \colhead{MED} & \colhead{HID}
}
\startdata
Analysis window\tablenotemark{c} (ms) & 42.9 & 22.5 & 17.5 \\
FSL particles ($10^6$) & 2.02 & 2.10 & 2.03 \\
$\log_{10}(\tau_\mathrm{i}/\tau_\mathrm{f})\ge 1$ (\% of FSL) & 0.91 & 4.7 & 8.8 \\
$\rho_\mathrm{f}/\rho_\mathrm{i}\le 0.9$, near-isobaric\tablenotemark{a} (\% of FSL) & 0.56 & 0.21 & 0.21 \\
\quad same, (\% of particles with $\log_{10}(\tau_\mathrm{i}/\tau_\mathrm{f})\ge 1$) & 61 & 4.6 & 2.4 \\
$\rho_\mathrm{f}/\rho_\mathrm{i}\le 0.98$, near-isobaric\tablenotemark{a} (\% of FSL) & 6.2 & 4.8 & 2.4 \\
\quad same, (\% of particles with $\log_{10}(\tau_\mathrm{i}/\tau_\mathrm{f})\ge 1$) & 100 & 80 & 27 \\
FSL median $\rho_\mathrm{f}/\rho_\mathrm{i}$ & 0.999 & 1.004 & 0.989 \\
FSL median $T_\mathrm{f}/T_\mathrm{i}$ & 1.000 & 1.002 & 0.999 \\
FHP median $\rho_\mathrm{f}/\rho_\mathrm{i}$ & 0.82 & 0.85 & 0.82 \\
FHP median $P_\mathrm{f}/P_\mathrm{i}$ & 1.007 & 1.011 & 0.979 \\
FHP median $T_\mathrm{f}/T_\mathrm{i}$ & 1.32 & 1.58 & 2.55 \\
FHP median $\Delta\ln T/\Delta\ln\rho$\tablenotemark{b} & $-1.38$ & $-2.90$ & $-4.93$ \\
$\log_{10}(\tau_\mathrm{i}/\tau_\mathrm{f})\ge 1$: median $\rho_\mathrm{f}/\rho_\mathrm{i}$ & 0.895 & 0.951 & 0.965 \\
$\log_{10}(\tau_\mathrm{i}/\tau_\mathrm{f})\ge 1$: median $T_\mathrm{f}/T_\mathrm{i}$ & 1.19 & 1.22 & 1.33 \\
$\log_{10}(\tau_\mathrm{i}/\tau_\mathrm{f})\ge 1$: median $\Delta\ln T/\Delta\ln\rho$\tablenotemark{b} & $-1.55$ & $-3.86$ & $-8.31$ \\
\enddata
\tablenotetext{a}{Density ratio at or below the stated value and $|\Delta\ln P|\le 0.25\,|\Delta\ln\rho|$. FSL particles are the tracers that remain unburned at the last snapshot ($\phi<0.03$) and have a valid ignition time at both snapshots; the FHP sample comprises 500 of them.}
\tablenotetext{b}{Median over the particles of the group whose density decreases.}
\tablenotetext{c}{Time spanned by the 15 snapshots used in this analysis. For MED, and marginally for HID, it is shorter than the analyzed interval in Table~\ref{t:models}, because the trajectory sample ends before the last snapshot of the FUEL run.}
\end{deluxetable}
\paragraph{Uncertainty of the coefficients.} The above assessment is also borne out quantitatively by the accuracy of the coefficients themselves. We estimated 95\% confidence intervals for every correlation coefficient in this analysis via a cluster-aware bootstrap. In the accuracy analysis we used 1000 resamples, drawn at the level of whole pressure-defined particle clusters rather than individual particles, following the same pressure-based clustering as in the spatial-compactness test of Section~\ref{s:clustering}. We note that resampling whole clusters, rather than individual particles, is necessary here because the small, 500-particle FUEL-based populations are themselves spatially clustered, as shown in Section~\ref{s:clustering}. In consequence, a naive particle-level bootstrap or analytic standard error may understate the true uncertainty by treating non-independent particles as independent.

For the single most-discussed coefficient in this analysis, density-ignition time, the resulting intervals differ sharply in width between the small, selection-restricted FUEL populations and the much larger FLT population. LOD SHP analysis yields a correlation coefficient of $-0.21$, but a 95\% confidence interval is $[-0.52,+0.93]$ admits either sign. In contrast, the LOD's own FLT value, $+0.79$, has a tight interval of $[+0.77,+0.92]$. The same pattern holds for the other two models. This reflects the difference in sample size and the selection criteria: FLT's eligible population is the full particle field near the flame ($2$--$3\times10^5$ particles per model), while each FUEL-based population is a fixed, 500-particle subset drawn from a much larger pool and, for FHP and SHP specifically, selected on an extreme tail of the outcome variable itself. Pressure illustrates the dependence on the population. Its correlation with the ignition time is weak in the FHP and FLT populations in all three models, which is consistent with the near-isobaric character of the evolution of the strongly preconditioned fuel established above: the reduction of the ignition time proceeds with little accompanying change of pressure. In the random FSL sample, however, the correlation is moderate and consistently negative, and in SHP it is weak to moderate and varies between the models. In the random FSL sample the pressure follows the hydrostatic stratification, and hence the height and the distance to the flame front, so this correlation is most likely mediated by $d_\mathrm{f}$ and does not indicate a pressure-driven change of the ignition time.

\paragraph{Physical interpretation.} A number of correlations point to well-understood dependencies of the physical quantities or their model representations. For example, the flame progress variable, $\phi$, is strongly correlated with temperature, $T$, but strongly anti-correlated with the distance to the flame front proper, $d_\mathrm{f}$, in the FLT population (see below); these two dependencies are the clearest and most consistent in the dataset. Within the FHP population specifically, the raw $\rho$--$\tau$ correlation is weakly positive for LOD and MED ($+0.19$, $+0.20$) but already negative for HID ($-0.30$) -- the sign implied by Equation~\ref{e:igtm} throughout. For LOD and MED, a suppression-effect argument applies directionally: within the FHP sample, $\rho$ and $T$ are not independent -- both are driven by the same underlying cause, proximity to the flame ($d_\mathrm{f}$): closer to the flame, $\rho$ decreases (the near-isobaric expansion established in Section~\ref{s:lagrangian}) while $T$ increases (the heating mechanism itself).

Because $\tau$ depends on $T$ far more steeply than on $\rho$ at fixed composition (cf.\ Equation~\ref{e:igtm}), the raw bivariate $\rho$--$\tau$ correlation is partly offset by $T$'s opposite-signed effect riding along on the shared $d_\mathrm{f}$ dependence. Controlling for $d_\mathrm{f}$ moves the correlation further toward the sign implied by Equation~\ref{e:igtm} in all three models: from $+0.19$, $+0.20$, and $-0.30$ (raw) to $+0.07$, $-0.02$, and $-0.29$ (partial) in LOD, MED, and HID, respectively -- MED crosses to the expected sign under this control, LOD moves substantially toward it without fully crossing, and HID, already correctly signed in the raw value, changes little. This pattern is specific to the FHP population; the FLT population (discussed further below), which is not selected on the ignition-time outcome and so does not carry this same confounding structure, shows a consistent, strongly negative $\tau$--$T$ correlation throughout ($-0.97$, $-0.80$, $-0.83$ for LOD, MED, HID) and, unlike FHP, does not require this suppression argument to explain its own sign. The temperature is also positively correlated with the velocity divergence, $\bm{\nabla \cdot v}$, in both the FHP and FLT populations and all three models -- the departure from the typically observed effect of the $P \mathrm{d}V$ work, and an independent confirmation, from a separate analysis path than the trajectory plots or the ODCD experiment, that expansion rather than compression accompanies the heating. These correlation analysis results are consistent with the proposed fuel heating mechanism, in which the heating is not caused by fuel compression; rather, fuel expansion perturbs the fuel away from hydrostatic equilibrium, and the restoration of pressure balance under the isobaric conditions enforced by hydrostatic equilibrium heats it and thereby decreases the fuel ignition time.
\paragraph{Relation to Paper~I.}Finally, we comment on the relation of the present picture to the interpretation offered in Paper I, which hypothesized that the ignition-time shortening arose from adiabatic compression of fuel by flame-induced buoyant acceleration. That interpretation was motivated by the identification of a new effect and by an instantaneous Eulerian velocity-divergence and flame-orientation analysis, rather than by a dedicated Lagrangian investigation of parcel histories. Paper I explicitly noted that its figure could not establish the time evolution of individual fuel parcels:
\begin{quote} Although time evolution of fuel ignition times cannot be inferred from the figure, the relative orientation of the flame front with respect to gravity distinguishes fuel initially compressed by the accelerating front from fuel associated with flame segments that are later overturned and decelerated.
\end{quote} 
The present analysis provides the missing time-integrated perspective. Particles experiencing the strongest ignition-time reduction show a net near-isobaric expansion, while the positive $T$--$\nabla\cdot\bm{v}$ correlation shows that their heating is associated with expansion rather than compression. The spatial association is nevertheless consistent between the two studies: the strongest effect occurs near the tops of rising flame bubbles, where the present analysis finds fuel expansion, while sinking spikes are associated with compression (Figure~\ref{f:fuelDivV}). Thus the present results do not require the instantaneous compression signal identified in Paper I to be absent; rather, they show that the hypothesized compressional interpretation does not describe the integrated thermodynamic history of the fuel that experiences the strongest ignition-time reduction. We therefore view the present result as a refinement of the mechanism hypothesized in Paper I, replacing an inference from instantaneous kinematics with a directly measured Lagrangian thermodynamic history. The larger FLT population, which samples the heating layer without selection on the magnitude of ignition-time reduction, gives a substantially stronger and more consistent view of this region than the outcome-selected FHP sample, providing an additional check that this interpretation is not confined to it; by construction FLT covers only the near-front portion of each parcel's history, the relevant window for a near-front mechanism, though a parcel that compressed early and expanded only near the front would register here as expansion-only.
\subsection{Potential consequences for the SN Ia explosion process}
The preconditioning mechanism identified here is one of several ways in which a propagating flame can alter the state of the fuel ahead of it, and other routes to ignition have been discussed in the literature. Fuel ignition times can in principle be shortened by a radiative preheat \citep{Karlin11}, which is not significant under our conditions. The effects of flame-fuel interactions were extensively discussed in the past by several groups, including the effects of flame thermal expansion on the upstream fuel in the terrestrial combustion context \citep{Sabelnikov+19,Sabelnikov+22}. Acoustic perturbations generated by the unsteady flame may be sufficiently amplified to ignite fuel and potentially lead to detonations \citep{Liberman+10,Poludnenko+19}. The intrinsic flame instabilities are frequently invoked as precursors of detonations, provided the instability-induced wave is sufficiently strong and either reactivity gradients \citep{Clavin25} or upstream turbulence \citep{Poludnenko+19} favor its strengthening. However, studies concerned with small scale flame instabilities \citep[e.g.][]{Poludnenko+19} require a DNS approach and fail to capture the flame thermal expansion that occurs on mesoscales. Related DNS work characterizes small-scale intermittency arising from premixed flame-front cusps \citep{Roy+23}, a further small-scale feature outside the reach of the present mesoscale analysis. Finally, we recall that nuclear burning is disabled in the fuel analyzed in this work: no ignition occurs in our models, and only ignition time scales are computed. The mechanism identified here is therefore a preconditioning mechanism, and its consequences, including any link to ignition or detonation, remain to be established with burning-enabled models.

\citet{Zeldovich+70} first recognized a role of reactivity gradients in formation of detonations in premixed combustion. The following theoretical formulation \citep{Zeldovich80} laid the groundwork for the delayed-detonation explosion scenario of \citet{Khokhlov1991a}, further developed by \citet{Khokhlov1997b}, with later research focused on turbulent preconditioning of partially burned thermonuclear fuel \citep{Woosley+11,CiaraldiSchoolmann+13,Brooker+25}.

Turbulent preconditioning and radiative preheat do not seem of significance in our application. The fuel in our model is initially assumed laminar and remains so even in the FSL of the flame front. The observed shortening of the FSL ignition times could, in a burning-enabled calculation, permit ignition of fuel disconnected from the primary flame. Whether this occurs, and how such events feed back on the flame dynamics, cannot be determined from the present models, in which the flame is the only source of nuclear energy. Warmer fuel may also change a laminar flame speed as temperature gradients will be shallower.

The turbulence Mach numbers measured directly on the mesh are low in all three models and support the view that the fuel velocity field is a weak perturbation of its thermodynamic state. The time-averaged fuel turbulent RMS Mach number in the region initially occupied by the tracer particles is less than $0.1$ in all models (Table~\ref{t:models}). These values are upper bounds for the perturbations experienced within fuel parcels, because the sinking fuel spikes move nearly ballistically, with essentially no internal perturbation, at speeds comparable to or exceeding RMS velocity fluctuation, and this coherent motion contributes to the turbulent velocity fluctuations. The RMS turbulent velocities reported in Paper~I for fuel embedded in the turbulent post-flame region are higher by a factor of $2$--$3$, but also well below the sound speed. The two sets of numbers are not directly comparable, primarily because they differ in the region sampled. The comparison is therefore qualitative. Since both sets of values are small, the conclusions drawn here do not depend on the difference.

Collisions of flame segments produce flame cusps. These structures can also be seen as fuel RT spikes. The elongated shape of those narrow regions combined with shortened ignition times resembles the preconditioned-fuel channels identified in low-density turbulence models in which DDT occurred via the Zel'dovich mechanism \citep{Brooker+21}.

The turbulent post-flame region, which separates the carbon and oxygen flame fronts and grows in extent as the deflagration advances \citep{Woosley+11}, may provide an additional site for DDT. Its increasing volume raises the probability that acoustic perturbations or isolated FSL ignition events produce conditions within the TPF region locally sufficient for oxygen detonation initiation. Temporal correlations between ignition events \citep{Lee+78} may further increase this probability super-linearly. 

We emphasize that no quantitative DDT criterion is evaluated in this work: the Chapman--Jouguet length scale relevant to our conditions ($\sim 2\times 10^{3}$ cm; Paper I) lies below the $62.5$ m mesh resolution of the present models, and the perturbation amplitudes reported here  (density decreases of at most about $30$ per cent along the FHP tracer trajectories, and average maximum Mach numbers of $\lesssim 0.02$ in the ODCD model of Section~\ref{s:odcd}) do not by themselves constitute detonation conditions. The statements above identify plausible pathways whose viability must be tested with burning-enabled models.
\subsection{The energy transfer in RT-unstable deflagrations}
In the adopted physical model, nuclear energy generated by the flame has two major effects. First, it expands the initially cold fuel, converting it into hot burning products. The flame thermal expansion accelerates fuel upstream of the flame, perturbing it away from hydrostatic equilibrium. This initiates a system response that aims at restoring hydrostatic pressure with the help of weak sound waves. Second, thermal expansion of the initially dense, cold fuel results in generation of gravitational potential energy. The resulting buoyancy force sustains the RT instability. The newly created gravitational potential energy is added to the pool of available potential energy and, combined with the RT instability, partially converted into turbulent motions. These motions transport kinetic energy along the cascade, create pressure perturbations, mix plasma, and are ultimately dissipated at small scales.

\begin{figure}
    \centering
    \includegraphics[width=\linewidth]{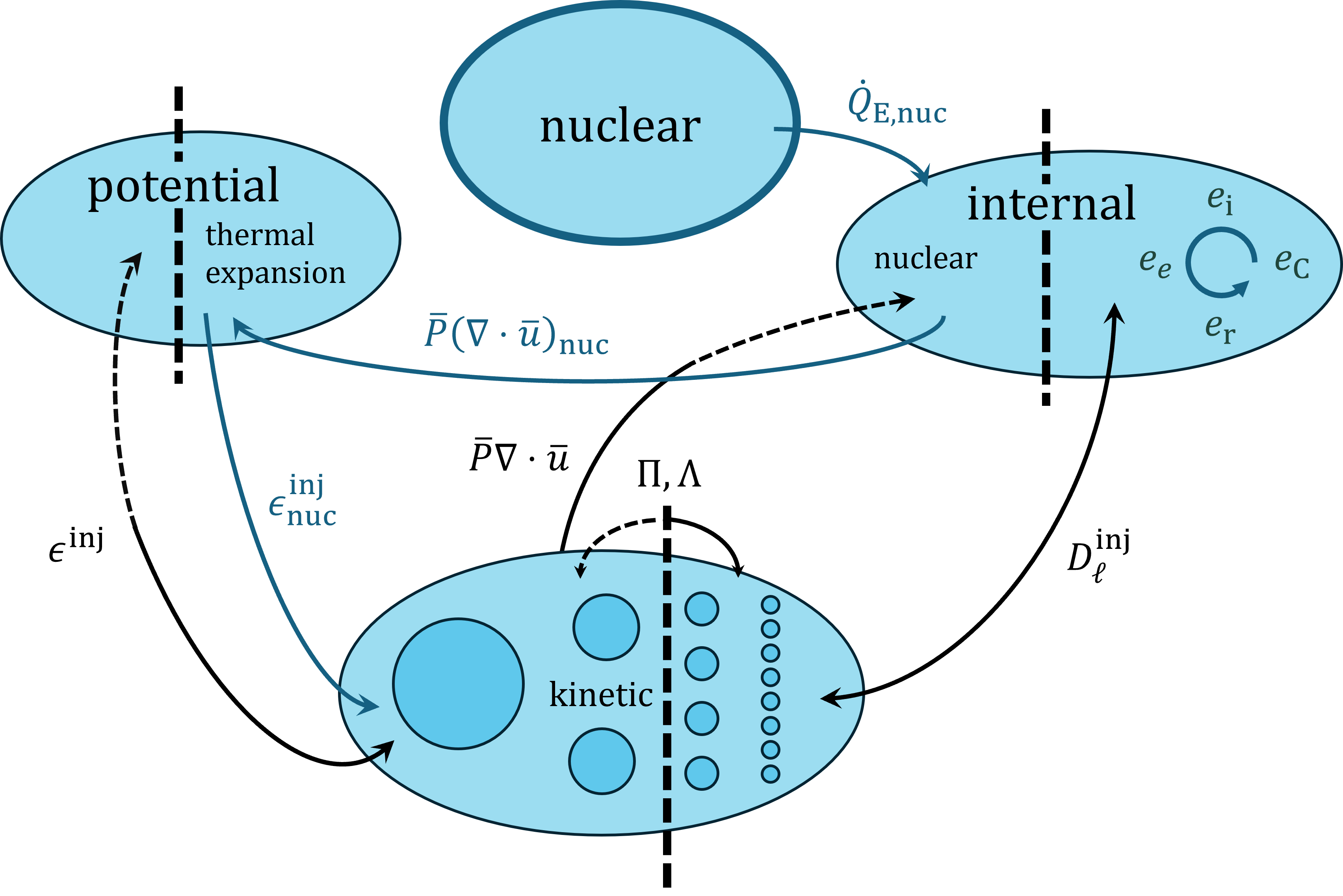}
    \figcaption{Conceptual model of generation and transport of various energy components in RTI-unstable deflagration. The reservoirs for potential, kinetic, internal, and nuclear energy are indicated as ovals  with transport processes and their directions marked with connecting arrows. The solid arrows correspond to major energy transport modes dependent on the flow dynamics while the dashed arrows indicate plausible advective redistribution of energy. The circular arrow within the internal reservoir indicates redistribution among its own plasma components. The non-reactive terms and processes match those discussed in \citet{Zhao+22}, while the nuclear energy generation, $\dot{Q}_\mathrm{E,nuc}$, is a new process which serves as a source of additional internal energy. That additional energy is subsequently distributed in the form of the buoyant energy, $P(\bm{\nabla \cdot v})_\mathrm{nuc}$, created by the thermal expansion of plasma. Ultimately, the added nuclear energy is responsible for injection of additional energy, $\epsilon^\mathrm{inj}_\mathrm{nuc}$, into turbulent motions.
    \label{f:energyTransfer}}
\end{figure}

These processes and their interactions are summarized schematically in Figure~\ref{f:energyTransfer},
which is an adaptation of the non-reactive RTI conceptual model \citep{Zhao+22}. In the figure, four energy reservoirs are represented by ovals: nuclear (top center, smaller), internal (top right), potential (left), and kinetic (bottom center, largest). The kinetic oval is divided by a vertical dashed line, with large filled circles on the left representing the energy-containing scales of the turbulent cascade and progressively smaller dotted circles on the right representing the dissipative range.

Connections fall into two groups distinguished by color. The black arrows reproduce the non-reactive RTI framework of \citet{Zhao+22}: $\epsilon^\mathrm{inj}$ transfers energy from the potential to the kinetic reservoir via buoyancy-driven injection; $\bar{P}\nabla\cdot\bar{u}$ couples the kinetic and internal reservoirs through pressure-dilatation work; $\Pi$ and $\Lambda$ represent pressure transport and dissipation within the cascade; and $D_\ell^\mathrm{inj}$ feeds energy from the internal reservoir into the kinetic reservoir at small scales. The dashed vertical lines dividing the potential and internal ovals indicate the presence of energy components absent from the non-reactive framework: a nuclear contribution to gravitational potential energy arising from flame thermal expansion, and a direct nuclear contribution to internal energy.

The blue arrows represent the new reactive additions. $\dot{Q}_\mathrm{E,nuc}$ transfers energy from the nuclear reservoir into the nuclear portion of the internal reservoir. $\bar{P}(\nabla\cdot\bar{u})_\mathrm{nuc}$ carries the buoyant work generated by flame thermal expansion into the nuclear portion of the potential reservoir, connecting nuclear burning directly to gravitational potential energy generation. $\epsilon^\mathrm{inj}_\mathrm{nuc}$ then transfers that additional potential energy into the kinetic reservoir, augmenting the standard buoyancy-driven injection. The blue circular arrow within the internal reservoir represents the quasi-isobaric redistribution of internal energy among plasma components. This redistribution is driven by the pressure deficit that develops in cold fuel expanded by the approaching flame: large-scale pressure gradients, communicated by weak acoustic waves, act to restore hydrostatic equilibrium and in doing so shift the internal energy balance from degenerate to thermal components --- the process directly responsible for the shortening of fuel ignition times identified in this work.
\section{Conclusions and future work}\label{s:conclusions}
We performed a computational study of RT-unstable thermonuclear deflagrations for conditions characteristic of Type Ia supernova explosions of centrally ignited massive white dwarfs. We dedicated our analysis of the simulation results to identifying the physics responsible for recently reported shortening of fuel ignition times. We used Lagrangian tracer particles to represent fuel parcels and recorded their thermodynamic histories as they are approached by the flame front.

The main findings of this project can be summarized as follows:
\begin{itemize}
\item Ignition times decrease systematically as the flame approaches the fuel, following an approximate power-law dependence on distance over the resolved transition region. This rate of decrease steepens strongly with fuel density, with ignition times reduced far more sharply in the higher-density models -- conceivably related to their stronger gravity, greater buoyancy, and consequently more strongly driven RTI.
\item The corresponding fuel temperature increases by about $13$\%, $24$\%, and $63$\% in the innermost reliably resolved bins. Evaluating Equation~\ref{e:igtm} with the measured thermodynamic state reproduces the ignition-time reduction, while the temperature-only reconstruction accounts for essentially all of it.
\item Lagrangian trajectories show that the most strongly preconditioned fuel is perturbed away from hydrostatic equilibrium by a modest near-isobaric density decrease, which acts as the initiating perturbation for the heating process. This perturbation is small, about $15$--$20$ per cent of the density in the extreme tail and only a few per cent for most strongly preconditioned tracers in the denser models. The restoration of hydrostatic pressure by weak acoustic perturbations redistributes internal energy from the degenerate to the thermal components, with degeneracy amplifying the resulting temperature rise. This energy redistribution, rather than the expansion itself, causes heating.
\item The result is a mesoscale fuel-preconditioning mechanism: flame-driven thermal expansion perturbs nearby unburned fuel away from hydrostatic equilibrium, and the acoustically mediated restoration of pressure balance preheats it before direct flame arrival. Because nuclear burning is disabled in the analyzed fuel, whether this preconditioning produces secondary ignition or contributes to DDT remains to be tested in simulations with burning enabled.
\item Mesoscale modeling is essential for capturing flame-fuel interactions and accounts for physics missed both in DNS flame models and large-scale SN Ia explosion simulations. Confined to domains far smaller than the buoyancy-driven deflagration structures, DNS flame models depend on prescribed upstream fuel conditions that only the mesoscale regime can supply.
\end{itemize}
The principal result of this study is that RT-unstable thermonuclear flames modify the thermodynamic state of nearby unburned fuel prior to flame arrival. Flame-driven expansion perturbs the fuel away from hydrostatic equilibrium, and the acoustically mediated restoration of pressure balance redistributes internal energy among its components. In consequence, fuel temperatures increase and ignition times decrease before direct flame interaction. This mesoscale fuel-preconditioning mechanism provides a physical explanation for the ignition-time shortening observed in our simulations. 

We stress that nuclear burning is disabled in the analyzed fuel and no ignition occurs in the present models; whether the identified preconditioning produces actual ignition events, and whether such events can contribute to DDT, are open questions that motivate the study outlined below.

The natural extension of this work is a hybrid model in which the nuclear burning source term is enabled in fuel-rich regions of the FSL. Such a study in 2D would provide initial insights into whether the identified preconditioning is sufficient to produce actual ignition events in fuel disconnected from the main flame. If the anticipated ignitions are observed, the proposed study would allow evaluation of the increase in flame unsteadiness and the strengthening of the acoustic background. These effects are expected to amplify the fuel preconditioning mechanism described here, potentially driving it toward a runaway. In this scenario, the resulting multi-site burning geometry will be defined by isolated deflagrating regions embedded in preconditioned fuel. The region may harbor conditions qualitatively different from those of a single propagating flame, and could produce the reactivity gradients required for DDT via the Zel'dovich mechanism \citep{Zeldovich80} under favorable conditions.

A concerted effect of nearby ignitions, each producing their own acoustic perturbation and communicated by acoustic waves, may have two distinct consequences. First, the waves could further lower ignition times in the surrounding fuel, potentially triggering its own ignition or, in the most extreme case, a detonation via the SWACER mechanism \citep{Lee+78,Lee+80,Khokhlov1991b,Brooker+21}. Because each ignition itself generates an acoustic perturbation, this effect can accumulate across multiple nearby ignitions: fuel originally too far from its own ignition threshold for preconditioning alone to reach may be pushed over that threshold by their combined contribution, with no single ignition needing to produce as strong a perturbation as an isolated event would require. Pre-existing acoustic perturbations or pressure gradients, whatever their initial strength, may therefore be sufficient to seed this cumulative process, since each ignition adds its own contribution as the process proceeds. Second, and independently, the accumulated acoustic energy could instead trigger oxygen detonations in the carbon ash. The turbulent post-flame region \citep{Brooker+25} is a natural candidate for this second path, as the TPF region --- defined as the separation between the carbon and oxygen flame fronts \citep{Woosley+11} --- increases in size with the advancing deflagration, potentially raising the probability of DDT. Positive results in 2D would justify continuation of the study in the full 3D configuration, where the interaction of flame-surface geometry, buoyancy-driven turbulence, and acoustic feedback can be captured self-consistently.

More broadly, the present results motivate further small-scale studies of flame--fuel interactions with self-heating enabled. Small-scale DNS such as that of \citet{Poludnenko+19}, which resolves the flame structure but not the mesoscale flame thermal expansion, is a natural complement to the mesoscale approach adopted here.
\begin{acknowledgments}
AZ is grateful to Art Bochevarov for providing computer hardware used for this work. This research used resources of the National Energy Research Scientific Computing Center, which is supported by the Office of Science of the U.S.\ Department of Energy under Contract No.\ DE-AC02-05CH11231. The software used in this work was in part developed by the DOE Flash Center at the University of Chicago. Data visualization was performed in part using VisIt \citep{HPV:VisIt}. This research has made use of NASA's Astrophysics Data System Bibliographic Services.
\end{acknowledgments}
\section*{Data availability}
The data underlying this article will be shared on reasonable request to the corresponding author.
\facilities{NERSC, ADS}

\software{\Proteus\ \citep{Brooker+25},
          \FLASH\ \citep{fryxell+00},
          \MATLAB\ \citep{matlab2022},
          \SciPy\ \citep{Virtanen+20}}
%
%
%
%
%
\appendix
\section{FLASH/Proteus Tracer Particle Module Verification}
\label{app:particle_verification}
Given the central role that passive tracer particles play in the analysis and conclusions developed in this paper, formal verification of the accuracy of the plasma properties sampled by tracer particles is a prerequisite for their use. As described in Section \ref{s:methods}, our implementation of the tracer particle module within \FLASH/\Proteus\ used for this Lagrangian analysis consists of two parts; the Heun's (a variant 2nd order Runge-Kutta) time integrator to advance the particles in time, and a multi-dimensional interpolation scheme to estimate the plasma properties at each particle location at each time step.

In this section, we outline a rigorous series of test problems featuring various dimensionalities and flow complexities (the verification test suite) in Section~\ref{app:testproblem_defs}, and then summarize the performance of our tracer particle module in a series of convergence tests across the problems in that test suite in Section~\ref{app:convergence} to demonstrate that our implementation of the passive tracer particle module matches the theoretically expected order of accuracy for both the time integrator and spatial interpolation subcomponents.
\subsection{Verification Test Suite}\label{app:testproblem_defs}
In order to verify the tracer particle module across dimensionalities and flow configurations, we developed a suite of 5 test problems, ranging in complexity from simple advection of smooth profiles in one dimension, to complex differentially rotating vortices in 2 and 3 dimensions. It is expected that this complexity share some of the essential characteristics of our turbulent field, namely the regions of strong vorticity and intense shear. The full series of problems that compose the test suite are summarized in Table \ref{t:testproblems}
\begin{deluxetable}{lccccc}[h]
\tablewidth{0pt}
\tablecaption{Test-problem summary.
\label{t:testproblems}}
\tablehead{
  \colhead{Test problem} &
  \colhead{Domain} &
  \colhead{$L$} &
  \colhead{$v_\mathrm{char}$} &
  \colhead{$t_\mathrm{max}$\tablenotemark{a}} 
}
\startdata
1D sinusoidal velocity field (SIN1D) & $[0,2\pi]$ & $2\pi$ & $1.000$ & $3.142$  \\[4pt]
2D isentropic vortex (IV2D) & $[-5,5]^2$ & $10$ & $0.7958$ & $6.283$  \\
2D Taylor--Green flow (TG2D) & $[0,2\pi]^2$ & $2\pi$ & $1.000$ & $3.142$  \\[4pt]
3D differentially-rotating vortex, tapered (IV3Ddr) & $[-5,5]^3$ & $10$ & $0.4615$ & $10.833$  \\
3D Taylor--Green flow (TG3D) & $[0,2\pi]^3$ & $2\pi$ & $2.000$ & $1.571$ \\
\enddata
\tablenotetext{a}{$50\%$ of $T_\mathrm{cross}=L/v_\mathrm{char}$ for each test problem, chosen from a dedicated trend analysis showing both temporal and spatial order plateau by this point for every test problem examined.}
\end{deluxetable}
and described in details below, with the spatial and temporal convergence results for each are discussed in the subsequent section. 
\subsubsection{1D sinusoidal velocity field (SIN1D).} A single-mode, time-independent field,
\begin{equation}\label{e:sin1d}
u(x) = u_0 \sin(k_{1d} x), \qquad u_0 = k_{1d} = 1,
\end{equation}
with $y$, $z$ fixed. The reference trajectory is obtained by direct integration of $\dot{x}=u(x)$, using an $\arctan$ branch-unwrapping correction.

\subsubsection{2D isentropic vortex (IV2D).} A steady, axisymmetric swirl centered at $(x_c, y_c) = (0,0)$,
\begin{equation}\label{e:vortex2d}
\begin{split}
u_\theta(r) &= \frac{\beta}{2\pi}\, r\, e^{(1-r^2)/2}\, \tau(r), \\
r &= \sqrt{(x-x_c)^2+(y-y_c)^2},
\end{split}
\end{equation}
$\beta=5$, with $\tau(r)$ a raised-cosine taper (exactly $1$ for $r \le R_2-W$, exactly $0$ for $r \ge R_2$):
\begin{equation}\label{e:taper}
\tau(r) = \begin{cases}
1 & r \le R_2 - W \\[4pt]
\frac{1}{2}\left[1+\cos\left(\pi\,\frac{r-(R_2-W)}{W}\right)\right] & R_2-W < r < R_2 \\[4pt]
0 & r \ge R_2,
\end{cases}
\end{equation}

$R_2$ set to half the domain size, $W=1$. The closed-form reference trajectory ($r(t)=r_0$, $\theta(t)=\theta_0+\Omega(r_0)(t-t_0)$, $\Omega=u_\theta/r$) uses the field's full angular speed, including both $\tau$ and the shared boundary-safety taper $T$ below.

\textbf{The boundary-safety taper, $T$.} Shared by IV2D, IV3Ddr, and any field needing a domain-edge margin: flat at $1$ out to $r_1$, a Gaussian shifted to peak at $r_1$ (zero-derivative there) and rescaled to reach exactly $0$ at $r_2$,
\begin{align}
T(r) &= \begin{cases}
1 & r \le r_1 \\[4pt]
\dfrac{e^{-(r-r_1)^2/2\sigma_T^2}-c}{1-c} & r_1 < r < r_2 \\[4pt]
0 & r \ge r_2,
\end{cases} \label{e:tapernew} \\
\sigma_T &= \frac{r_2-r_1}{\sqrt{-2\ln c}}, \nonumber
\end{align}
$c=10^{-4}$, $(r_1,r_2)$ expressed as margins from the domain's own half-width wherever this shared form is used (e.g., IV2D above); IV3Ddr's own taper uses a different functional form entirely, described below.

\subsubsection{3D differentially-rotating vortex, tapered (IV3Ddr).} A spherically-structured field built from three independent, non-concentric components:
\begin{enumerate}
\item genuine $v_z$: a clipped spherical Gaussian ``blob'' centered at $(x_v,y_v,z_v)$;
\item differential rotation: the swirl \emph{rate}, modulated by a $z$-only clipped Gaussian centered at $z_\Omega$;
\item a taper applied to the whole raw velocity vector, centered at $(x_\tau,y_\tau,z_\tau)$.
\end{enumerate}
using the shared clipped-Gaussian building block
\begin{equation}\label{e:gauss3ddr}
G(d) = \max\!\big(0,\ e^{-d^2/2\sigma^2} - c\big), \qquad \sigma=0.8,\ c=0.05,
\end{equation}
giving
\begin{equation}\label{e:vortex3ddr}
\begin{split}
u_\theta(r_\perp,z) &= \frac{\beta}{2\pi}\, r_\perp\, e^{(1-r_\perp^2)/2}\, \big[1+\epsilon_\Omega\,G(z-z_\Omega)\big], \\
v_z &= \frac{\beta}{2\pi}\, G\!\big(d_v\big), \\
d_v &= \sqrt{(x-x_v)^2+(y-y_v)^2+(z-z_v)^2}, \\
(v_x,v_y,v_z)\Big|_\mathrm{final} &= (v_x,v_y,v_z)\,T_\mathrm{3ddr}(d_\tau), \\
d_\tau &= \sqrt{(x-x_\tau)^2+(y-y_\tau)^2+(z-z_\tau)^2},
\end{split}
\end{equation}
with $\beta=1$, $\epsilon_\Omega=1$, $(x_v,y_v,z_v)=(-0.3,0.2,-0.1)$, $z_\Omega=0.3$, and $(x_\tau,y_\tau,z_\tau)=(0.1,-0.2,0.3)$. Unlike the shared boundary-safety taper above, $T_\mathrm{3ddr}$ uses a septic (seventh-order) polynomial smoothstep -- a different functional form, not a re-parameterization of the same one -- smooth (vanishing first through third derivatives) at both endpoints by construction:
\begin{equation}
T_\mathrm{3ddr}(r) = \begin{cases}
1 & r \le r_1 \\[4pt]
1 - 35\xi^4 + 84\xi^5 - 70\xi^6 + 20\xi^7 & r_1 < r < r_2 \\[4pt]
0 & r \ge r_2,
\end{cases}
\qquad \xi = \frac{r-r_1}{r_2-r_1}, \label{e:taper3ddr}
\end{equation}
flat to $r_1=3.0$, zero at $r_2=4.0$.

\begin{figure}
    \centering
    \includegraphics[width=0.465\textwidth]{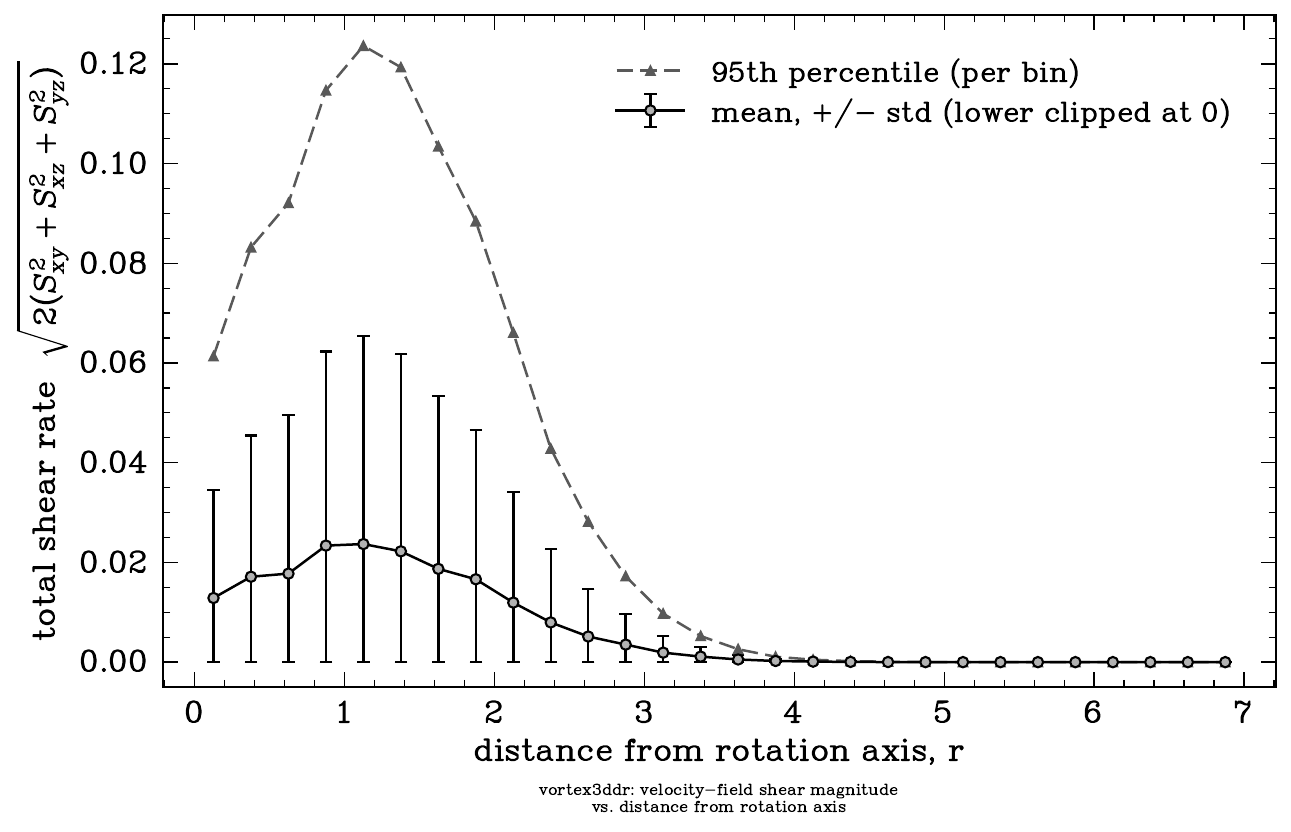}
    \figcaption{IV3Ddr test problem velocity-field shear magnitude (total shear rate of the strain-rate tensor, $\sqrt{2(S_{xy}^2+S_{xz}^2+S_{yz}^2)}$) as a function of distance from the field's own rotation axis, evaluated on $6\times10^4$ points sampled uniformly across the domain. Solid line: per-bin mean, with the lower error bar clipped at the physical bound of zero; dashed line: per-bin 95th percentile. The peak at $r\approx1$--$1.5$ coincides with the swirl profile's own characteristic radius.
    \label{f:shear_iv3ddr}}
\end{figure}

We note that no closed-form solution trajectory exists, and thus verification tests in the subsequent section use self-convergence, comparing every job's own final particle positions against a finer-resolution reference run. IV3Ddr's field is strongly spatially inhomogeneous, quantified via the strain-rate tensor $S_{ij}=\tfrac{1}{2}(\partial u_i/\partial x_j + \partial u_j/\partial x_i)$ -- off-diagonal components are shear strain rates. Figure~\ref{f:shear_iv3ddr}
shows the total shear rate as a function of distance from the field's rotation axis; it peaks at $r\approx1$--$1.5$, coincident with the swirl profile's own characteristic radius, and is negligible beyond $r\approx4$, which directly affects this test problem's own $L_\infty$ convergence behavior (Section~\ref{app:convergence}).
\subsubsection{3D Taylor--Green flow (TG3D).} The standard triply-periodic field:
\begin{align}
v_x &= g(t)\,U_0\,A\,\cos(kx)\sin(ky)\sin(kz), \nonumber \\
v_y &= g(t)\,U_0\,B\,\sin(kx)\cos(ky)\sin(kz), \label{e:tg3d} \\
v_z &= g(t)\,U_0\,C\,\sin(kx)\sin(ky)\cos(kz), \nonumber
\end{align}
$k=U_0=1$, $(A,B,C)=(1,1,-2)$ (for $A+B+C=0$, incompressibility). $g(t)$ is a modulation prefactor also referenced in Section~\ref{app:interp_isolation}; here $g(t)\equiv1$, a steady field. Once again, no closed-form solution  exists and our subsequent verification uses self-convergence, as for IV3Ddr above. Section~\ref{app:tg_norev} discusses this test problem's own spatial-order measurement.
\subsubsection{2D Taylor--Green flow (TG2D).}
The two-dimensional reduction of Equation~\ref{e:tg3d}:
\begin{align}
v_x &= g(t)\,U_0\,\cos(kx)\sin(ky), \nonumber \\
v_y &= -g(t)\,U_0\,\sin(kx)\cos(ky), \label{e:tg2d}
\end{align}
same $k$, $U_0$, $g(t)\equiv1$, restricted to $x,y$. As its exact orbit requires Jacobi elliptic functions, we make use of self-convergence like its 3D analog TG3D above. 
\subsection{Empirical Convergence Results}\label{app:convergence}
We present the results of the above defined test suite of problems as follows; first, in Section~\ref{app:interp_isolation} we assess the spatial mesh-to-particle interpolation scheme's accuracy  in a static, zero-time-evolution step and compare the interpolated velocity against each fields known analytic formula. This zero-time-evolution approach allows for the clear isolation of spatial error behavior from temporal and mixed terms, and serves as a baseline reference for time-evolving cases. Once this baseline behavior is established, we consider the full, time-evolving system in Section~\ref{app:full_time_verification}, measuring both temporal order $q$ (the integrator's own accuracy) and spatial order $p$ (the interpolator's own accuracy under actual particle advection) together.

As discussed in the problem descriptions in Section~\ref{app:testproblem_defs}, these time-dependent tests use one of two distinct reference mechanisms, depending on whether a closed-form solution exists for the field in question. SIN1D and IV2D admit an exact analytic trajectory and so the reference is that trajectory, evaluated directly, with no separate reference run needed. IV3Ddr, TG3D, and TG2D do not: their own dynamics are either genuinely non-integrable (IV3Ddr's three independent, non-concentric field components) or a textbook case of Lagrangian chaos in a steady Eulerian field (TG3D/TG2D), so no closed-form trajectory exists for any of them. These three test problems instead use self-convergence: a dedicated reference run, using a field-agnostic ODE integrator (adaptive 8th-order Runge--Kutta, tolerance $10^{-13}$) that solves the governing ODE directly and provides an effectively exact position at each recorded checkpoint, against which every mesh-based comparison job is measured.

\subsubsection{Zero-Time-Evolution Spatial Verification}
\label{app:interp_isolation}

For this spatial only verification, the velocity of the 8000 randomly located passive tracer particles as computed by the spatial interpolation routines calculation using the cell-averaged velocity values is compared with the analytically known value of the velocity at the particle position at $t=0s$. The particles were randomly binned into 20 ensembles and the $L_1$, $L_2$ and $L_\infty$ norms were computed for each ensembles. The mean norms and their standard deviation across the 20 independent particle ensembles, for the spatial order of convergence for each of the test problems for this verification test, are given in Table \ref{t:standalone_interp}.
\begin{deluxetable}{lcccccc}
\tablecaption{Standalone (zero-time-evolution) interpolation order $p$, by field, with monotonization off vs.\ on. Convergence order uncertainties correspond to a standard deviation across the 20 independent ensembles.
\label{t:standalone_interp}}
\tablehead{
  \colhead{Field} &
  \multicolumn{2}{c}{$L_1$} &
  \multicolumn{2}{c}{$L_2$} &
  \multicolumn{2}{c}{$L_\infty$} \\
  \colhead{} &
  \colhead{off} & \colhead{on} &
  \colhead{off} & \colhead{on} &
  \colhead{off} & \colhead{on}
}
\startdata
SIN1D & $3.00\pm0.00$ & $2.96\pm0.01$ & $2.99\pm0.00$ & $2.51\pm0.00$ & $2.98\pm0.01$ & $1.99\pm0.00$ \\[4pt]
IV2D & $2.82\pm0.03$ & $2.88\pm0.03$ & $2.69\pm0.03$ & $2.80\pm0.05$ & $2.29\pm0.08$ & $2.41\pm0.12$ \\
TG2D\tablenotemark{a} & $3.00\pm0.00$ & $3.25\pm0.01$ & $2.99\pm0.00$ & $2.92\pm0.02$ & $2.95\pm0.01$ & $1.98\pm0.01$ \\[4pt]
IV3Ddr & $2.61\pm0.05$ & $2.68\pm0.04$ & $2.47\pm0.05$ & $2.57\pm0.04$ & $2.15\pm0.04$ & $2.21\pm0.10$ \\
TG3D\tablenotemark{a} & $3.00\pm0.00$ & $3.24\pm0.04$ & $3.00\pm0.00$ & $3.18\pm0.03$ & $2.98\pm0.02$ & $2.22\pm0.09$ \\
\enddata
\tablecomments{Design value $p=3$ throughout. All five fields use the same six-resolution sweep as their own regular fixed-space tests.}
\tablenotetext{a}{TG2D/TG3D give an identical result regardless of the field's own reversal modulation $g(t)$ ($g(0)=1$ by construction, Equation~\ref{e:tg3d}, and this test never advances time).}
\end{deluxetable}
Unmonotonized, SIN1D and both Taylor--Green fields land within uncertainty of $p=3$ across all three norms. Their monotonized variants exhibit the expected degradation to order $p=2$ as derived in the literature \citep{Huynh+93}. Poorer performance is noted in  IV2D and IV3Ddr ($p\approx2.1$--$2.9$) under both monotonization settings. We trace this degraded performance to the strong anisotropy in velocity field sheer magnitude across differing distances from the fields rotational axes. Investigation of individual particle residuals reveals the largest errors present in regions with the highest shear rates. 
\subsubsection{Full Time-Dependent Spatial and Temporal Verification}
\label{app:full_time_verification}
For the verification of both spatial and temporal convergence in a time evolving system, we apply the same random binning and averaging technique of particles as in the zero-time-evolution case, but now allow the simulation to advance to $50\%$ of the the domains crossing time as documented Table \ref{t:testproblems}. Order estimates were estimated according to the composite error model 
\begin{equation}\label{e:composite}
E(\Delta\xi,\Delta t) = C_t\,\Delta t^{q} + C_s\,\Delta\xi^{p} + C_m\,\Delta t\,\Delta\xi^{p-1},
\end{equation}
where $q$ represents the observed order of accuracy of the particle time-advancing scheme $p$ represents the spatial order of accuracy of interpolation scheme. This composite error model additionally includes the leading-order coupling term. The temporal and spatial orders $q$ and $p$ are estimated in a decoupled fashion; $q$ is estimated by holding the spatial resolution, $\Delta \xi$, fixed and computing the selection of norms across $6$ values of the time-step, $\Delta t$, and likewise, $p$ is estimated by holding $\Delta t$ fixed and computing the norms across $6$ values of $\Delta \xi$. 

\begin{deluxetable}{lcccccc}
\tablewidth{0pt}
\tablecaption{Empirical values of temporal convergence order $q$ (pure $\Delta t$, fixed-mesh time-step sweep) and spatial convergence order $p$ (pure $\Delta\xi$, fixed-$\Delta t$ mesh-resolution sweep, the classical measurement) in all three norms for production (monotonized) configuration. Uncertainties denote the standard deviation across the 20 independent ensembles described in Section~\ref{app:interp_isolation}. Reference design values are $q=2$ and $p=3$.
\label{t:convergence}}
\tablehead{
  \colhead{Test problem} &
  \multicolumn{2}{c}{$L_1$} &
  \multicolumn{2}{c}{$L_2$} &
  \multicolumn{2}{c}{$L_\infty$} \\
  \colhead{} &
  \colhead{$q$} & \colhead{$p$} &
  \colhead{$q$} & \colhead{$p$} &
  \colhead{$q$} & \colhead{$p$}
}
\startdata
SIN1D & $1.94\pm0.01$ & $3.11\pm0.01$ & $1.89\pm0.02$ & $3.14\pm0.03$ & $1.80\pm0.05$ & $2.93\pm0.03$ \\[4pt]
IV2D & $2.07\pm0.01$ & $3.13\pm0.03$ & $2.09\pm0.02$ & $2.92\pm0.03$ & $2.03\pm0.03$ & $2.43\pm0.06$ \\
TG2D & $1.92\pm0.02$ & $3.57\pm0.03$ & $1.91\pm0.02$ & $3.33\pm0.02$ & $1.84\pm0.04$ & $2.92\pm0.04$ \\[4pt]
IV3Ddr & $1.92\pm0.02$ & $2.76\pm0.05$ & $1.92\pm0.02$ & $2.53\pm0.11$ & $1.90\pm0.02$ & $2.23\pm0.11$ \\
TG3D & $1.92\pm0.02$ & $3.68\pm0.07$ & $1.89\pm0.03$ & $3.54\pm0.03$ & $1.84\pm0.04$ & $2.93\pm0.06$ \\
\enddata
\end{deluxetable}
Table~\ref{t:convergence} summarizes the empirical order of convergence measured for each of five test problems, using the production configuration of the Runge-Kutta evolution and mesh-interpolation schemes. For each test problem, the temporal order $q$ is obtained from a mesh-independent, integrator-evolution time-step sweep against the field's exact analytic trajectory, or, when such trajectory is unavailable, a self-convergence reference. The spatial order $p$ is obtained from a classical, decoupled mesh-resolution sweep, $\Delta t$ held fixed, only $\Delta\xi$ varied. Both results are fit with an ordinary power law to independently estimate the exponents in the first two terms of the error model. This decoupled construction is used because a fixed-ratio $\Delta t/\Delta\xi$ diagonal sweep is subject to a known accumulated-error degeneracy that can suppress the fitted spatial order below its true, design value; \citet{KashiwabaraTakemura26} establish this rigorously for a related semi-Lagrangian interpolation scheme, and the note at the end of this section applies their result to this project's own scheme and confirms it empirically. For this reason, we do not measure or interpret the  leading order coupling term featuring $\Delta t \Delta \xi$, and defer further discussion of this term to the end of the this section. 

We note that $L_1$ is the most consistently well-behaved metric across all five test problems, with measured spatial orders of $2.76$--$3.68$, close to the nominal design value of 3. $L_2$ follows a similar, slightly more mixed pattern. $L_\infty$, by contrast, sits meaningfully below $L_1$/$L_2$ for two test problems specifically, IV2D and IV3Ddr. Investigation of this issue identified it as consistent with the behavior observed in the zero-time-evolution spatial verification test, where a collection of outlier particles associated with the peak velocity shear profile as depicted in Figure~\ref{f:shear_iv3ddr}. As the error associated with particles experiencing this peak shear is resolution-independent, they degrade the convergence behavior of the scheme when using outlier sensitive metrics such as the $L_\infty$ norm. 

A diagonal-sweep, monotonized-vs-unmonotonized comparison (SIN1D, IV2D, IV3Ddr) gives values overlapping within uncertainty for all three test problems, confirming monotonization is not a materially contributing factor to the diagonal-sweep spatial order for any of them; this comparison has not yet been repeated for the classical measurement reported as this table's own primary result.

\begin{figure*}
    \centering
    \includegraphics[width=\textwidth]{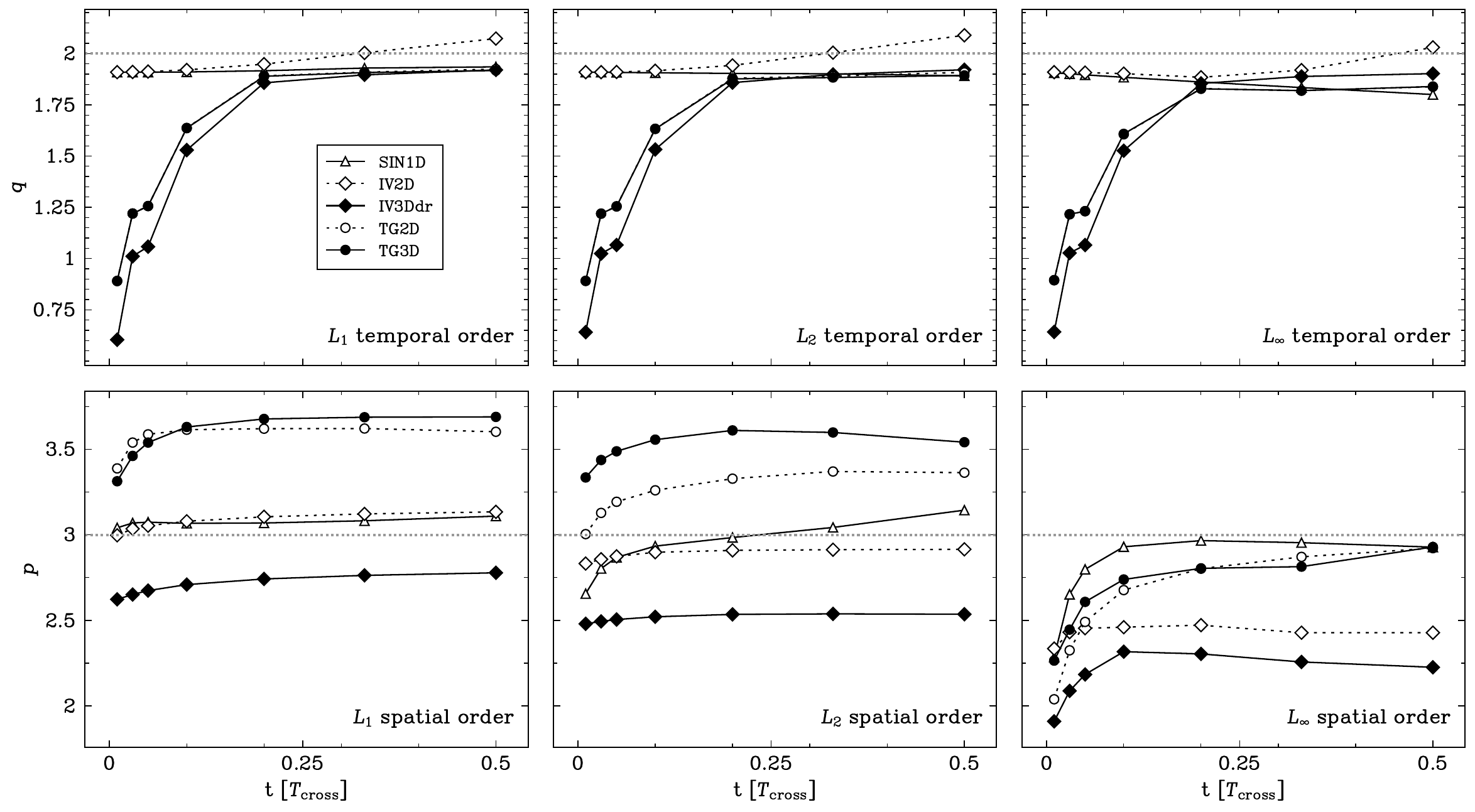}
    \figcaption{Empirical order of convergence vs.\ $t\,[T_\mathrm{cross}]$, all three norms (columns) and all five test problems (distinguished by marker/line style, see legend), for temporal order $q$ (top row) and spatial order $p$ (bottom row). Dashed horizontal lines mark the design values ($q=2$, $p=3$). SIN1D and IV2D (closed-form reference) are sampled at seven fractions from $1\%$ to $50\%$; IV3Ddr, TG3D, and TG2D (self-convergence reference) are sampled at seven checkpoints from $2\%$ to $100\%$ of their own $T_\mathrm{cross}$. TG2D and TG3D's own estimated temporal order is nearly identical at every sampled point (differing by at most $0.015$ across all three norms) and is not visually distinguishable in the top row; their markers coincide essentially exactly there, unlike the bottom row, where the two separate clearly.
    \label{f:grid_allnorms}}
\end{figure*}

To ensure our convergence tests properly sample the asymptotic regime of the test problems, we consider the dependence of the spatial and temporal order estimates according to our procedure above on the final evolution time of the system, as depicted in Figure~\ref{f:grid_allnorms}.
This figures shows the full order-vs-$t\,[T_\mathrm{cross}]$ trend behind Table~\ref{t:convergence}'s own $t_\mathrm{max}/T_\mathrm{cross}=50\%$ endpoint values: both temporal order $q$ (top row) and spatial order $p$ (bottom row) rise from a pre-asymptotic regime at small $t\,[T_\mathrm{cross}]$ toward a stable plateau, confirming $50\%$ sits within that plateau for every test problem and every norm. 
\subsubsection*{Note: TG3D/TG2D and the Diagonal-Sweep Spatial-Order Degeneracy}
\label{app:tg_norev}
TG3D and TG2D's own spatial order does not reach design under a fixed-ratio $\Delta t/\Delta\xi$ sweep, as noted above (Section~\ref{app:convergence}). This is explained by a known accumulated-error mechanism: \citet{KashiwabaraTakemura26} prove that such a sweep accumulates per-step interpolation error over $N=t_\mathrm{max}/\Delta t$ steps for a related semi-Lagrangian scheme, collapsing the fitted spatial order below the interpolant's true design value; applying the same accumulation argument to this project's own quadratic interpolant and RK2 integrator predicts exactly the one-unit shortfall observed here, and predicts recovery under a fixed-$\Delta t$, varying-$h$ sweep instead. This was tested directly: the classical sweep reported as Table~\ref{t:convergence}'s own primary measurement recovers design order in $L_1$/$L_2$ and essentially design order in $L_\infty$ for both test problems. Repeating both the classical and fixed-ratio measurements with monotonization disabled gives results unchanged at every tested resolution, confirming the effect is a property of the interpolant and sweep construction, not of monotonization.
\subsection{Summary}
\label{app:tracer_summary}
The present verification suite adopts a quantitative, test-problem-driven verification posture, applied to a block-structured AMR quadratic mapping rather than a structured logical-space Jacobian or a stochastic dispersion model -- with the important caveat, following \citet{Genel+13}, that this verification establishes integrator-plus-interpolation accuracy against a known field, not fidelity of velocity-field tracers to true Lagrangian mass transport in a discontinuous flow.

The empirical results of Section~\ref{app:convergence} (Table~\ref{t:convergence}) show this pipeline converging to a well-defined order for every test problem: temporal order close to design throughout, and an $L_1$ spatial order of $2.76$--$3.68$, close to the design value of 3, for every test problem. $L_\infty$ shows two mechanistically explained shortfalls (IV2D, IV3Ddr), each traced to a physically grounded cause.
\section{Local-Slope Diagnostic for the Ignition-Time--Distance Power-Law Fit}
\label{app:localslope}
We characterize a dominant trend relating ignition time and the flame front distance (cf.\ Section~\ref{s:results}) using a power-law fit, with the fit window for the $\gamma$ exponent chosen from a local-slope diagnostic, $\mathrm{d}\log_{10}\tau/\mathrm{d}\log_{10}d_\mathrm{f}$, computed as the numerical gradient of the binned mean $\log_{10}\tau$ with respect to $\log_{10}d_\mathrm{f}$ across the same 50 log-uniform bins used for the fit itself. A window restricting the fit region is necessary in order to eliminate a possible bias due to proximity of tracer particles to the near-front region where mesh-to-particle interpolation stencils extend beyond the flame front proper ($\phi = 3 \times 10^{-2}$, Section~\ref{s:methods}).

\begin{figure*}
    \centering
    \begin{tabular}{ccc}
        \includegraphics[height=0.265\textwidth]{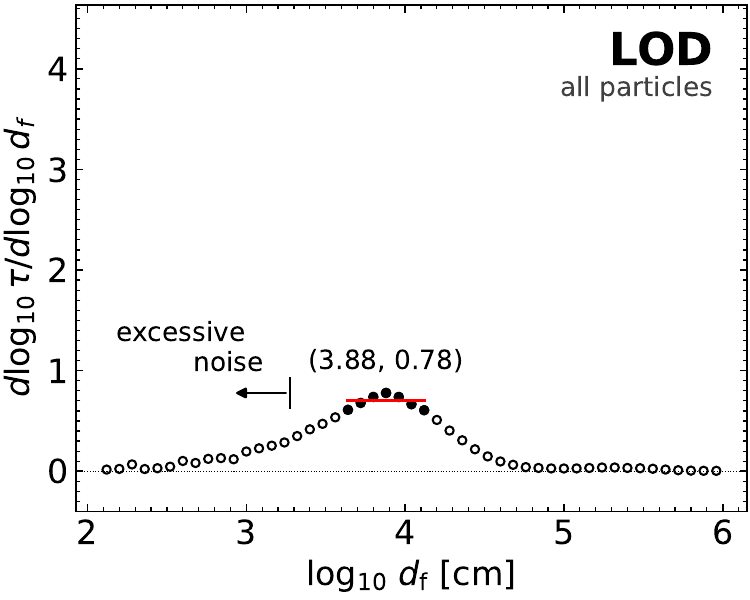}   &
        \includegraphics[height=0.265\textwidth]{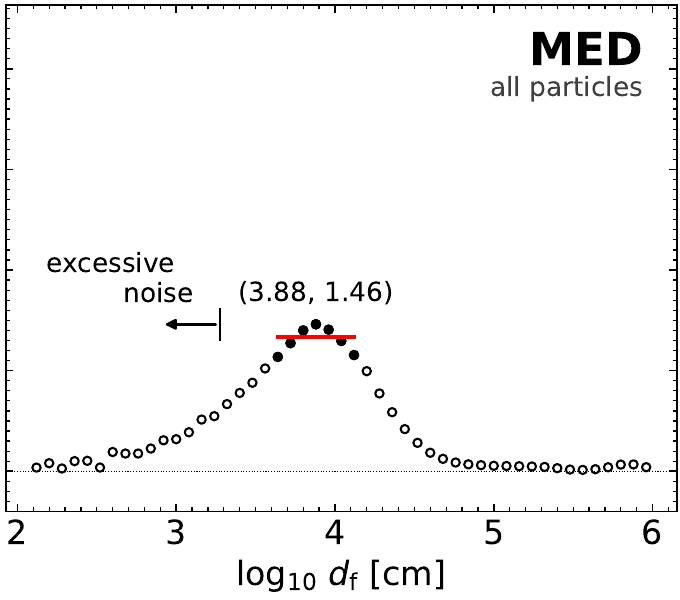} &
        \includegraphics[height=0.265\textwidth]{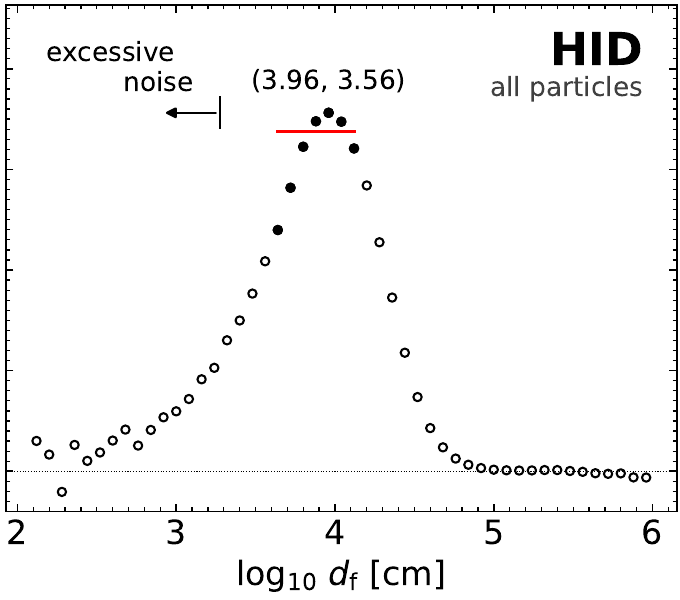}
    \end{tabular}
    \figcaption{Local-slope diagnostic ($\mathrm{d}\log_{10}\tau/\mathrm{d}\log_{10}d_\mathrm{f}$ vs.\ $\log_{10}d_\mathrm{f}$) used to choose the $\gamma$-fit window (Section~\ref{s:results}), for the LOD, MED, and HID models (left to right). Open circles: all bins. Filled disks: bins inside the confirmed fit window. Red line: the fitted $\gamma$, spanning the fit window. Labeled point: each model's own peak, wherever it falls. Arrow: the near-front region excluded from the fit (cf.\ Section~\ref{s:results}).
    \label{f:localslope}}
\end{figure*}

Figure~\ref{f:localslope}
shows this diagnostic for all three models. Open circles mark every distance bin's own local slope; filled disks mark the bins actually used in the $\gamma$ fit (the confirmed $3.64\leq\log_{10}d_\mathrm{f}\leq4.12$ window). The horizontal red line, spanning the fit window at $y=\gamma$, shows directly how well a single, constant exponent describes the curve over that range -- on this diagnostic's own axis, a true power law is by definition flat. Each model's actual peak is marked and labeled (LOD: $3.88$~dex, $0.78$; MED: $3.88$~dex, $1.46$; HID: $3.96$~dex, $3.56$), wherever it falls, not restricted to the fit window. The near-front region flagged as unsuitable for fitting (Section~\ref{s:results}) is marked separately.

In all three models, the fit window clearly brackets a real, well-defined peak in the local-slope curve -- rising into it and, for LOD and MED, falling away from it again within the same window -- rather than sitting in a flat, noise-dominated plateau. The window is visibly separated from the flagged near-front region by many bins in every model. The bins defining $\gamma$ are therefore free from a potential near-front interpolation bias, and the fit window tracks real curve structure. HID's peak is visibly narrower and steeper than LOD's or MED's broad, gentle peaks, consistent with the local-slope diagnostic already described in Section~\ref{s:results}.
\bibliographystyle{aasjournal}
\bibliography{references}

@article{Bak2010,
  author        = {{Bak}, Stanley and {McLaughlin}, Joyce and {Renzi}, Daniel},
  title         = {{Some Improvements for the Fast Sweeping Method}},
  journal       = {SIAM Journal on Scientific Computing},
  volume        = {32},
  number        = {5},
  pages         = {2853--2874},
  year          = {2010},
  doi           = {10.1137/090749645}
}

@article{Brooker+21,
  author        = {{Brooker}, E. and {Plewa}, T. and {Fenn}, D.},
  title         = {{Type Ia supernovae deflagration-to-detonation transition explosions powered by the Zel'dovich reactivity gradient mechanism}},
  journal       = {\mnras},
  volume        = {501},
  number        = {1},
  pages         = {L23--L27},
  year          = {2021},
  doi           = {10.1093/mnrasl/slaa141},
  adsurl        = {https://ui.adsabs.harvard.edu/abs/2021MNRAS.501L..23B}
}

@article{Brooker+25,
  author        = {Brooker, E and Zhiglo, A and Plewa, T},
  title         = {Mesoscale turbulence in type ia supernova deflagrations: buoyancy-driven fuel heating and prospects for delayed-detonations},
  journal       = {\mnras},
  volume        = {543},
  number        = {4},
  pages         = {3486--3502},
  year          = {2025},
  doi           = {10.1093/mnras/staf1683},
  eprint        = {https://academic.oup.com/mnras/article-pdf/543/4/3486/64486894/staf1683.pdf}
}

@incollection{HPV:VisIt,
  author        = {Hank Childs and Eric Brugger and Brad Whitlock and Jeremy Meredith and Sean Ahern and David Pugmire and Kathleen Biagas and Mark Miller and Cyrus Harrison and Gunther H. Weber and Hari Krishnan and Thomas Fogal and Allen Sanderson and Christoph Garth and E. Wes Bethel and David Camp and Oliver R\"{u}bel and Marc Durant and Jean M. Favre and Paul Navr\'{a}til},
  editor        = {E. Wes Bethel and Hank Childs and Charles Hansen},
  title         = {VisIt: An End-User Tool For Visualizing and Analyzing Very Large Data},
  booktitle     = {High Performance Visualization--Enabling Extreme-Scale Scientific Insight},
  pages         = {357},
  year          = {2012},
  publisher     = {CRC Press/Francis--Taylor Group},
  address       = {Boca Raton}
}

@article{CiaraldiSchoolmann+13,
  author        = {{Ciaraldi-Schoolmann}, F. and {Seitenzahl}, I.~R. and {R{\"o}pke}, F.~K.},
  title         = {{A subgrid-scale model for deflagration-to-detonation transitions in Type Ia supernova explosion simulations. Numerical implementation}},
  journal       = {\aap},
  volume        = {559},
  pages         = {A117},
  year          = {2013},
  doi           = {10.1051/0004-6361/201321480},
  eprint        = {1307.8146},
  archivePrefix = {arXiv},
  primaryClass  = {astro-ph.SR},
  adsurl        = {https://ui.adsabs.harvard.edu/abs/2013A&A...559A.117C}
}

@article{Clavin25,
  author        = {{Clavin}, Paul},
  title         = {{Physics of the transition to detonation of gaseous flames}},
  journal       = {European Physical Journal Plus},
  volume        = {140},
  number        = {3},
  pages         = {258},
  year          = {2025},
  doi           = {10.1140/epjp/s13360-025-06192-5},
  adsurl        = {https://ui.adsabs.harvard.edu/abs/2025EPJP..140..258C}
}

@book{Clayton83,
  author        = {{Clayton}, D. D.},
  title         = {Principles of stellar evolution and nucleosynthesis},
  year          = {1983},
  publisher     = {University of Chicago Press},
  address       = {Chicago},
  adsurl        = {https://ui.adsabs.harvard.edu/abs/1983psen.book.....C}
}

@book{DavisonHinkley1997,
  author        = {Davison, A. C. and Hinkley, D. V.},
  title         = {Bootstrap Methods and their Application},
  year          = {1997},
  publisher     = {Cambridge University Press},
  address       = {Cambridge},
  isbn          = {9780521574716}
}

@article{Dubey+12,
  author        = {{Dubey}, A. and {Daley}, C. and {ZuHone}, J. and {Ricker}, P.~M. and {Weide}, K. and {Graziani}, C.},
  title         = {{Imposing a Lagrangian Particle Framework on an Eulerian Hydrodynamics Infrastructure in FLASH}},
  journal       = {\apjs},
  volume        = {201},
  number        = {2},
  pages         = {27},
  year          = {2012},
  doi           = {10.1088/0067-0049/201/2/27},
  adsurl        = {https://ui.adsabs.harvard.edu/abs/2012ApJS..201...27D}
}

@article{Dunn1961,
  author        = {{Dunn}, Olive Jean},
  title         = {{Multiple Comparisons Among Means}},
  journal       = {Journal of the American Statistical Association},
  volume        = {56},
  number        = {293},
  pages         = {52--64},
  year          = {1961},
  doi           = {10.1080/01621459.1961.10482090}
}

@article{DursiTimmes2006,
  author        = {{Dursi}, L. Jonathan and {Timmes}, F.~X.},
  title         = {{Local Ignition in Carbon-Oxygen White Dwarfs. I. One-Zone Ignition and Spherical Shock Ignition of Detonations}},
  journal       = {\apj},
  volume        = {641},
  number        = {2},
  pages         = {1071--1086},
  year          = {2006},
  doi           = {10.1086/500638},
  eprint        = {astro-ph/0510367},
  archivePrefix = {arXiv},
  primaryClass  = {astro-ph},
  adsurl        = {https://ui.adsabs.harvard.edu/abs/2006ApJ...641.1071D}
}

@article{fryxell+00,
  author        = {{Fryxell}, B. and {Olson}, K. and {Ricker}, P. M. and {Timmes}, F. X. and {Zingale}, M. and {Lamb}, D. Q. and {MacNeice}, P. and {Rosner}, R. and {Truran}, J. W. and {Tufo}, H.},
  title         = {FLASH: An Adaptive Mesh Hydrodynamics Code for Modeling Astrophysical Thermonuclear Flashes},
  journal       = {\apjs},
  volume        = {131},
  pages         = {273--334},
  year          = {2000}
}

@article{Genel+13,
  author        = {{Genel}, S. and {Vogelsberger}, M. and {Nelson}, D. and {Sijacki}, D. and {Springel}, V. and {Hernquist}, L.},
  title         = {{Following the flow: tracer particles in astrophysical fluid simulations}},
  journal       = {\mnras},
  volume        = {435},
  pages         = {1426--1442},
  year          = {2013},
  doi           = {10.1093/mnras/stt1383},
  adsurl        = {https://ui.adsabs.harvard.edu/abs/2013MNRAS.435.1426G}
}

@article{GotelliMcCabe2002,
  author        = {Gotelli, Nicholas J. and McCabe, Darren J.},
  title         = {Species Co-Occurrence: A Meta-Analysis of {J. M. Diamond's} Assembly Rules Model},
  journal       = {Ecology},
  volume        = {83},
  number        = {8},
  pages         = {2091--2096},
  year          = {2002},
  doi           = {10.1890/0012-9658(2002)083[2091:SCOAMA]2.0.CO;2}
}

@article{Huynh+93,
  author        = {{Huynh}, H. T.},
  title         = {{Accurate Monotone Cubic Interpolation}},
  journal       = {SIAM Journal on Numerical Analysis},
  volume        = {30},
  number        = {1},
  pages         = {57--100},
  year          = {1993},
  doi           = {10.1137/0730004}
}

@article{Karlin11,
  author        = {{Karlin}, Vladimir},
  title         = {{Radiation Preheating can Trigger Transition from Deflagration to Detonation}},
  journal       = {Flow, Turbulence and Combustion},
  volume        = {87},
  number        = {2-3},
  pages         = {511--523},
  year          = {2011},
  doi           = {10.1007/s10494-010-9317-9},
  eprint        = {1003.5888},
  archivePrefix = {arXiv},
  primaryClass  = {physics.flu-dyn},
  adsurl        = {https://ui.adsabs.harvard.edu/abs/2011FTC....87..511K}
}

@article{KashiwabaraTakemura26,
  author        = {{Kashiwabara}, Takahito and {Takemura}, Haruki},
  title         = {{Error estimates of the cubic interpolated pseudo-particle scheme for one-dimensional advection equations}},
  journal       = {Japan Journal of Industrial and Applied Mathematics},
  volume        = {43},
  number        = {9},
  year          = {2026},
  doi           = {10.1007/s13160-025-00750-1},
  eprint        = {2402.11885},
  archivePrefix = {arXiv}
}

@article{Khokhlov1991a,
  author        = {{Khokhlov}, A.~M.},
  title         = {{Delayed detonation model for type IA supernovae}},
  journal       = {\aap},
  volume        = {245},
  number        = {1},
  pages         = {114--128},
  year          = {1991},
  adsurl        = {https://ui.adsabs.harvard.edu/abs/1991A&A...245..114K}
}

@article{Khokhlov1991b,
  author        = {{Khokhlov}, A.~M.},
  title         = {{Mechanisms for the initiation of detonations in the degenerate matter of supernovae}},
  journal       = {\aap},
  volume        = {246},
  number        = {2},
  pages         = {383--396},
  year          = {1991},
  adsurl        = {https://ui.adsabs.harvard.edu/abs/1991A&A...246..383K}
}

@article{Khokhlov1995,
  author        = {{Khokhlov}, Alexei M.},
  title         = {{Propagation of Turbulent Flames in Supernovae}},
  journal       = {\apj},
  volume        = {449},
  pages         = {695},
  year          = {1995},
  doi           = {10.1086/176091},
  adsurl        = {https://ui.adsabs.harvard.edu/abs/1995ApJ...449..695K}
}

@article{Khokhlov1997b,
  author        = {{Khokhlov}, A.~M. and {Oran}, E.~S. and {Wheeler}, J.~C.},
  title         = {{Deflagration-to-Detonation Transition in Thermonuclear Supernovae}},
  journal       = {\apj},
  volume        = {478},
  number        = {2},
  pages         = {678--688},
  year          = {1997},
  doi           = {10.1086/303815},
  eprint        = {astro-ph/9612226},
  archivePrefix = {arXiv},
  primaryClass  = {astro-ph},
  adsurl        = {https://ui.adsabs.harvard.edu/abs/1997ApJ...478..678K}
}

@article{Lee+78,
  author        = {{Lee}, J.~H. and {Knystautas}, R. and {Yoshikawa}, N.},
  title         = {{Photochemical initiation of gaseous detonations}},
  journal       = {Acta Astronautica},
  volume        = {5},
  number        = {11},
  pages         = {971--982},
  year          = {1978},
  doi           = {10.1016/0094-5765(78)90003-6},
  adsurl        = {https://ui.adsabs.harvard.edu/abs/1978AcAau...5..971L}
}

@article{Lee+80,
  author        = {{Lee}, J.~H.~S. and {Moen}, I.~O.},
  title         = {{The mechanism of transition from deflagration to detonation in vapor cloud explosions}},
  journal       = {Progress in Energy and Combustion Science},
  volume        = {6},
  pages         = {359--389},
  year          = {1980},
  doi           = {10.1016/0360-1285(80)90011-8}
}

@article{Liberman+10,
  author        = {{Liberman}, M.~A. and {Ivanov}, M.~F. and {Kiverin}, A.~D. and {Kuznetsov}, M.~S. and {Chukalovsky}, A.~A. and {Rakhimova}, T.~V.},
  title         = {{Deflagration-to-detonation transition in highly reactive combustible mixtures}},
  journal       = {Acta Astronautica},
  volume        = {67},
  number        = {7},
  pages         = {688--701},
  year          = {2010},
  doi           = {10.1016/j.actaastro.2010.05.024},
  adsurl        = {https://ui.adsabs.harvard.edu/abs/2010AcAau..67..688L}
}

@article{NorthCurtisSham2002,
  author        = {North, Bernard V. and Curtis, David and Sham, Pak C.},
  title         = {A Note on the Calculation of Empirical {$P$} Values from {Monte Carlo} Procedures},
  journal       = {American Journal of Human Genetics},
  volume        = {71},
  number        = {2},
  pages         = {439--441},
  year          = {2002}
}

@article{Poludnenko+19,
  author        = {Poludnenko, Alexei Y. and Chambers, Jessica and Ahmed, Kareem and Gamezo, Vadim N. and Taylor, Brian D.},
  title         = {A unified mechanism for unconfined deflagration-to-detonation transition in terrestrial chemical systems and type Ia supernovae},
  journal       = {Science},
  volume        = {366},
  number        = {6465},
  pages         = {eaau7365},
  year          = {2019},
  doi           = {10.1126/science.aau7365}
}

@article{Roy+23,
  author        = {{Roy}, Amitesh and {Picardo}, Jason R. and {Emerson}, Benjamin and {Lieuwen}, Tim C. and {Sujith}, R.~I.},
  title         = {{Small-scale intermittency of premixed turbulent flames}},
  journal       = {Journal of Fluid Mechanics},
  volume        = {957},
  pages         = {A21},
  year          = {2023},
  doi           = {10.1017/jfm.2023.63},
  eprint        = {2202.12066},
  archivePrefix = {arXiv},
  primaryClass  = {physics.flu-dyn},
  adsurl        = {https://ui.adsabs.harvard.edu/abs/2023JFM...957A..21R}
}

@article{Sabelnikov+22,
  author        = {{Sabelnikov}, Vladimir A. and {Lipatnikov}, Andrei N. and {Nikitin}, Nikolay V. and {Hern{\'a}ndez-P{\'e}rez}, Francisco E. and {Im}, Hong G.},
  title         = {{Effects of thermal expansion on moderately intense turbulence in premixed flames}},
  journal       = {Physics of Fluids},
  volume        = {34},
  number        = {11},
  pages         = {115127},
  year          = {2022},
  doi           = {10.1063/5.0123211},
  adsurl        = {https://ui.adsabs.harvard.edu/abs/2022PhFl...34k5127S}
}

@article{Sabelnikov+19,
  author        = {Sabelnikov, V. A. and Lipatnikov, A. N. and Nishiki, S. and Hasegawa, T.},
  title         = {Investigation of the influence of combustion-induced thermal expansion on two-point turbulence statistics using conditioned structure functions},
  journal       = {Journal of Fluid Mechanics},
  volume        = {867},
  pages         = {45},
  year          = {2019},
  doi           = {10.1017/jfm.2019.128}
}

@article{Sethian1996,
  author        = {J A Sethian},
  title         = {A fast marching level set method for monotonically advancing fronts.},
  journal       = {Proceedings of the National Academy of Sciences},
  volume        = {93},
  number        = {4},
  pages         = {1591--1595},
  year          = {1996},
  doi           = {10.1073/pnas.93.4.1591},
  eprint        = {https://www.pnas.org/doi/pdf/10.1073/pnas.93.4.1591}
}

@misc{matlab2022,
  author        = {{The MathWorks Inc.}},
  title         = {{MATLAB} version 9.12 (R2022a)},
  year          = {2022},
  publisher     = {The MathWorks Inc.},
  address       = {Natick, Massachusetts},
  url           = {https://www.mathworks.com}
}

@article{TimmesSwesty2000,
  author        = {{Timmes}, F.~X. and {Swesty}, F. Douglas},
  title         = {{The Accuracy, Consistency, and Speed of an Electron-Positron Equation of State Based on Table Interpolation of the Helmholtz Free Energy}},
  journal       = {\apjs},
  volume        = {126},
  number        = {2},
  pages         = {501--516},
  year          = {2000},
  doi           = {10.1086/313304},
  adsurl        = {https://ui.adsabs.harvard.edu/abs/2000ApJS..126..501T}
}

@article{Virtanen+20,
  author        = {Virtanen, Pauli and Gommers, Ralf and Oliphant, Travis E. and Haberland, Matt and Reddy, Tyler and Cournapeau, David and Burovski, Evgeni and Peterson, Pearu and Weckesser, Warren and Bright, Jonathan and {van der Walt}, St{\'e}fan J. and Brett, Matthew and Wilson, Joshua and Millman, K. Jarrod and Mayorov, Nikolay and Nelson, Andrew R. J. and Jones, Eric and Kern, Robert and Larson, Eric and Carey, C J and Polat, {\.I}lhan and Feng, Yu and Moore, Eric W. and {VanderPlas}, Jake and Laxalde, Denis and Perktold, Josef and Cimrman, Robert and Henriksen, Ian and Quintero, E. A. and Harris, Charles R. and Archibald, Anne M. and Ribeiro, Ant{\^o}nio H. and Pedregosa, Fabian and {van Mulbregt}, Paul and {SciPy 1.0 Contributors}},
  title         = {{SciPy} 1.0: Fundamental Algorithms for Scientific Computing in {Python}},
  journal       = {Nature Methods},
  volume        = {17},
  pages         = {261--272},
  year          = {2020},
  doi           = {10.1038/s41592-019-0686-2}
}

@book{Weiss+04,
  author        = {{Weiss}, A. and {Hillebrandt}, W. and {Thomas}, H.-C. and {Ritter}, H.},
  title         = {{Cox and Giuli's Principles of Stellar Structure}},
  year          = {2004},
  publisher     = {Cambridge Scientific Publishers},
  address       = {Cambridge},
  adsurl        = {https://ui.adsabs.harvard.edu/abs/2004cgps.book.....W}
}

@article{Woosley+11,
  author        = {{Woosley}, S.~E. and {Kerstein}, A.~R. and {Aspden}, A.~J.},
  title         = {{Flames in Type Ia Supernova: Deflagration-detonation Transition in the Oxygen-burning Flame}},
  journal       = {\apj},
  volume        = {734},
  number        = {1},
  pages         = {37},
  year          = {2011},
  doi           = {10.1088/0004-637X/734/1/37},
  adsurl        = {https://ui.adsabs.harvard.edu/abs/2011ApJ...734...37W}
}

@article{Zeldovich80,
  author        = {{Zeldovich}, Ya. B.},
  title         = {{Regime classification of an exothermic reaction with nonuniform initial conditions}},
  journal       = {Combustion and Flame},
  volume        = {39},
  number        = {2},
  pages         = {211--214},
  year          = {1980},
  doi           = {10.1016/0010-2180(80)90017-6},
  adsurl        = {https://ui.adsabs.harvard.edu/abs/1980CoFl...39..211Z}
}

@article{Zeldovich+70,
  author        = {Zel'dovich, Ya. B. and Librovich, V. B. and Makhviladze, G. M. and Sivashinskil, G. I.},
  title         = {On the onset of detonation in a nonuniformly heated gas},
  journal       = {J. Appl. Mech. Tech. Phys.},
  volume        = {11},
  number        = {2},
  pages         = {264--270},
  year          = {1970},
  doi           = {10.1007/BF00908106}
}

@article{Zhao2004,
  author        = {Zhao, Hongkai},
  title         = {A fast sweeping method for Eikonal equations},
  journal       = {Mathematics of Computation},
  volume        = {74},
  number        = {250},
  pages         = {603--627},
  year          = {2004}
}

@article{Zhao+22,
  author        = {{Zhao}, Dongxiao and {Betti}, Riccardo and {Aluie}, Hussein},
  title         = {{Scale interactions and anisotropy in Rayleigh-Taylor turbulence}},
  journal       = {Journal of Fluid Mechanics},
  volume        = {930},
  pages         = {A29},
  year          = {2022},
  doi           = {10.1017/jfm.2021.902},
  eprint        = {2006.04301},
  archivePrefix = {arXiv},
  primaryClass  = {physics.flu-dyn},
  adsurl        = {https://ui.adsabs.harvard.edu/abs/2022JFM...930A..29Z}
}

@article{Zhiglo07,
  author        = {{Zhiglo}, Andrey V.},
  title         = {{Flame-capturing Technique. I. Adaptation to Gas Expansion}},
  journal       = {\apjs},
  volume        = {169},
  number        = {2},
  pages         = {386--400},
  year          = {2007},
  doi           = {10.1086/511525},
  adsurl        = {https://ui.adsabs.harvard.edu/abs/2007ApJS..169..386Z}
}
\end{document}